# Roadmap on Quantum Magnetic Materials

Antonija Grubišić-Čabo[1], Marcos H. D. Guimarães[1], Dmytro Afanasiev[2], Jose H. Garcia Aguilar[3], Irene Aguilera[4, 5], Mazhar N. Ali[6], Semonti Bhattacharyya[7], Yaroslav M. Blanter[6], Rixt Bosma[1], Zhiyuan Cheng[7], Zhiying Dan[1], Saroj P. Dash[8], Joaquín Medina Dueñas[3, 9], Joaquín Fernandez-Rossier[10], Marco Gibertini[11], Sergii Grytsiuk[2], Maurits J. A. Houmes[6], Anna Isaeva[4, 12], Chrystalla Knekna[1,4], Arnold H. Kole[5, 13], Samer Kurdi[1,14], Jose Lado[15], Samuel Mañas-Valero[6], J. Marcelo J. Lopes[16], Damiano Marian[17], Mengxing Na[2], Falk Pabst[4], Sergio Barquero Pierantoni[4], Mexx Regout[18], Riccardo Reho[5, 13], Malte Rösner[2], David Sanz[20], Toeno van der Sar[6], Jagoda Sławińska[1], Matthieu J. Verstraete[5, 18, 19], Muhammad Waseem[1], Herre S. J. van der Zant[6], Zeila Zanolli[5, 13], David Soriano[20]

1. Zernike Institute for Advanced Materials, University of Groningen, Groningen, The Netherlands
2. Institute for Molecules and Materials, Radboud University, The Netherlands
3. Institut Català de Nanociència i Nanotecnologia, Spain
4. University of Amsterdam, The Netherlands
5. European Theoretical Spectroscopy Facility (ETSF)
6. Kavli Institute of Nanoscience, Delft University of Technology, The Netherlands and Department of Quantum Nanoscience, Faculty of Applied Sciences, Delft University of Technology, The Netherlands
7. Huygens-Kamerlingh Onnes Laboratory, Leiden University, The Netherlands
8. Department of Microtechnology and Nanoscience, Chalmers University of Technology, Sweden
9. Universitat Autònoma de Barcelona, Spain
10. International Iberian Nanotechnology Laboratory, Portugal
11. University of Modena and Reggio Emilia, Italy
12. Technical University of Dortmund, Germany
13. Chemistry Department, Debye Institute for Nanomaterials Science, Condensed Matter and Interfaces, Utrecht University, The Netherlands.
14. Institute of Photonics and Quantum Sciences, SUPA, Heriot-Watt University, Edinburgh, United Kingdom
15. Department of Applied Physics, Aalto University, Finland
16. Paul-Drude-Institut für Festkörperelektronik, Germany
17. Università di Pisa, Italy
18. NanoMat/Q-Mat Université de Liège, and European Theoretical Spectroscopy Facility, Belgium
19. ITP, Physics Department, Utrecht University, The Netherlands
20. Departamento de Física Aplicada, Universidad de Alicante, Spain

# Table of Contents

# 1. Introduction

Marcos H. D. Guimarães[1], Antonija Grubišić-Čabo[1], Zeila Zanolli[2], David Soriano[3]
[1] Zernike Institute for Advanced Materials, University of Groningen, Groningen, The Netherlands
[2] Chemistry Department, Debye Institute for Nanomaterials Science and ETSF, Condensed Matter and Interfaces, Utrecht University, PO Box 80.000, 3508 TA Utrecht, The Netherlands.
[3] Departamento de Física Aplicada, Universidad de Alicante, Spain

**Abstract**

Fundamental research on two-dimensional (2D) magnetic systems based on van der Waals materials has been gaining traction rapidly since their recent discovery. With the increase of recent knowledge, it has become clear that such materials have also a strong potential for applications in devices that combine magnetism with electronics, optics, and nanomechanics. Nonetheless, many challenges still lay ahead. Several fundamental aspects of 2D magnetic materials are still unknown or poorly understood, such as their often-complicated electronic structure, optical properties, and magnetization dynamics, and their magnon spectrum. To elucidate their properties and facilitate integration in devices, advanced characterization techniques and theoretical frameworks need to be developed or adapted. Moreover, developing synthesis methods which increase critical temperatures and achieve large-scale, high-quality homogeneous thin films is crucial before these materials can be used for real-world applications. Therefore, the field of 2D magnetic materials provides many challenges and opportunities for the discovery and exploration of new phenomena, as well as the development of new applications. This Roadmap presents the background, challenges, and potential research directions for various relevant topics in the field on the fundamentals, synthesis, characterization, and applications. We hope that this work can provide a strong starting point for young researchers in the field and provide a general overview of the key challenges for more experienced researchers.

**Introduction**

Magnetic van der Waals (vdW) materials are one of the newest members of the two-dimensional (2D) materials family. While 2D insulators, semiconductors, (semi)metals, and superconductors have been around for over one decade [1], 2D magnetic materials are only around 7 years old. Using the same sticky tape technique that was used to isolate graphene 20 years ago, the first 2D layers from vdW ferromagnetic and antiferromagnetic crystals were obtained in 2016 and 2017 [2-4]. Since then, we have experienced a rapid growth in the field, with an increasing number of new materials being developed and characterized every year, continuously improving in quality. The fast pace in this field has been fueled by the strong potential of vdW magnets for applications in non-volatile data storage devices, and new types of spintronic and magnonic circuits [5,6]. Since 2D magnetic materials maintain a high magnetic anisotropy down to atomic thicknesses while providing atomically sharp and high-quality interfaces, these systems offer the potential for downscaling magnetic devices without common problems which plague conventional materials, such as chemical intermixing in heterostructures and the degradation of the magnetic properties in ultrathin layers.

While 2D magnetic materials hold great promise for applications, several challenges must be addressed for them to achieve their full potential, see Table 1.1. Firstly, advances in the synthesis of 2D magnets are essential, particularly large-scale growth, together with novel strategies to increase the Curie/Néel critical temperatures for metallic and semiconducting systems. Although centimeter scale growth of some materials - e.g. $Fe_3GeTe_2$ - has been shown, the highest quality homogeneous films so far have employed molecular beam epitaxy, which makes it difficult to industrialize the production of such films. The Curie temperature of metallic layers has shown a dramatic increase since the first demonstrations, with record $T_c$ values of over 400 K. Nonetheless, 2D magnetic semiconductors are still lagging, only achieving (close to) room temperature values upon high charge carrier doping concentrations - i.e. degenerate limit - hindering its desirable semiconducting properties. Further details on the challenges and synthesis approaches associated with some of the most promising and intriguing 2D magnets are provided in Table 1.1.

Significant progress is still needed in the fundamental understanding of 2D magnets. Their complex band structure, involving strongly correlated electrons and topological features, makes the accurate description of their properties challenging and computationally expensive. The development of new theoretical tools exploiting first-principles simulations and artificial intelligence [7], along with the optimization of existing tools, might lead to important breakthroughs in understanding the microscopic properties of 2D magnetic systems. Understanding emergent phenomena, such as proximity effects arising from creating heterostructures combining 2D magnets with other vdW materials [8-11] and the corresponding moiré patterns, add another layer of complexity, but it is essential for harnessing the full potential out of these materials. For example, the interface physics in heterostructures of 2D magnetic materials with superconductors is particularly intricate, requiring a simultaneous treatment of quantum atomistic interactions and superconductivity. This complexity has recently led to the development of joint first-principles/Bogoliubov de Gennes codes [12-14] which will enable the next-generation simulations of 2D magnetic/superconducting heterostructures.

Characterization of the magnetic, optical and electronic properties of vdW magnets poses yet another challenge. Surface science techniques often require materials with large-scale homogeneity, which is often lacking in mechanically exfoliated crystals. To address this, improving and adapting conventional techniques and developing new growth methods is imperative, and should lead to a deeper understanding of the vdW magnetic materials. Similarly, the magnetic properties of these systems are rather difficult to address, since standard magnetometry techniques are not capable of resolving the small magnetic moment of the atomically thin layers. Therefore, various advanced characterization techniques have been developed or adapted for use on 2D magnetic layers. Examples include the study of nanomechanical resonances, scanning diamond nitrogen vacancy (NV) magnetometry [15], and magneto-optic microscopy techniques [16]. Each of these approaches comes with their own set of challenges. Optical techniques, such as the widely used magneto-optic Kerr effect, offer large signals and provide valuable information on the magnetic properties of 2D magnets. However, deeper understanding of the light-matter interaction in these materials is needed in order to disentangle effects due to the magneto-optical efficiency from the ones arising from the magnetic order. Electronic structure characterisation is also problematic - samples are often not uniform or degrade under ambient conditions, making it difficult to study fine details arising from magnetic order. Accurate theoretical description of these materials is also challenging: interactions between light (photons), lattice vibrations (phonons), and matter (electrons, holes, excitons) need to be described using many-body theory (GW approximation and Bethe-Salpeter Equation), which is computationally extremely demanding for realistic systems (including defects, or moiré heterostructures). Furthermore, the description of the relaxation processes following laser excitation, requires explicit inclusion of interactions between excitons and phonons, which is accessible only for limited-size systems.

Finally, before 2D magnetic materials can transition from the academic research into the real-world applications, several key challenges need to be addressed. For this, it is crucial to develop devices which harness some of the unique properties of vdW magnets, potentially achieving much better performance, or even enable completely new device architectures that are not possible using conventional systems. For instance, 2D magnetic materials could find applications in quantum devices that operate at low temperature, such as qubit architectures based on Majorana zero-energy modes resulting from van der Waals superconductor/ferromagnet heterostructures [17]. Devices based on spin-waves (magnonics) and spin-orbit torques are examples of directions which could exploit peculiar properties of 2D magnets, such as using topological magnons [18] to transfer information, or the high sensitivity of the magnetic properties to electrostatic gating, lowering the energy barrier for magnetic switching [19, 20]. The high sensitivity to electrostatic gating together with the possibility of transferring information through magnons or skyrmions in absence of charge currents, positions magnetic 2D materials as the game-changers in the design and development of future low-energy consumption devices.

Achieving these advancements requires a rapid and efficient feedback loop where progress in device architectures, theoretical predictions, and the synthesis of new magnetic materials optimized for specific applications reinforce one another. This is by no means an easy feat. It is our hope that this Roadmap will provide a starting point for new researchers and a guide to more experienced ones in the field of 2D magnetic quantum materials, helping to coordinate research efforts towards the further advancement in the field. To further facilitate this, a table listing some of the promising materials, as well as their most interesting properties is provided as well, Table 1. This Roadmap was inspired by discussions during a workshop hosted at the Lorentz Center in Leiden, the Netherlands, and it includes contributions written by experts in the field offering different perspectives, tackling many of the key challenges listed above. As this is a fast-paced field, our goal is to provide a single access point offering forward-looking insights on selected promising materials and a perspective on future research directions, rather than presenting a comprehensive overview of all quantum magnetic materials and potential avenues.

| Material | Synthesis | Interest & Challenges |
|---|---|---|
| $Fe_{5-x}GeTe_2$ (x < 2) [21-28] | CVT, solid state reaction, Te self-flux (bulk crystals)  MBE on graphene/SiC or $Al_2O_3$, sputtering (thin films) | FM metal with high Tc (220 - 390 K), high tunability (e.g. via Fe composition; Ni or Co doping); stacking faults |
| $Fe_3GaTe_2$ [29] | Te self-flux | FM metal, PMA, high Tc ~350-380 K |
| $CrGeTe_3$ [30-32] | Self-flux (bulk crystal), MBE (thin film) | FM semiconductor, PMA, Tc < 100 K; stacking faults, twin domain formation |
| $CrI_3$ [33] | CVT with $I_2$ | Highest reported $T_C$ = 68 K and strongest magnetic anisotropy among trihalides; site disorder, reverse-obverse twinning, chemical disorder from stacking faults |
| $RuI_3$ [34] | Solid state reaction at 6 GPa from amorphous $RuI_3$ | Metallic Kitaev quantum spin liquid candidate; stacking disorder |
| $RuCl_3$ [35,36] | CVT of amorphous $RuCl_3$ | Kitaev quantum spin liquid candidate; in-plane sliding stacking faults, structural disorder during crystal growth, postgrowth processing, or upon cooling through the first order structural transition |
| $MnBi_2Te_4$ [37,38] | Solid state reaction | Magnetic TI, robust AFM; metastable phase, narrow temperature window for crystal growth |
| $Mn_{1-x}Sb_{2+x}Te_4$ [39-41] | Solid state reaction (bulk crystals), MBE (thin films) | 2D magnetic TI candidate, difficult to control stoichiometry, very important for properties, changes TC significantly, FM/FiM/AFM |
| $XPS_3$ (X = Mn, Fe, Ni) [4,42-44] | Solid state reaction, vapor sublimation, CVT with $Cl_2$ or excess sulfur (X = Ni) | Mn; Metallic, alternating AFM, $T_N$ = 78 K<br><br>Fe: semiconductor, out-of-plane zigzag type AFM, $T_N$ = 123K<br><br>Ni: semiconductor, canted out-of-plane zigzag type AFM, $T_N$ = 155 K<br><br>Stacking faults |
| $CsV_3Sb_5$ [45] | Sb self-flux | Kagome metal, CDW and superconducting transition; no vdW material requires more complex exfoliation |
| $Nb_3Br_8$ [46] | CVT with $NH_4Br$ | Strongly correlated system |
| $CrTe_2$ [47,48], $Cr_{1+\delta}Te_2$ [49] | MBE (thin films), oxidation of $KCrTe_2$ (bulk) | $CrTe_2$; FM metal with $T_C$ = 310 K in bulk or $T_C$ up to 300 K in thin films, air stable<br><br>$Cr_{1+x}Te_2$; FM semimetal with $T_C$ = 350 K, many stable stoichiometries depending on the Cr self intercalation, air stable |

| CrOCl [50] | CVT from $Cr_2O_3$ and $CrCl_3$ powder mix | AFM insulator with $T_N$ = 13.6 K, strong in-plane anisotropy |

Table 1.1: A summary of selected 2D magnetic quantum materials, detailing their synthesis methods, key properties of interest, and associated challenges. The table highlights the versatility of these materials in terms of their magnetic and electronic behaviours, as well as the synthesis complexities and structural issues that impact their study and applications.

**Abbreviations**

2D – Two dimensional

AFM – antiferromagnet

ARPES – Angle-resolved photoemission spectroscopy

CDW – charge density wave

CGT – $Cr_2Ge_6Te_6$

CVD/CVT – chemical vapor deposition/transport

DFT – Density functional theory

DMI – Dzyaloshinskii-Moriya interaction

DMRG – Density matrix renormalization group

ESR – Electron spin resonance

FGT – $Fe_3GeTe_2$

FM – ferromagnet

FiM – Ferrimagnet

hBN – Hexagonal boron nitride

IP – In plane

KPFM – Kelvin probe force microscopy

LKAG – Liechtenstein-Katsnelson-Antropov-Gubanov

LLG – Landau-Lifshitz-Gilbert

LSWT – Linear spin wave theory

MBE – Molecular Beam Epitaxy

MCB/MCD – Magnetic circular birefringence/dichroism

MBPT – Many-body perturbation theory

MF – Multiferroic

ML – Monolayer

MLB/MLD – Magnetic linear birefringence/dichroism

MO – Magneto-optical

MPS – Matrix product states

MOKE – Magneto-optical Kerr effect

MOF – Metal-organic framework

MZM – Majorana Zero Modes

NV – Nitrogen vacancy

OOP – Out of plane

PMA – Perpendicular magnetic anisotropy

QAHE – Quantum anomalous Hall effect

SHG – Second harmonic generation

SPM – Scanning probe microscopy

SOC/SOT – Spin-orbit coupling/torque

TDDFT – Time-dependent density functional theory

TI – topological insulator

TMD – Transition metal dichalcogenide

UHV – Ultra-high vacuum

vdW – van der Waals

vHs – van Hove singularity

XMCD – X-ray magnetic circular dichroism

YIG – Yttrium iron garnet

YSR – Yu-Shiba-Rusinov

**Acknowledgments**

We are thankful to the Lorentz Centre in Leiden, the Netherlands, for hosting the workshop on Quantum Magnetic Materials (Oct. 2023), which provided the opportunity to carry the initial discussions which lead to the writing of this article. AGC, ZZ and MHDG acknowledge the research program "Materials for the Quantum Age" (QuMat) for financial support. This program (registration number 024.005.006) is part of the Gravitation program financed by the Dutch Ministry of Education, Culture and Science (OCW). MHDG acknowledges the financial support of the European Union through grant ERC, 2D-OPTOSPIN, 101076932, the Dutch Research Council (NWO—OCENW.XL21.XL21.058). MHDG and AGC acknowledge the financial support of the Zernike Institute for Advanced Materials. ZZ has received funding from the European Union's Horizon Europe research and innovation programme under grant agreement No 101130384 (QUONDENSATE). DS acknowledges financial support from Generalitat Valenciana through the CIDEGENT program (CIDEGENT/2021/052) and the Advanced Materials program by MCIN with funding from European Union NextGenerationEU (MFA/2022/045). AGC acknowlegdes help from Dr. Falk Pabst in the preparation of the table of materials.

**References**

[1] K. S. Novoselov, D. Jiang, F. Schedin, A. K. Geim, "Two-dimensional atomic crystals", *PNAS*, vol. 102, pp. 10451-10453, 2005

[2] B. Huang *et al.*, "Layer-dependent ferromagnetism in a van der Waals crystal down to the monolayer limit", *Nature*, vol. 546, pp. 270-273, 2017

[3] C. Ging *et al.*, "Discovery of intrinsic ferromagnetism in two-dimensional van der Waals crystals", Nature, vol. 546, pp. 265-269, 2017

[4] J.-U. Lee et al., "Ising-type magnetic ordering in atomically thin $FePS_3$", *Nano Letters*, vol. 16, pp. 7433-7438, 2016

[5] S. Qi *et al.*, "Giant electrically tunable magnon transport anisotropy in a van der Waals antiferromagnetic insulator", *Nature Communications*, vol. 14, pp. 2526, 2023

[6] T. S. Ghiasi, A. A. Kaverzin, A. H. Dismukes, D. K. de Wal, X. Roy & B. J. van Wees, "Electrical and thermal generation of spin currents by magnetic bilayer graphene", *Nature Nanotechnology,* vol. 16, pp. 788–794, 2021

[7] D. Soriano, "Uncovering magnetic interactions in moiré magnets", *Nature Computational Science*, vol. 3, pp. 282-284, 2023

[8] Z. Zanolli, "Graphene-multiferroic interfaces for spintronics applications" *Scientific Reports,* vol. 6, 31346, 2016

[9] C. Cardoso, D. Soriano, N. A. García-Martínez, J. Fernández-Rossier, "Van der Waals spin valves", *Physical Review Letters*, vol. 121, pp. 067701, 2018

[10] Z. Zanolli, C. Niu, G. Bihlmayer, Y. Mokrousov, P. Mavropoulos, M. J. Verstraete, and S. Blügel, "Hybrid quantum anomalous Hall effect at graphene-oxide interfaces", Phys. Rev. B vol. 98, 155404 2018

[11] T. P. Lyons *et al.*, "Interplay between spin proximity effect and charge-dependent exciton dynamics in MoSe$_2$/CrBr$_3$ van der Waals heterostructures", *Nature Communications*, vol. 11, pp. 6021, 2020

[12] R. Reho, N. Wittemeier, A. H. Kole, P. Ordejón, Z. Zanolli, "Density functional Bogoliubov-de Gennes theory for superconductors implemented in the SIESTA code", *Physical Review B*, vol. 110, 134505, 2024

[13] P. Rüssmann, S. Blügel, "Density functional Bogoliubov-de Gennes analysis of superconducting Nb and Nb(110) surfaces", *Physical Review B*, vol. 105, pp. 125143, 2022

[14] G. Csire, B. Újfalussy, J. Cserti, B. Győrffy, "Multiple scattering theory for superconducting heterostructures", *Physical Review B*, vol. 91, pp. 165142, 2015

[15] T. S. Ghiasi *et al.*, "Nitrogen-vacancy magnetometry of CrSBr by diamond membrane transfer", *npj 2D materials and applications*, vol. 7, pp. 62, 2023

[16] S. Liu, A. Malik, V. L. Zhang, T. Yu, "Lightning the spin: Harnessing the potential of 2D magnets in opto-spintronics", *Advanced Materials*, pp. 2306920, 2023

[17] S. Kezilebieke, M. N. Huda, V. Vaňo, M. Aapro, S. C. Ganguli, O. J. Silveira, S. Głodzik, A. S. Foster, T. Ojanen & P. Liljeroth, "Topological superconductivity in a van der Waals heterostructure", *Nature*, vol. 588, pp. 424–428, 2020

[18] A. T. Costa, D. L. R. Santos, N. M. R. Peres, J. Fernández-Rossier, "Topological magnons in CrI3 monolayers: an itinerant fermion description", *2D materials*, vol. 7, pp. 045031, 2020

[19] S. Jiang, J. Shan, K. F. Mak, "Electric-field switching of two-dimensional van der Waals magnets", *Nature Materials*, vol. 17, pp. 406-410, 2018

[20] B. Huang *et al.*, "Electrical control of 2D magnetism in bilayer CrI$_3$", *Nature Nanotechnology*, vol. 13, pp. 544-548, 2018

[21] J. Stahl, E. Shlaen, and D. Johrendt, "The van der Waals Ferromagnets Fe5−δGeTe2 and Fe5−δ−xNixGeTe2 – Crystal Structure, Stacking Faults, and Magnetic Properties," *Zeitschrift für anorganische und allgemeine Chemie*, vol. 644, no. 24, pp. 1923–1929, 2018

[22] M. Ribeiro, G. Gentile, A. Marty, D. Dosenovic, H. Okuno, C. Vergnaud, J.-F. Jacquot, D. Jalabert, D. Longo, P. Ohresser, A. Hallal, M. Chshiev, O. Boulle, F. Bonell, and M. Jamet, "Large-scale epitaxy of two-dimensional van der Waals room-temperature ferromagnet Fe5GeTe2," *NPJ 2D Mater Appl*, vol. 6, no. 1, p. 10, 2022

[23] H. Lv, A. da Silva, A. I. Figueroa, C. Guillemard, I. Fernández Aguirre, L. Camosi, L. Aballe, M. Valvidares, S. O. Valenzuela, J. Schubert, M. Schmidbauer, J. Herfort, M. Hanke, A. Trampert, R. Engel-Herbert, M. Ramsteiner and J. M. J. Lopes, "Large-Area Synthesis of Ferromagnetic Fe5−xGeTe2/Graphene van der Waals Heterostructures with Curie Temperature above Room Temperature", *Small*, vol. 19, no. 39, p. 2302387, 2023

[24] A. F. May, D. Ovchinnikov, Q. Zheng, R. Hermann, S. Calder, B. Huang, Z. Fei, Y. Liu, X. Xu, and M. A. McGuire, "Ferromagnetism Near Room Temperature in the Cleavable van der Waals Crystal Fe5GeTe2," *ACS Nano*, vol. 13, no. 4, pp. 4436–4442, 2019

[25] Q. Zhao, C. Xia, H. Zhang, B. Jiang, T. Xie, K. Lou, and C. Bi, "Ferromagnetism of Nanometer Thick Sputtered Fe3GeTe2 films in the Absence of Two-Dimensional Crystalline Order: Implications for Spintronics Applications," *ACS Applied Nano Materials*, vol. 6, no. 4, pp. 2873–2882, 2023

[26] A. F. May, S. Calder, C. Cantoni, H. Cao, and M. A. Mcguire, "Magnetic structure and phase stability of the van der Waals bonded ferromagnet Fe3−xGeTe2," *Physical Review B*, vol. 93, p. 14411, 2016

[27] J. M. J Lopes, D. Czubak, E. Zallo, A. I. Figueroa, C. Guillemard, M. Valvidares, J. Rubio-Zuazo, J. López-Sanchéz, S. O. Valenzuela, M. Hanke, and M. Ramsteiner, "Large-area van der Waals epitaxy and magnetic characterization of Fe3GeTe2films on graphene," *2D Materials*, vol. 8, no. 4, p. 041001, 2021

[28] J. Ke, M. Yang, W. Xia, H. Zhu, C. Liu, R. Chen, C. Dong, W. Liu, M. Shi, Y. Guo, J. Wang, "Magnetic and magneto-transport studies of two-dimensional ferromagnetic compound Fe3GeTe2," *Journal of Physics: Condensed Matter*, vol. 32, no. 40, p. 405805, 2020

[29] G. Zhang, F. Guo, H. Wu, X. Wen, L. Yang, W. Jin, W. Zhang, and H. Chang, "Above-room-temperature strong intrinsic ferromagnetism in 2D van der Waals Fe3GaTe2 with large perpendicular magnetic anisotropy," *Nature Communications*, vol. 13, no. 1, p. 5067, 2022

[30] G. T. Lin, H. L. Zhuang, X. Luo, B. J. Liu, F. C. Chen, J. Yan, Y. Sun, J. Zhou, W. J. Lu, P. Tong, Z. G. Sheng, Z. Qu, W. H. Song, X. B. Zhu, and Y. P. Sun, "Tricritical behavior of the two-dimensional intrinsically ferromagnetic semiconductor CrGeTe3," *Physical Review B*, vol. 95, p. 245212, 2017

[31] M. Mogi, A. Tsukazaki, Y. Kaneko, R. Yoshimi, K. S. Takahashi, M. Kawasaki, and Y. Tokura, "Ferromagnetic insulator Cr2Ge2Te6 thin films with perpendicular remanence," *APL Materials*, vol. 6, no. 9, p. 91104, 2018

[32] C. Gong, L. Li, Z. Li, H. Ji, A. Stern, Y. Xia, T. Cao, W. Bao, C. Wang, Y. Wang, Z. Q. Qiu, R. J. Cava, S. G. Louie, J. Xia, and X. Zhang, "Discovery of intrinsic ferromagnetism in two-dimensional van der Waals crystals," *Nature*, vol. 546, no. 7657, pp. 265–269, 2017

[33] M. A. McGuire, H. Dixit, V. R. Cooper, and B. C. Sales, "Coupling of Crystal Structure and Magnetism in the Layered, Ferromagnetic Insulator CrI3," *Chemistry of Materials*, vol. 27, no. 2, pp. 612–620, 2015

[34] D. Ni, X. Gui, K. M. Powderly, and R. J. Cava, "Honeycomb-Structure RuI3, A New Quantum Material Related to α-RuCl3," *Advanced Materials*, vol. 34, no. 7, 2022

[35] J. Sears, Y. Shen, M. J. Krogstad, H. Miao, J. Yan, S. Kim, W. He, E. S. Bozin, I. K. Robinson, R. Osborn, S. Rosenkranz, Y. J. Kim, and M. P. M. Dean, "Stacking disorder in α-RuCl3 investigated via x-ray three-dimensional difference pair distribution function analysis," *Physical Review B*, vol. 108, no. 14, p. 144419, 2023

[36] R. D. Johnson, S. C. Williams, A. A. Haghighirad, J. Singleton, V. Zapf, P. Manuel, I. I. Mazin, Y. Li, H. O. Jeschke, R. Valentí, and R. Coldea, "Monoclinic crystal structure of α-RuCl3 and the zigzag antiferromagnetic ground state," *Physical Review B*, vol. 92, no. 23, p. 235119, 2015

[37] D. S. Lee, T.-H. Kim, C.-H. Park, C.-Y. Chung, Y. S. Lim, W.-S. Seo, and H.-H. Park, "Crystal structure, properties and nanostructuring of a new layered chalcogenide semiconductor, Bi2MnTe4," *CrystEngComm*, vol. 15, no. 27, p. 5532, 2013

[38] A. Zeugner, F. Nietschke, A.U.B. Wolter, S. Gaß, R.C. Vidal, T.R.F. Peixoto, D. Pohl, C. Damm, A. Lubk, R. Hentrich, S.K. Moser, C. Fornari, C.H. Min, S. Schatz, K. Kißner, M. Ünzelmann, M. Kaiser, F. Scaravaggi, B. Rellinghaus, A. Isaeva, "Chemical Aspects of the Candidate Antiferromagnetic Topological Insulator MnBi2Te4," *Chemistry of Materials*, vol. 31, no. 8, pp. 2795–2806, 2019

[39] L. C. Folkers, L. T. Corredor, F. Lukas, M. Sahoo, A. U. B. Wolter, and A. Isaeva, "Occupancy disorder in the magnetic topological insulator candidate Mn1-xSb2+xTe4," *Zeitschrift für Kristallographie - Crystalline Materials*, vol. 237, no. 4–5, pp. 101–108, 2022

[40] M. Sahoo, M.C. Rahn, E. Kochetkova, O. Renier, L.C. Folkers, A. Tcakaev, M.L. Amigó, F.M. Stier, V. Pomjakushin, K. Srowik, V.B. Zabolotnyy, E. Weschke, V. Hinkov, A. Alfonsov, V. Kataev, B. Büchner, A.U.B. Wolter, J.I. Facio, L.T. Corredor, A. Isaeva, "Tuning strategy for Curie-temperature enhancement in the van der Waals magnet Mn1+x Sb2 xTe4," *Materials Today Physics*, vol. 38, p. 101265, 2023

[41] S. Wimmer, J. Sánchez-Barriga, P. Küppers, A. Ney, E. Schierle, F. Freyse, O. Caha, J. Michalička, M. Liebmann, D. Primetzhofer, M. Hoffman, A. Ernst, M.M. Otrokov, G. Bihlmayer, E. Weschke, B. Lake, E. Chulkov, M. Morgenstern, G. Bauer, O. Rader, "Mn-Rich MnSb2Te4 : A Topological Insulator with Magnetic Gap Closing at High Curie Temperatures of 45–50 K," *Advanced Materials*, vol. 33, no. 42, p. 2102935, 2021

[42] B. E. Taylor, J. Steger, and A. Wold, "Preparation and properties of some transition metal phosphorus trisulfide compounds," *Journal of Solid State Chemistry*, vol. 7, no. 4, pp. 461–467, 1973

[43] C.-T. Kuo, M. Neumann, K. Balamurugan, H. J. Park, S. Kang, H. W. Shiu, J. H. Kang, B. H. Hong, M. Han, T. W. Noh, and J.-G. Park, "Exfoliation and Raman Spectroscopic Fingerprint of Few-Layer NiPS3 Van der Waals Crystals", *Scientific Reports*, 2016, vol. 6, no. 1, p. 20904, 2016

[44] P. A. Joy and S. Vasudevan, "Magnetism in the layered transition-metal thiophosphates MPS3 (M =Mn, Fe, and Ni)," *Physical Review B*, vol. 46, 1992

[45] B. Song, T. Ying, X. Wu, W. Xia, Q. Yin, Q. Zhang, Y. Song, X. Yang, J. Guo, L. Gu, X. Chen, J. Hu, A. P. Schnyder, H. Lei, Y. Guo, and S. Li, "Anomalous enhancement of charge density wave in kagome superconductor CsV3Sb5 approaching the 2D limit," *Nature Communications*, vol. 14, no. 1, p. 2492, 2023

[46] H. Wu, Y. Wang, Y. Xu, P. K. Sivakumar, C. Pasco, U. Filippozzi, S. S. P. Parkin, Y. J. Zeng, T. McQueen, and M. N. Ali, "The field-free Josephson diode in a van der Waals heterostructure," *Nature* vol. 604, no. 7907, pp. 653–656, 2022

[47] X. Zhang, Q. Lu, W. Liu, W. Niu, J. Sun, J. Cook, M. Vaninger, P.F. Miceli, D.J. Singh, W.-W. Lian, T.-R. Chang, X. He, J. Du, L. He, R. Zhang, G. Bian, and Y. Xu, "Room-temperature intrinsic ferromagnetism in epitaxial CrTe$_2$ ultrathin films," *Nature Communications*, vol. 12, no. 1, pp. 2492, 2021

[48] D.C. Freitas, R. Weht, A. Sulpice, G. Remenyi, P. Strobel, F. Gay, J. Marcus and M. Núñez-Regueiro, "Ferromagnetism in layered metastable 1*T*-CrTe$_2$," *Journal of Physics: Condensed Matter,* vol. 27, no. 17, pp. 176002, 2015

[49] Y. Fujisawa, M. Pardo-Almanza, J. Garland, k. Yamagami, X. Zhu, X. Chen, K. Araki, T. Takeda, M. Kobayashi, Y. Takeda, C.H. Hsu, F.C. Chuang, R. Laskowski, K.H. Khoo, A. Soumyanarayanan, A. and Y. Okada, "Tailoring magnetism in self-intercalated Cr$_{1+\delta}$Te$_2$ epitaxial films." *Physical Review Materials*, vol. 4, no. 11, pp. 114001, 2020

[50] T. Zhang, Y. Wang, H. Li, F. Zhong, J. Shi, M. Wu, Z. Sun, W. Shen, B. Wei, W. Hu, X. Liu, L. Huang, C. Hu, Z. Wang, C. Jiang, S. Yang, Q.-M. Zhang, Z. Qu, "Magnetism and Optical Anisotropy in van der Waals Antiferromagnetic Insulator CrOCl," *ACS Nano*, vol. 13, no. 10, pp. 11353–11362, 2019

## 2. Topology and Strong Correlation in 2D Magnetic Quantum Materials

Mazhar N. Ali[1,2], Semonti Bhattacharyya[3], Saroj P. Dash[4,], Antonija Grubišić-Čabo[5], Sergii Grytsiuk[6], Malte Rösner[6]

[1.] Kavli Institute of Nanoscience, Delft University of Technology, [2.] Department of Quantum Nanoscience, Faculty of Applied Sciences, Delft University of Technology, [3.] Huygens-Kamerlingh Onnes Laboratory, Leiden University, [4.] Department of Microtechnology and Nanoscience, Chalmers University of Technology, [5.] Zernike Institute for Advanced Materials, University of Groningen, [6.] Institute for Molecules and Materials, Radboud University

**Status**

Two dimensional van der Waals magnets provide an exceptional platform for investigating intertwined effects of magnetism, electron correlation, and topology, possibly yielding various quantum phases, as depicted in Figure 2.1. Introducing magnetism into 2D topological systems breaks time-reversal symmetry, leading to exotic momentum-space topology and to new emergent quantum phases such as the QAHE, Chern Insulator phase, and fractional quantum anomalous Hall phases [1]. The QAHE, characterised by dissipation-free, topologically protected edge states, is both fundamentally exciting and potentially useful in a wide range of applications, from measurement standards to low-power electronics. Initially discovered in magnetically doped topological insulators [2], the QAHE in these materials is limited to mK temperature ranges due to presence of intrinsic disorder, with higher temperatures achievable in intrinsic magnetic TIs[3], still below the liquid nitrogen temperature. Of further particular interest are materials with triangular and triangularly-derived transition metal nets, as can be seen in Figure 2.1. The triangular sublattice (Figure 2.1, bottom right) is found in the 2D magnet CrSBr, while the honeycomb lattice (Figure 2.1, bottom left) is seen in 2D ferromagnets $CrX_3$ (X = Cl, Br, I). Another pivotal structure for examining electron correlation and quantum phases is the kagome lattice, or trihexagonal tiling (Figure 2.1, top). Composed of hexagons and triangles, the kagome lattice naturally hosts geometrical frustration, leading to phenomena such as magnetic frustration and quantum spin ice and spin liquid states [4]. The interplay among its three sublattices produces notable electronic band features, including vHSs, and flat bands as depicted in the representative band structure in Figure 2.1. Combined with reduced screening effects in 2D materials, and thus naturally enhanced local and non-local Coulomb interactions, these vHSs and flat bands, with their high densities of states and enhanced effective masses, can promote strong correlation effects. Consequently, van der Waals kagome materials can provide an excellent platform for studying the interplay between topology, magnetism, and electron correlation. Finally, van der Waals heterostructures of 2D materials could play a significant role in achieving the so far elusive QAHE at room temperature [5], and they can also host real-space topological structures such as skyrmions [6], potentially open new avenues of research in topological magnetism and its spintronic applications.

**Current and Future Challenges**

Initial observations of the QAHE in topological systems were made in magnetically doped topological insulators [2], however, this approach leads to a high level of disorder, limiting the realisation of QAHE to mK temperature ranges. An alternative is to identify intrinsic magnetic topological insulators [3] which are inherently disorder-free stoichiometric materials that can host QAHE at relatively higher temperatures of several Kelvin, and use them as test beds to study the interplay of magnetism and topology (Figure 2.1a). Unfortunately, the limited availability of these intrinsic magnetic topological insulators limits the possibility of significantly raising the QAHE temperature. Studies using 2D kagome systems also face structural challenges such as stacking disorder. Furthermore, substrate effects in both systems must be understood and accounted for. It is also not yet clear whether magnetism in 2D quantum magnets competes with other ordered phases, such as charge density wave phases, or if they are complementary [7,8]. Additionally, understanding the thickness dependence effects on the long range out-of-plane magnetism in these systems remains incomplete. Finally, theoretical descriptions need to be further developed for systems exhibiting magnetism and correlation (Figure 2.2b). Around the vHSs filling, the inclusion of the local Hubbard interaction ($U_0$) and nearest neighbour site interaction ($U_1$) is proposed to induce a variety of electronic phases depending on the correlation strength compared with hopping energy [9]. The appearance of these ordered states is complex, with different studies predicting different states such as magnetism with strong spin fluctuations, CDWs, and superconductivity. Therefore, many challenges in this field need to be addressed, from developing atomistic theoretical descriptions to fundamentally understanding the interplay of (competing) phases to achieve the QAHE at room temperature. In the following chapter we briefly discuss several advances needed to tackle these challenges.

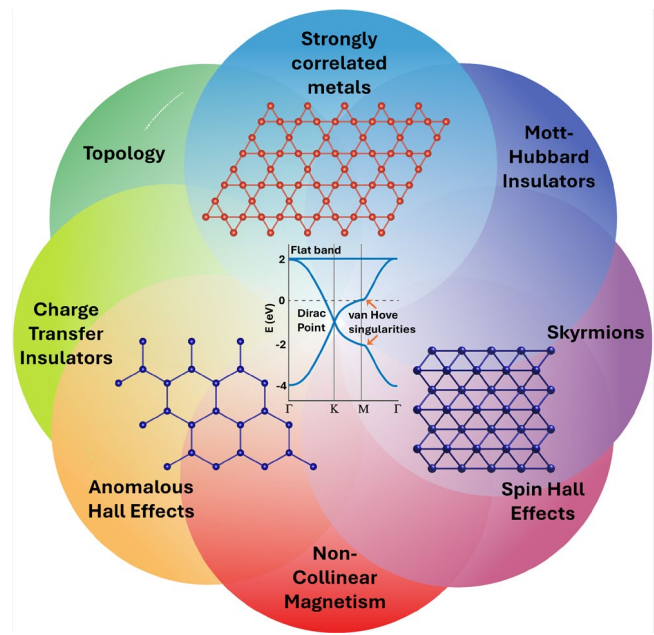

**Figure 2.1.** Schematic showcasing the intertwining of various physical phenomena in 2D magnetic materials. Three example transition metal sublattices are shown: top is the vanadium nearest neighbour (NN) Kagome lattice in the $AV_3Sb_5$ (A = K, Rb, Cs) compounds, bottom right is the buckled triangular chromium NN lattice in CrSBr, and bottom left is the honeycomb chromium NN lattice in the $CrX_3$ (X = Cl, Br, I) compounds. A characteristic Kagome band structure is shown in the center, with flat band, van Hove singularities, and Dirac point (DP), labelled.

**Advances in Science and Technology to Meet Challenges**

One of the main experimental challenges is synthesising materials with low structural disorder and impurities in order to disentangle effects of magnetism, topology, and correlations. A promising approach is to use intrinsic magnetic topological and correlated systems [3] to avoid any additional disorder introduced by magnetic doping [2]. Alternatively, magnetism can be introduced into topological systems via proximity effect from an adjacent magnetic material [9]. The method is attractive because (i) the topological insulator and the magnetic material can be chosen independently, enhancing versatility, and (ii) magnetism is introduced through proximity, allowing the use of ferromagnets, ferrimagnets, and even antiferromagnets of type A. With this approach the QAHE was achieved in MBE grown heterostructures of magnetic materials and TIs at higher temperatures than other methods [10]. However, MBE relies on lattice matching of consecutive bulk materials, limiting the number of possible material combinations. Van der Waals stacking can dramatically increase the combinatorial possibilities, but research in this area has been limited [1] and further studies of van der Waals heterostructures using 2D magnetic materials will play a significant role in obtaining the QAHE at room temperature. 2D magnets are also relevant for studies of topology in real space: with strong spin-orbit and magnetic exchange interactions, topological protection is often stabilized by an antisymmetric exchange interaction, called Dzyaloshinskii-Moriya interaction. This plays a significant role in generating non-collinear magnetic textures spurring observations of real-space topological spin textures such as skyrmions (Figure 2.2c) and more exotic states with varying topology, such as antiskyrmions, merons and hopfions. Recent studies on 2D magnets have uncovered and enabled the manipulation of different types of skyrmions in 2D quantum magnets [6,11]. Finally, advances in theoretical modelling are catching up with the rapid pace of experimental findings: the combination of first principle calculations with minimal model handling within quantum embedding frameworks hold great promise for the ab initio description of strong correlation effects within correlated kagome layered magnetic materials [8].

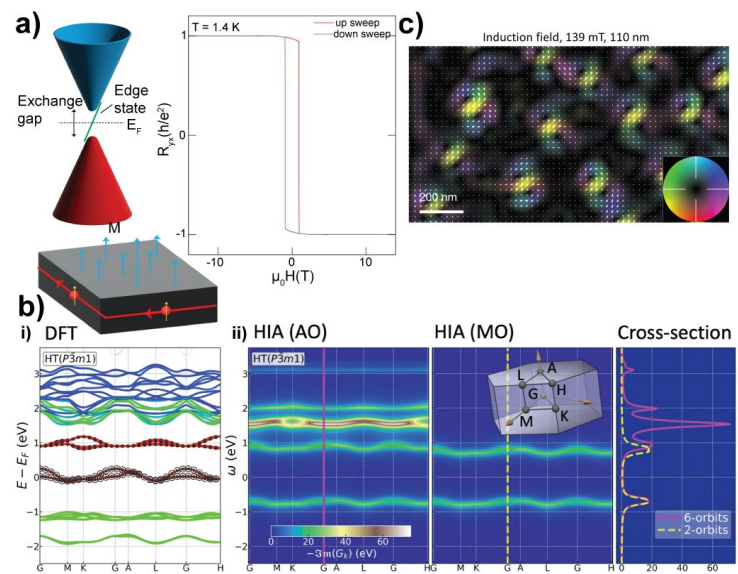

**Figure 2.2.** Examples of interplay of magnetism, topology and correlation in 2D quantum magnets. (a) Quantum anomalous Hall effect: Magnetic field opens a gap in the surface states of topological insulators and leads to conductive 1D edge states. The data shows Hall resistance $R_{xy}$ that gets quantized at a value of 0.998 e2/h. (b) Example of the effect that strong Hubbard interaction can have on breathing-mode kagome materials (here Nb3Cl8). i) DFT electronic structure that predicts metallic behaviour of the material. ii) Interacting spectral functions that show Mott insulating behaviour. Left and middle panels show results from atomic-orbital and molecular-orbital models, respectively, right panel shows cross-sections of those spectral functions at one selected momentum. (c) Real space topological structures skyrmions imaged through magnetic induction field map of in a 110-nm-thick$(Fe_{0.5}Co_{0.5})_5GeTe_2$ flakes obtained using 4D- Lorentz scanning tunnelling electron microscopy along with an electron microscope pixel array detector, with applied fields of 139 mT. The color and arrows indicate induction field components perpendicular to the beam propagation direction of Neél skyrmions at an 18° tilt. (a) Top right corner taken from Ref. [3], (b) taken from Ref. [8] and (c) taken from Ref. [10].

**Concluding Remarks**

Two-dimensional van der Waals magnets represent a cutting-edge platform for investigating the intricate interplay between electronic correlations, topology, and various quantum phases. Introducing magnetism in 2D topological systems disrupts time-reversal symmetry, leading to emergent quantum phases such as the quantum anomalous Hall effect, Chern insulator phase, and fractional quantum anomalous Hall effect. These systems can naturally exhibit strong Coulomb correlations due to reduced screening effects, making them ideal for studying electron interactions. Structures like the kagome lattice, composed of hexagons and triangles, are particularly significant for examining these phenomena, as they display key electronic features such as van Hove singularities and flat bands.

Despite challenges in material synthesis and understanding phase interactions, substantial progress has been made. Van der Waals heterostructures hold promise for achieving room-temperature QAHE and hosting real-space topological structures like skyrmions, which can be manipulated for advanced spintronic applications. Overcoming these challenges through improved theoretical models and high-quality material synthesis will unlock the full potential of topologically non-trivial correlated 2D magnetic materials, paving the way for innovative applications in low-power electronics and beyond.

**Acknowledgements**
SPD acknowledges funding from European Commission EIC project 2DSPIN-TECH (No. 101135853) and 2D TECH VINNOVA competence centre (No. 2019-00068). MNA acknowledges support from the NWO Talent Programme financed by the NWO VI.Vidi.223.089 and the Kavli Institute Innovation Award 2023. AGC acknowledges the financial support of the Zernike Institute for Advanced Materials. MNA, SB, AGC and MR acknowledge the research program "Materials for the Quantum Age" (QuMat) for financial support. This program (registration number 024.005.006) is part of the Gravitation program financed by the Dutch Ministry of Education, Culture and Science (OCW). MR acknowledges support from the Dutch Research Council (NWO) via the 'TOPCORE' consortium.

**References**

[1] S. Bhattacharyya, G. Akhgar, M. Gebert, J. Karel, M.T. Edmonds, M.S. Fuhrer, "Recent Progress in Proximity Coupling of Magnetism to Topological Insulators," *Adv. Mater.* Vol. 33, pp. 2007795, 2021

[2] C.-Z. Chang, J. Zhang, X. Feng, J. Shen, Z. Zhang, M. Guo, K. Li, Y. Ou, P. Wei, L.-L. Wang, Z.-Q. Ji, Y. Feng, S. Ji, X. Chen, J. Jia, X. Dai, Z. Fang, S.-C. Zhang, K. He , Y. Wang, L. Lu, X.-C. Ma, and Q.-K. Xue, " Experimental Observation of the Quantum Anomalous Hall Effect in a Magnetic Topological Insulator," *Science* vol. 340, pp. 160 – 170, 2013

[3] Y. Deng, Y. Yu, M.Z. Shi, Z. Guo, Z. Xu, J. Wang, X.H. Chen and Y. Zhang, "Quantum anomalous Hall effect in intrinsic magnetic topological insulator $MnBi_2Te_4$," *Science* vol. 367, pp. 895-900, 2020

[4] M.N. Ali, Y. Wang, H. Wu, G.T. McCandless and J.Y. Chan, "Quantum states and intertwining phases in kagome materials," *Nat. Rev. Phys.* vol. 5, pp. 635–658, 2023

[5] S.K. Chong, K.B. Han, A. Nagaoka, R. Tsuchikawa, R. Liu, H. Liu, Z.V. Vardeny, D.A. Pesin, C. Lee, T.D. Sparks, and V.V. Deshpande, "Topological Insulator-Based van der Waals Heterostructures for Effective Control of Massless and Massive Dirac Fermions," *Nano Lett.* vol. 18, pp. 8047–8053, 2018

[6] L. Powalla, M. T. Birch, K. Litzius, S. Wintz, F. S. Yasin, L. A. Turnbull, F. Schulz, D. A. Mayoh, G. Balakrishnan, M. Weigand, X. Yu, K. Kern, G. Schütz, M. Burghard, "Seeding and Emergence of Composite Skyrmions in a van der Waals Magnet," *Adv. Mater*. vol. 35, pp. 2208930, 2023

[7] N. Ru, J.-H. Chu, and I. R. Fisher, "Magnetic properties of the charge density wave compounds $R$Te$_3$ ($R$=Y, La, Ce, Pr, Nd, Sm, Gd, Tb, Dy, Ho, Er, and Tm)," *Phys. Rev. B* vol. 78, pp. 012410, 2008

[8] S. Grytsiuk, M.I. Katsnelson, E.G.C.P van Loon and Malte Rösner, "Nb3Cl8: a prototypical layered Mott-Hubbard insulator," *npj Quantum Mater.* vol. 9, pp. 8, 2024

[9] J.-X. Yin, B. Lian and M.Z. Hassan, "Topological kagome magnets and superconductors," *Nature* vol. 612, pp. 647-657, 2022

[10] J. Jiang, D. Xiao, F. Wang, J.-H. Shin, D. Andreoli, J. Zhang, R. Xiao, Y.-F. Zhao, M. Kayyalha, L. Zhang, K. Wang, J. Zang, C. Liu, N. Samarth, M.H.W. Chan and C.-Z. Chang, "Concurrence of quantum anomalous Hall and topological Hall effects in magnetic topological insulator sandwich heterostructures," *Nat. Mater.* vol. 19, pp. 732–737, 2020

[11] H. Zhang, D. Raftrey, Y.-T. Chan, Y.-T. Shao, R. Chen, X. Chen, X. Huang, J.T. Reichanadter, K. Dong, S. Susarla, L. Caretta, Z. Chen, J. Yao, P. Fischer, J.B. Neaton, W. Wu, D.A. Muller, R.J. Birgeneau and R. Ramesh, "Room-temperature skyrmion lattice in a layered magnet (Fe$_{0.5}$Co$_{0.5}$)$_5$GeTe$_2$," *Sci. Adv.* vol. 8, pp. eabm7103, 2022

# 3. Unified computational treatment of superconductivity, magnetism and topology

Arnold H. Kole[1], Zeila Zanolli[1]

[1] Chemistry Department, Debye Institute for Nanomaterials Science and ETSF, Condensed Matter and Interfaces, Utrecht University, PO Box 80.000, 3508 TA Utrecht, The Netherlands.

**Status**

In recent years there has been an increased interest in materials with topological properties. These materials are characterized by a topological phase, which is separated from the normal phase by a discontinuous transformation. Topological phases and their related properties are therefore protected from disorder. Topological superconductors are a subclass of topological materials where a superconducting material exhibits a topological phase. They might exhibit exciting new forms of physics such as edge states that behave like Majorana fermions [1]. In addition to their theoretical interest, this would also make them potential candidates for building topological quantum computers. It has been theoretically demonstrated that one route to topological superconductivity is the combination of s-wave superconductivity with spin-orbit coupling and magnetism [1]. Topological superconductivity would manifest as in-gap states with a zero-bias conductance peak and characteristic transport features such as quantized Hall conductivity and anomalous Josephson effects [1]. This can be experimentally realized in composite materials consisting of superconductors and low-dimensional magnetic materials, such as 0D impurities, 1D chains or 2D islands [1-3]. Experiments on these systems show in-gap states with zero energy which could be attributed to Majorana Zero Modes (MZM) [2, 3]. However, other states that mimic Majorana signatures exist. The presence of magnetic impurities that are exchanged coupled to the Cooper pair electrons can also induce in-gap states known as Yu-Shiba-Rusinov (YSR) states [2]. Alternatively, in-gap states can be Andreev bound states, which form near boundaries and vortex cores [1]. Therefore, conclusive proof of the existence of MZMs remains lacking [3].

While the study of superconductor-based materials using tight-binding models provided valuable insights [1-3], the development of theoretical and computational tools that treat superconductivity and material-specific details on the same footing is needed to fully explain the experimental findings. Several such tools now exist, differing in the underlying methodology and in the treatment of the superconducting pairing interaction, either fully ab-initio (superconducting DFT, SCDFT [4]) or semi-phenomenologically [4-8]. Here we consider the semi-phenomenological DFT-BdG approaches, KKR-BdG [5-6] and SIESTA-BdG [7]. KKR-BdG has been used to model magnetic impurities and atomic chains on superconducting surfaces, Yu-Shiba-Rusinov and Majorana-like in-gap states and their response to changes in material specific details (magnetic ordering, type of magnetic impurities, spin-orbit coupling effects) [9,10]. Moreover, the DFT-BdG methods can predict the existence (or absence) of zero energy states and analyze their nature, a powerful new method to study the topological nature of superconductivity [9,10].

**Current and Future Challenges**

Despite the advances in the computational and theoretical description of superconductor-magnet interfaces, many questions remain unanswered. We identify four major types of challenges to overcome, related to (i) the description of magnetism, (ii) the description of superconductivity, (iii) the unambiguous identification of topological versus trivial states, and (iv) computational bottlenecks.

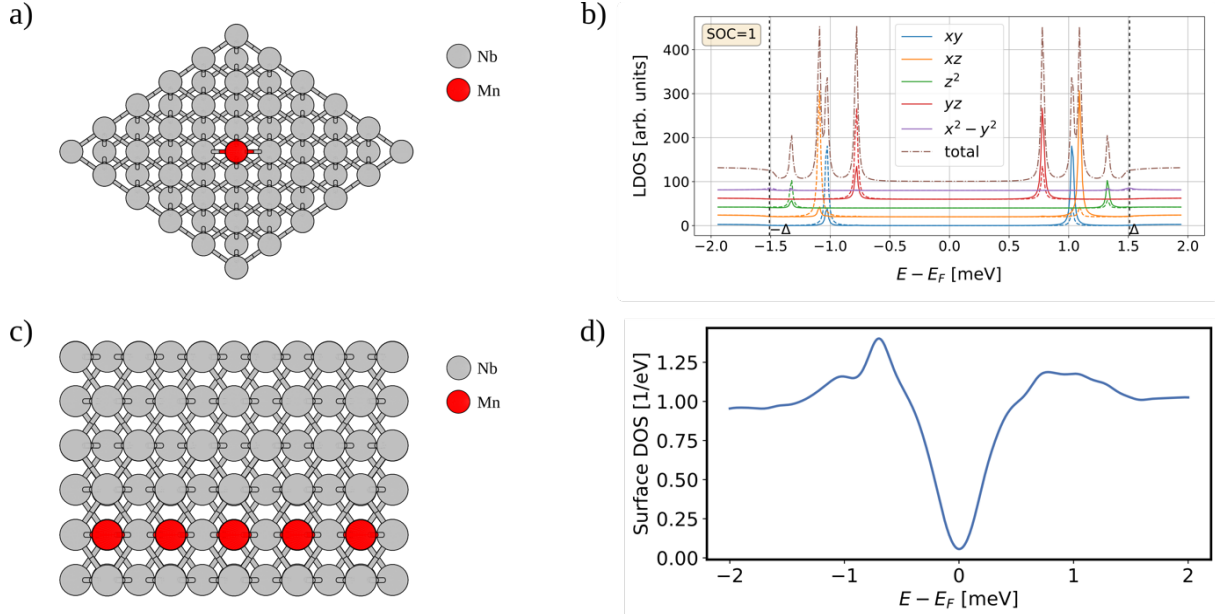

**Figure 3.1.** Examples of superconductor-magnetic hybrid materials. Mn adatom on Nb(110) surface: (a) ball-and-stick model and (b) local DOS computed with KKR-BdG. Reproduced with permission from [9]. Mn chain on Nb(110) surface: (c) ball-and-stick model and (d) DOS computed with SIESTA-BdG. Reprinted figure with permission from [9]. Copyright 2021 by the American Physical Society.

An accurate *description of magnetism* is essential to correctly predict the topological character of a system. While several accurate methods exist, DFT+U, TDDFT, DMFT just to name a few, it remains difficult to model magnetic interactions. Due to the small (of the order of 10-100 meV) energy differences among magnetic phases, it is difficult to unambiguously identify the magnetic ground state. The inclusion of spin-orbit coupling and non-collinear magnetism, an essential ingredient for topologically non-trivial behaviour, complicates things further. The intrinsically long-ranged nature of magnetic interactions makes its treatment even more challenging in periodic boundary condition simulations: accurate convergence studies must ensure that the computational supercell is large enough to prevent artefacts due to periodicity.

Just as important as magnetism are the *pairing interactions* that give rise to superconductivity. Conventional superconductivity is explained in terms of electron-phonon interaction, but a similar understanding of more complicated pairings is still lacking. Tight-binding models and DFT-BdG codes use a semi-phenomenological value for pairing interaction [5-7], but do not compute it. The introduction of interfaces and magnetism raises even more questions: is the pairing in the bulk transferrable to interfaces? How is the pairing interaction affected when magnetism is introduced through proximity? What is the spatial extent of the proximity effect? How does proximity induced superconductivity decay?

The *unambiguous identification of topological zero-energy modes* is one of the most difficult challenges. Current works focus on the search for Majorana Zero Modes, yet it remains difficult to disentangle them from other in-gap states with similar signatures [1-3]. In a recent work the local density of states in real-space along an atomic chain is used as an analogue for the band structure, looking for band inversion in such an analogue is a promising new approach [10]. Nevertheless, a richer toolset for answering this question is needed. The definitive answer about the nature of the zero-energy states can only come from non-local transport, both in experiments and simulations.

One should also consider the *computational challenges* in modelling these systems. Energy scales for superconducting gaps are about 0.1-1 meV, even smaller than for magnetism, requiring much higher accuracy than normal-state first-principles simulations [6]. Large supercells with many atoms are often required to properly describe the interface between the superconductor and the magnet. Consequently, DFT-BdG calculations of realistic systems are very demanding, both in terms of computational time and use of memory. A profound knowledge of the underlying superconductor physics and a systematic investigation are necessary to properly converge the simulation parameters.

**Advances in Science and Technology to Meet Challenges**

Accurate modelling of magnetism in first-principles simulations (challenge *i*) can be made easier by the combined development of algorithms and methods that either speed up convergence of magnetic calculations or accelerate the search of the magnetic ground state. New methods for mapping magnetic interactions would be particularly useful. Most notably, proper implementation of simulations with constrained spin-orbit coupling [11] is essential to systematically explore the landscape of magnetic states.

A better understanding of the superconducting pairing interaction (challenge *ii*) can come from two sides. First and foremost, there is a need for a collaborative effort to develop models and tools to describe and compute the pairing interactions for a particular system, bridging the gap between semi-phenomenological descriptions of superconductivity, strongly correlated and fully ab-initio methods. The ultimate approach is using quantum algorithms and quantum computers to explicitly compute the pairing potential, considering that SCDFT is an exact theory [4]. On the other hand, one should not forgo the practical approach of learning from computational experiments: DFT-BdG methods offer the possibility of exploring different types of pairing and identify the critical ones. This is implemented in a particularly user-friendly fashion in [7].

The extension of superconducting codes to non-local transport would provide robust new ways of classifying the topological nature of zero-energy modes (challenge *iii*). A proof-of-concept simulation of normal/superconductor junction is reported in [7] by combining scattering theory and DFT-BdG formalism. Active development is needed to build upon this proof-of-concept to allow full treatment of non-local transport and superconductivity at the BdG level. Combined with the already available tools, researchers will have a rigorous approach of analyzing the topological character of superconductor-magnet hybrids.

Finally, to be able to model more classes of heterostructures, improvements in computational efficiency (challenge *iv*) are more than welcome. On the software side, by optimizing the memory usage of codes and by exploring all possible avenues of parallelisation, including the use of specialised hardware like GPUs. In the spirit of open science, these improvements are ideally incorporated in shared libraries (such as ELSI and ELPA) such that any code can benefit from this. On the hardware side, by development of faster processors and including these in powerful (exascale) supercomputers that can be accessed by researchers worldwide. Close collaboration between experts in science, software and hardware developers will be indispensable here.

We would also like to state the importance for these tools to be widely accessible in every sense of the word. Not only should the codes and data be openly accessible, but they should be well documented and user friendly. The presence of FAIR (findable, accessible, interoperable, reusable) curated data sets of superconducting pairing interactions and magnetic parameters, as well as the development of automated workflows will drastically impact the field. For instance, workflows would enable high-throughput studies to screen hundreds or even thousands of superconductor-magnet combinations, identifying promising candidates for topological superconductivity. The result is a playing field that is FAIR to all researchers and the acceleration of new discoveries which could benefit everyone.

**Concluding Remarks**

A better understanding of the interplay between magnetism, topology, and superconductivity will allow the prediction of novel candidate materials for hosting topological superconductivity, and perhaps even the tuning of their properties. This could pave the way for powerful new quantum technologies that can help tackle the problems of the future. Furthermore, the advancements necessary to reach this understanding will also benefit the broader field of computational material science for magnets and superconductors and have the potential to yield exciting new discoveries on many fronts.

**Acknowledgements**

ZZ acknowledges the research program "Materials for the Quantum Age" (QuMat) for financial support. This program (registration number 024.005.006) is part of the Gravitation program financed by the Dutch Ministry of Education, Culture and Science (OCW). ZZ has received funding from the European Union's Horizon Europe research and innovation programme under grant agreement No 101130384 (QUONDENSATE). AK acknowledges financial support from Sector Plan Program 2019-23.

**References**

[1] M. Sato and Y. Ando, "Topological superconductors: a review," *Rep. Prog. Phys.*, vol. 80, no. 7, p. 076501, 2017

[2] L. Schneider, P. Beck, J. Neuhaus-Steinmetz, L. Rózsa, T. Posske, J. Wiebe, and R. Wiesendanger, "Precursors of Majorana modes and their length-dependent energy oscillations probed at both ends of atomic Shiba chains," *Nat. Nanotechnol.*, vol. 17, no. 4, Art. no. 4, 2022

[3] S. Kezilebieke, M. N. Huda, V. Vaňo, M. Aapro, S. C. Ganguli, O. J. Silveira, S. Głodzik, A. S. Foster, T. Ojanen, and P. Liljeroth, "Topological superconductivity in a van der Waals heterostructure," *Nature*, vol. 588, no. 7838, Art. no. 7838, 2020

[4] L. N. Oliveira, E. K. U. Gross, and W. Kohn, "Density-Functional Theory for Superconductors," *Phys. Rev. Lett.*, vol. 60, no. 23, pp. 2430–2433, 1988

[5] G. Csire, B. Újfalussy, J. Cserti, and B. Győrffy, "Multiple scattering theory for superconducting heterostructures," *Phys. Rev. B*, vol. 91, no. 16, p. 165142, 2015

[6] P. Rüßmann and S. Blügel, "Density functional Bogoliubov-de Gennes analysis of superconducting Nb and Nb(110) surfaces," *Phys. Rev. B*, vol. 105, no. 12, p. 125143, 2022

[7] R. Reho, N. Wittemeier, A. H. Kole, P. Ordejón, and Z. Zanolli, "Density functional Bogoliubov-de Gennes theory for superconductors implemented in the SIESTA code." *Physical Review B*, vol. 110, no.13, pp. 134505 2024

[8] D. Pashov, S. Acharya, W. R. L. Lambrecht, J. Jackson, K. D. Belashchenko, A. Chantis, F. Jamet, and M. van Schilfgaarde, "Questaal: A package of electronic structure methods based on the linear muffin-tin orbital technique," *Computer Physics Communications*, vol. 249, p. 107065, 2020

[9] B. Nyári, A. Lászlóffy, L. Szunyogh, G. Csire, K. Park, and B. Ujfalussy, "Relativistic first-principles theory of Yu-Shiba-Rusinov states applied to Mn adatoms and Mn dimers on Nb(110)," *Phys. Rev. B*, vol. 104, no. 23, p. 235426, 2021

[10] B. Nyári, A. Lászlóffy, G. Csire, L. Szunyogh, and B. Újfalussy, "Topological superconductivity from first principles. I. Shiba band structure and topological edge states of artificial spin chains," *Phys. Rev. B*, vol. 108, no. 13, p. 134512, 2023

[11] R. Cuadrado, M Pruneda, A García and P Ordejón, *J. Phys. Mater.* vol. 1, p. 015010, 2018

## 4. Theory of Magnons in 2D magnets

Irene Aguilera[1,2], Mexx Regout[3], David Sanz[4], Matthieu J. Verstraete[2,3,5]

[1]Institute for Theoretical Physics, University of Amsterdam 1098 XH Amsterdam, The Netherlands
[2]European Theoretical Spectroscopy Facility (ETSF)
[3]NanoMat/Q-Mat Université de Liège, and European Theoretical Spectroscopy Facility, B-4000 Liège, Belgium
[4]Departamento de Física Aplicada, Universidad de Alicante, 03690, San Vicente del Raspeig, Alicante, Spain
[5]ITP, Physics Department, Utrecht University 3508 TA Utrecht, The Netherlands

**Status**

Spin waves are propagating excitations within a spin texture (typically a ferromagnetic ground state). Spin excitations come in two main forms: magnons, which are quasiparticles of collective spin rotation, and "Stoner" single spin flip excitations. The concept of a magnon was introduced by Felix Bloch in 1930 to explain the temperature dependence of magnetization in ferromagnets: thermal energy excites spin waves, leading to a reduction in net magnetization. Holstein, Primakoff and Dyson furthered this work with a modern quantum mechanical representation of spin operators and considering interactions between the magnons of a bare Hamiltonian. In 1966 Mermin and Wagner found that there should be no long range order for magnets in 1 or 2 dimensions. In the following decades it was shown that FM and AFM order could be maintained in ever thinner magnetic films, down to a few atomic layers, thanks to the presence of magnetic anisotropy which breaks the rotation symmetry of the spins, but the conviction persisted that strictly 2D systems would not host magnetic order.

With the discovery and fabrication of 2D materials, such as graphene, a new avenue for exploring magnetism in reduced dimensions opened up. In 2D systems magnons exhibit unique properties, due to the reduced dimensionality and often enhanced quantum effects. Initial efforts induced magnetism in 2D layers by proximity with a FM substrate. The low spin-orbit coupling and huge mean free paths for spins in graphene have yielded many useful spintronics devices. Interest in 2D magnons surged further with the discovery of intrinsic magnetic ordering in monolayer materials such as $CrI_3$ and $Cr_2Ge_2Te_6$ and CrSBr. Depending on layer thickness, coupling, and dimensionality one can find both in plane (IP) and out of plane (OOP) spin textures, due to competing effects of stray fields, anisotropy and dipole-dipole interactions. The understanding and manipulation of 2D magnons enables applications in spintronics, memories, sensors, information transfer over long distances or times, and opto-magnetics.

Much progress has been made since the inception of magnon theory almost a century ago. Various formalisms have been established for the theoretical description of spin dynamics. Most studies employ the Heisenberg model, which separates magnetic degrees of freedom from the fast motion of electrons. This model is defined by exchange parameters that can be obtained, for instance, from constrained DFT or the LKAG Green's functions method. The exchange coupling can be augmented with the DMI, crystalline anisotropy and higher order spin terms [1]. The Heisenberg model can be solved using linear spin wave (perturbation) theory (LSWT), Monte Carlo stochastic methods, or the LLG equation. LSWT allows for efficient calculation of long-wavelength spin waves but neglects single-particle Stoner excitations and does not provide access to the linewidths of spin-wave resonances, (inverse magnon lifetimes). The Heisenberg model is strictly applicable only to systems with localized moments, such as those with rare-earth magnetic ions, and not to materials with spin-polarized itinerant electrons. While the model still produces reasonable results for long-wavelength excitations in itinerant-magnets, it is unsatisfactory for short-wavelength excitations.

Using these methods there has been intensive exploration of (thermo)dynamics, but also topology and chirality, in particular for 2D systems, e.g. in the form of magnon-skyrmion interactions and Chern magnons [2]. Chirality is visible in surface states, dipolar spin waves, and magnons in altermagnets.

**Current and Future Challenges**
### Realistic devices
To compare with experiments many additional non-magnetic details should be taken into account, such as nanostructuring/texturing, temperature effects, and the local environment (substrates, gating, doping). Fig. 1 shows the temperature evolution of the Seebeck coefficient (electronic transport property) through a magnetic phase transition of CrSBr, in good agreement with experiment [3].

### Interactions with other quasiparticles
Simulations of spin-lattice dynamics [4] and other combinations with first-principles approaches allow to go beyond linear spin wave theory and study interactions of magnons with other magnons, phonons, excitons, etc [1].

### Interactions with Skyrmions
Skyrmions are localized spin textures, which are wound up with a net topological charge [5]. The topology protects them against perturbations, making them potential information carriers in spintronics memories and devices. The nature of magnons in the presence of a skyrmion, and the interaction between the two has recently become a hot topic. Magnons can drag and push skyrmions through scattering, as a route to control information flow [6]. In spontaneous or moiré-induced lattices of skyrmions, additional topological and interaction effects can appear in the spin wave behavior.

### Topological Magnons
Topological magnon insulators feature an inverted gap in the bulk magnon dispersion relation that leads to magnon states localized at the edges of a sample. This holds potential for applications in magnon spintronics, including topologically protected travelling-wave amplifiers and magnon lasers [7]. Experimental evidence of signatures of topological magnons are scarce. One of the most promising candidates is $CrI_3$, whose monolayer has signatures suggesting non-trivial magnon topology [2]. The combination of magnons and topology gives rise to largely unexplored questions, such as the effects of the electron-magnon, exciton-magnon, or phonon-magnon scattering, or to which extent non-Hermitian effects are relevant for applications.

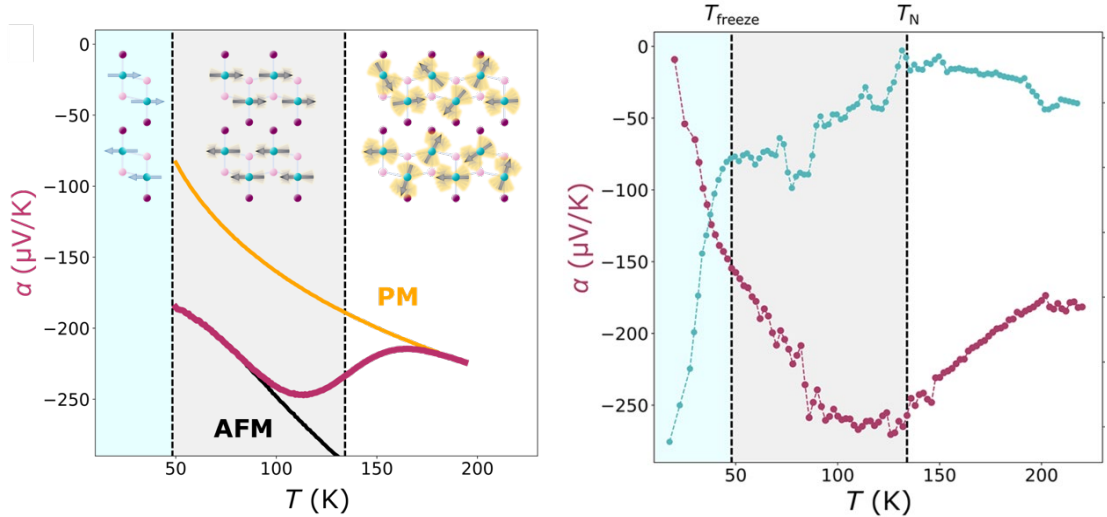

**Figure 4.1.** Temperature evolution of the Seebeck coefficient of monolayer CrSBr, through several magnetic transitions to paramagnetism. Right panel is experiment and left panel theory. Adapted from Ref [3].

**Beyond semiclassical and LSWT methods**.

LSWT has been extensively applied to describe the low-energy spin excitations of magnetic materials with strong correlations, i.e., magnons. However, various novel 2D vdW magnetic heterostructures and monolayers, such as alpha-$RuCl_3$, $RuO_2$ or $NiPS_3$, exhibit important quantum spin fluctuations for which LSWT fails. These fluctuations arise from magnetic frustration including Kitaev interactions, competing exchanges, and may lead to the emergence of quantum spin liquids. New methods that include many-body and entanglement effects have become necessary to accurately describe the spin dynamics of these materials. MPS combined with the DMRG offer a promising alternative.

**Van der Waals magnetic heterostructures.**

Magnetic heterostructures have emerged as a promising framework for new tunable and exotic magnetic phases, due to their strong dependence on stacking order, external perturbations or proximity effects. Their magnon dynamics are largely determined by the interlayer exchange coupling, which is significantly controllable as a consequence of its weak nature and plays a central role. Fig. 2. shows the distribution of the magnetic interlayer coupling yielding magnetic domains when two layers of $CrI_3$ undergo a relative shift, a phenomenon that can be extrapolated to twisted magnetic heterostructures. However, various challenges must be addressed to enable the engineering of magnetic phases in these systems. In particular, numerical methods for obtaining the interlayer coupling must be improved. Although ab initio and LKAG methods provide reasonable results, the dependence on stacking and the large sizes of twisted superlattices prevent from its accurate results.

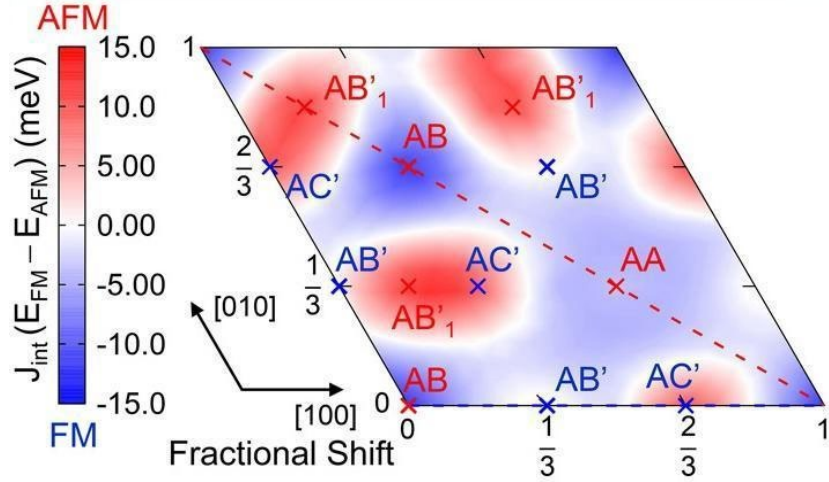

**Figure 4.2.** Variation of the magnetic exchange coupling in bilayer CrI$_3$, depending on layer shifts and local stacking. Adapted from Ref. [8].

**Advances in Science and Technology to Meet Challenges**

New theoretical approaches need to be developed that can include more complex effects and also treat larger systems.

### Improved fitting of magnetic model Hamiltonians.

There is a push in the community to homogenize and compare approaches to fit Heisenberg exchange parameters, whether from energy differences (4 state method), the generalized Bloch theorem, TDDFT, or perturbation theory (LKAG). This is an important avenue to enable robust and reproducible nanoscale interactions. On the other end of the scale, accurately simulating magnon interactions with large-scale structures like skyrmion lattices and long wavelength spiral ground states[9] will require advances in micromagnetic and continuum models.

### Post-Heisenberg models.

There are various approaches to go beyond the Heisenberg model. The most direct way is to add more magnetic interaction terms like DMI, crystalline/exchange anisotropy and higher order spin terms like the biquadratic interaction. Another avenue is the allowance of variation in the magnitude of the magnetic moment which can be modelled with DFT, and yields for example the Landau-Lifshitz-Bloch equations. Another possibility, which does not assume the localized moments, is the spin-cluster expansion where a set of one-spin basis functions is used.

### Many-body perturbation theory

New fully first principles methods based on MBPT and TDDFT are emerging. Both single-spin (Stoner) and collective spin excitations arise naturally as poles in the full magnetic susceptibility, which describes the correlated motion of electron-hole pairs influenced by a screened electron-electron interaction. Recent initiatives include self-consistent *GW* and *T*-matrix approximations [10]. This method naturally encompasses both magnon (with spin flip) and exciton (without) excitations. The latter play a key role in 2D materials, due to their reduced screening. Although the mathematics of the general method exist, current codes are restricted to the calculation of the transverse magnetic response (excluding excitons). Future extensions should also include electron-exciton scattering in the electron-magnon GT electronic self-energy [11]. Both MBPT and TDDFT have achieved good agreement with experimental magnon dispersions in 3d ferromagnets, but applications to antiferromagnets are scarce. Further theoretical developments are required to extend these methods to more complex materials, including antiferromagnets and altermagnets.

### MPS and DMRG

Where semiclassical spin dynamics theory falls short, quasi-exact many-body methods, such as MPS and time-dependent DMRG have proved invaluable. They provide a novel platform for the study of magnetic ground states and excitations in more exotic systems. Recently, these methods have been applied to the context of magnons in 2D materials, to address finite lifetimes from magnon-magnon interactions [12] or frustrated magnetic materials with spin liquid phases.

## 4. Concluding Remarks

In conclusion, existing theories for the calculation of magnons are fully applicable to 2D materials and heterostructures, but strong challenges have emerged. These come from the limitations of the Heisenberg model, of current methods to parameterize and solve it, from material complexity, and from coupling to the environment, thermal and particle baths, and external fields.

**Acknowledgements**

*D. S. acknowledges financial support from Generalitat Valenciana through the CIDEGENT program (CIDEGENT/2021/052). MJV and MR acknowledge ARC project DREAMS (G.A. 21/25-11) funded by Federation Wallonie Bruxelles and ULiege, and Excellence of Science project CONNECT number 40007563 funded by FWO and FNRS.*

**References**

[1] S. M. Rezende, *Fundamentals of Magnonics*. Springer Science+Business Media, 2020

[2] P. A. McClarty, "Topological Magnons: A Review," *Annual review of condensed matter physics*, vol. 13, no. 1, pp. 171–190, 2022

[3] A. Canetta, S. Volosheniuk, S. Satheesh, J. P. Alvarinhas Batista, A. Castellano, R. Conte, D. G. Chica, K. Watanabe, T. Taniguchi, X. Roy, H. S. J. van der Zant, M. Burghard, M. J. Verstraete, and P. Gehring, "Impact of Spin-Entropy on the Thermoelectric Properties of a 2D Magnet," *Nano letters*, vol. 24, no. 22, pp. 6513, 2024

[4] Y. Zhou, J. Tranchida, Y. Ge, J. Y. Murthy, and T. S. Fisher, "Atomistic simulation of phonon and magnon thermal transport across the ferromagnetic-paramagnetic transition," *Physical review B,* vol. 101, no. 22, 2020

[5] Q. Tong, F. Liu, J. Xiao and W. Yao, "Skyrmions in the Moiré of van der Waals 2D Magnets", *Nano Letters,* vol. 18, no. 11, pp. 7194-7199, 2018

[6] C. Schütte and M. Garst, "Magnon-skyrmion scattering in chiral magnets," *Physical Review B*, vol. 90, no. 9, 2014

[7] J. Cenker, B. Huang, N. Suri, P. Thijssen, A. Miller, T. Song, T. Taniguchi, K. Watanabe, M. A. McGuire, D. Xiao, and X. Xu, "Direct observation of two-dimensional magnons in atomically thin CrI3," *Nature Physics*, vol. 17, no. 1, pp. 20–25, 2021

[8] Nikhil Sivadas, S. Okamoto, X. Xu, C. J. Fennie, and D. Xiao, "Stacking-Dependent Magnetism in Bilayer $CrI_3$," *Nano Letters*, vol. 18, no. 12, pp. 7658–7664, 2018

[9] A. Mook, B. Göbel, J. Henk and I. Mertig, "Magnon transport in noncollinear spin textures: Anisotropies and topological magnon Hall effects", Physical review B, vol. 9, pp. 020401(R), 2017

[10] C. Friedrich, M. C. T. D. Müller, and S. Blügel, "Many-Body Spin Excitations in Ferromagnets from First Principles," *Springer eBooks*, pp. 919–956, 2020

[11] M. C. T. D. Müller, S. Blügel, and C. Friedrich, "Electron-magnon scattering in elementary ferromagnets from first principles: Lifetime broadening and band anomalies," *Physical review. B./Physical review. B*, vol. 100, no. 4, 2019

[12] P. A. McClarty, X.-Y. Dong, M. Gohlke, J. G. Rau, F. Pollmann, R. Moessner, and K. Penc, "Topological magnons in Kitaev magnets at high fields," *Physical review B*, vol. 98, no. 6, 2018

## 5. Magnetic and optical frontiers in 2D semiconductors

Riccardo Reho[1], Marcos H. D. Guimarães[2] and Zeila Zanolli[1]

[1] Chemistry Department, Debye Institute for Nanomaterials Science and ETSF, Condensed Matter and Interfaces, Utrecht University, PO Box 80.000, 3508 TA Utrecht, The Netherlands.
[2] Zernike Institute for Advanced Materials, University of Groningen, Groningen, The Netherlands

**Status**

Two-dimensional magnetic systems represent a fascinating frontier in materials science, uniquely combining the reduced dimensionality effects with magnetic interactions. Due to their low dimensionality, these materials possess physical phenomena distinct from their bulk counterparts, such as enhanced excitonic effects and highly electrically and optically tunable magnetic orders, which facilitate light-matter-spin coupling [1].

Chromium triiodide ($CrI_3$) was among the first 2D systems discovered to exhibit intrinsic ferromagnetism. 2D magnets are characterized by strong magnetic anisotropy; their physical properties are dominated by the spin's orientation along the easy axis. The magnetic order, ranging from FM to AFM, is dictated by the complex spin exchange interaction. Interestingly, 2D magnets are usually antiferromagnetic below Néel temperature. Both IP and OOP FM and AFM 2D magnets have been demonstrated [2]. Additionally, MLs of these materials can be assembled into vdW multilayer structures (Figure 5.1.a), introducing interlayer magnetic order and providing additional controlling knobs through geometric factors (stacking, straining, twisting and moiré modulations [3]) and proximity effects, such as coupling with a semiconductor operating in the visible range [4]. These features enable precise control over opto-spintronic excitations, with multilayer vdW systems exhibiting multiple distinguishable and easily switchable magnetic states including FM, AFM, FiM, and metastable states due to spin-flip and spin-canting phenomena. Hence, they provide a versatile platform for exploring new physics in low-dimensional systems and hold potential for revolutionary applications in spintronics and quantum computing.

Similar to opto-electronic systems, which control and detect light through electronic transitions and/or excitonic absorption/recombination, 2D magnets offer promising prospects for quantum technologies as transductors of light and spin interactions. The underlying physics involves the interplay between magnetic ordering and quasi-particles like excitons, phonons, and magnons, necessitating a comprehensive theoretical framework that incorporates external magnetic fields, intrinsic magnetism, spin-orbit coupling, and light-matter interactions. 2D magnets function across a spectrum from GHz to X-rays and engage in both linear and non-linear optical processes, such as SHG. Magneto-optical effects in 2D magnets provide crucial "readout" (detection) and "write-in" (control) capabilities for information processing, manipulating magnetic orders and spin precessions. Furthermore, these systems can be engineered for advanced spintronic applications, including magnetic tunnel junctions and magnetoresistance devices.

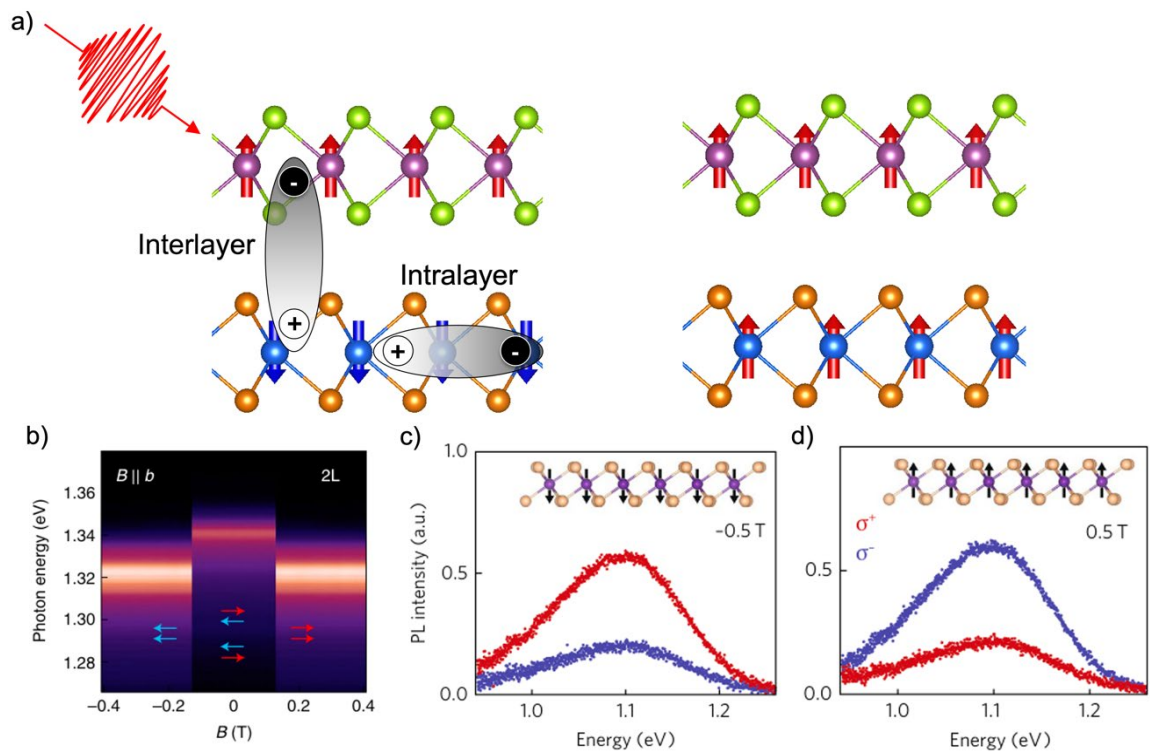

**Figure 5.1.** a) Schematic of a prototypical van der Waals bilayer heterostructure displaying: a ferromagnetic (FM) and antiferromagnetic (AFM) interlayer ordering, an interlayer and an intralayer exciton. b-c) Measured photoluminescence spectra for monolayer CrI$_3$ with circularly polarized σ$^+$(red) or σ$^-$ (blue) components. The polarization flips sweeping the external magnetic field from -0.5 T to 0.5 T, while it remains unchanged when the field is lowered to 0T. Reproduced from [5], with permission from Springer Nature. d) Measured photoluminescence spectra of bilayer CrSBr as a function of magnetic field (B) along the crystallographic b-axis. An abrupt change in intensity and peak position is observed at a critical field of B$_c$=0.134 ± 0.003 T. This result is due to a spin flip transition from AFM to FM interlayer ordering, which modifies the electronic and excitonic properties. Reproduced from [6], with permission from Springer Nature.

**Current and Future Challenges**

Recent studies, such as the exploration of excitonic properties in van der Waals AFM NiPS$_3$ [5], [6] or monolayer ferromagnets, as CrI$_3$ [7], highlight the significant interest in understanding and harnessing these interactions for advanced technological applications.

For instance, in NiPS$_3$, the AFM order enriches the excitonic states, creating coherent exciton-polariton many-body states that are observable even at room temperature. This coherence is a breakthrough, demonstrating the substantial impact of magnetic ordering on excitonic properties. The microscopic origin of this exciton-polariton state is still under debate between Zhan-Rice singlet/triplet transitions and Wannier-like excitons.

Furthermore, recent studies on CrSBr, CrI$_3$ and CrBr$_3$ [7, 2, 8] offer illustrative examples of how magnetic and excitonic properties interplay. CrSBr exhibits a unique ability to manipulate interlayer electronic coupling through its magnetic ordering (Figure 5.1.b). This capability allows for the modulation of excitonic transitions by altering the relative magnetic orientation between layers, providing a method to tune the electronic properties dynamically. The control of these transitions is crucial for applications in devices where the magnetization can be used to switch or modulate electronic states on demand.

On the other hand, CrI$_3$ demonstrates intriguing photoluminescence behaviour linked to its unique electronic structure. The ligand-field transitions in CrI$_3$ influence the excitonic behaviour, leading to distinct luminescence for which the polarization is highly sensitive to the magnetization direction (Figures 5.1.c and 5.1.d). This sensitivity is characterized by the circular polarization of the emitted light, which can be controlled by altering the magnetic state, applying gate voltages or magnetic fields. Such properties underscore the potential of 2D magnetic semiconductors to host new optoelectronic functionalities, where the magnetic state directly influences optical outputs, opening avenues for innovative light-based technologies and optospintronics.

**Advances in Science and Technology to Meet Challenges**

The field of 2D magnetic materials is advancing fast and advanced characterization techniques, and optical techniques in particular, have played a leading role in this. This stems from the very high light-matter interaction and magneto-optical responses, largely thanks to their excitonic behaviours. But still many questions remain on the optical and excitonic properties of these materials and how we can get control over them. Many works rely on the magneto-optic Kerr effect or magnetic circular dichroism amplitudes to characterize changes in the magnitude of the magnetization (saturation magnetization) with, for example, doping and electric fields [9]. However, it is still unclear if the observed changes arise from changes in the magnetization, or in the material's magneto-optical strength, or structural effects [10]. Additionally, the characterization of more complex excitonic states, such as charged excitons, bi-excitons and excited excitonic states, are largely unexplored. An increase in sample quality, both from the growth and from the van der Waals heterostructure perspectives, will be fundamental for further research in this direction, in a similar way as happened for 2D semiconductors. The operating wavelengths accessible to 2D magnets also present challenges. For instance, THz spectroscopy, requiring a beam size on the micrometre scale, remains unachievable, and in the GHz spectrum, the coupling between magnons and excitons is weak. In this direction, the further development of near-field techniques should allow for the exploration of 2D magnets in the THz domain with high spatial resolution.

The growth of large-scale high-quality 2D magnetic semiconductors with high operating (Curie/Néel) temperatures is one of the main challenges in the field. While 2D magnetic metals have currently overcome room-temperature operation for Fe-based compounds (e.g. $Fe_3GaTe_2$ and $Fe_5GeTe_2$), intrinsic 2D magnetic semiconductors are still far from it. Here, low carrier densities are fundamental for maintaining their desirable optical properties, such a strong excitonic behaviour. In this sense, a strong and concerted theoretical and experimental approach should be taken to find different possibilities for achieving this goal. Here, a few possibilities would be exploiting proximity effects with high spin-orbit coupling (insulating) substrates and other 2D materials, or combining different substitutional dopants to enhance the magnetic exchange and anisotropy while also maintaining a low carrier density. Meeting this challenge would fulfil one of the key goals for new semiconducting devices that combine semiconducting and spintronic functionalities for applications in, for example, in-memory computing.

From a theoretical standpoint, 2D magnetic semiconductors can be simulated using DFT. However, DFT faces significant challenges when applied to strongly correlated materials. These materials exhibit strong electron-electron interactions that are not well captured. The underlying reasons are, among others, the approximated exchange-correlation functional and the presence of local and fluctuating interactions (dynamic correlations). In order to address these issues, several methods have been developed such as DFT + U, hybrid functional, and GW approximations. DFT + U introduces a Hubbard

U term to account for on-site Coulomb interactions to account for the strong Coulomb interaction among localized electrons, but the choice of U is arbitrary, and it suffers from double counting problem. Moreover, DFT+U assumes a local self-energy and might not be suitable to model long-range magnetic ordering or spin fluctuations.

DFT in combination with many-body perturbation techniques such as the GW approximation and Bethe-Salpeter Equation provides a detailed quantum microscopic view that comprehensively captures magnetic (including spin-orbit coupling) and excitonic effects. The simulations' challenges are due to the millielectron volt energy scales associated with exciton-magnetic effects, which are at the edge of the accuracy of these methods. In addition, simulating van der Waals heterostructures, particularly those involving multiple layers with varied magnetic orders, requires large simulation cells (hundred atoms), hence becoming extremely demanding in terms of computational time and memory. These challenges can be overcome either with the advent of the exascale era ($10^{18}$ double precision floating point operations per second) or through the development of more cost-effective theoretical models that maintain high accuracy and precision, such as Wannierized models of the excitonic Hamiltonian.

The Wannierization technique offers a refined description of a quantum Hilbert space using a subset of localized basis functions, significantly reducing computational expenses. While many ground state DFT codes currently integrate the Wannierization process [11], the development of maximally localized Wannier functions for excitons is still in its early stages. A recent proof-of-concept study on LiF reproduced singlet and triplet exciton dispersions at a fraction of the computational cost of ab initio methods [12]. Further advancements in this technique could address challenges in predicting exciton dynamics, lifetime, (non-)linear optical response and exciton transport. For example, low-dimensional materials exhibit non-analytical exciton band structures, necessitating fine sampling of the Brillouin Zone. Efficient Fourier interpolation of second-principles exciton tight-binding models could overcome issues related to coarse grids and provide optimized alternatives for calculating exciton-phonon/exciton-exciton couplings. Additionally, this approach could pave the way for computing Berry curvature associated with exciton bands and exploring their topological properties.

**Concluding Remarks**
The research on 2D magnetic semiconductors promises to open new pathways for developing devices that leverage their unique magnetic and excitonic properties. There is a vast potential for creating tunable optoelectronic devices that can be manipulated using magnetic fields. Moreover, the ability to engineer the ligand fields in materials like $CrI_3$ suggests possibilities for customizing optical properties for specific applications, ranging from sensors to quantum computing. While challenges remain, the ongoing research is crucial for unlocking the full potential of these materials. Encouraging collaboration and sustained investigation will be key to overcoming obstacles and achieving breakthroughs in this exciting field.

**Acknowledgements**
ZZ and MHDG acknowledge the research program "Materials for the Quantum Age" (QuMat) for financial support. This program (registration number 024.005.006) is part of the Gravitation program financed by the Dutch Ministry of Education, Culture and Science (OCW). RR acknowledges financial support from Sector Plan Program 2019-2023. ZZ acknowledges the European Union's Horizon Europe research and innovation programme under grant agreement No 101130384 (QUONDENSATE). MHDG acknowledges the financial support of the European Union through grant ERC, 2D-OPTOSPIN, 101076932, the Dutch Research Council (NWO—OCENW.XL21.XL21.058), and the Zernike Institute for Advanced Materials.

**References**

[1] S. Liu, I. A. Malik, V. L. Zhang, and T. Yu, "Lightning the Spin: Harnessing the Potential of 2D Magnets in Opto-Spintronics," *Advanced Materials*, pp. 2306920, 2023

[2] N. P. Wilson *et al.*, "Interlayer electronic coupling on demand in a 2D magnetic semiconductor," *Nature Materials*, vol. 20, no. 12, pp. 1657–1662, 2021

[3] R. Reho, A. R. Botello-Méndez, D. Sangalli, M. J. Verstraete, Z. Zanolli "Excitonic response in TMD heterostructures from first-principles: impact of stacking order and interlayer distance,"*Physical Review B, vol.* 10, p. 035118, 2024

[4] Z. Zanolli, C. Niu, G. Bihlmayer, Y. Mokrousov, P. Mavropoulos, M. J. Verstraete, D. Blügel, "Hybrid quantum anomalous Hall effect at graphene-oxide interfaces," *Physical Review B*, vol. 98, no. 15, p. 155404, 2018

[5] S. Kang *et al.*, "Coherent many-body exciton in van der Waals antiferromagnet $NiPS_3$," *Nature*, vol. 583, no. 7818, pp. 785–789, 2020

[6] T. Klaproth *et al.*, "Origin of the Magnetic Exciton in the van der Waals Antiferromagnet $NiPS_3$," *Physical Review Letter*, vol. 131, no. 25, p. 256504, 2023

[7] K. L. Seyler *et al.*, "Ligand-field helical luminescence in a 2D ferromagnetic insulator," *Nature Physics*, vol. 14, no. 3, pp. 277–281, 2018

[8] Z. Zhang, J. Shang, C. Jiang, A. Rasmita, W. Gao, and T. Yu, "Direct Photoluminescence Probing of Ferromagnetism in Monolayer Two-Dimensional $CrBr_3$," *Nano Letters*, vol. 19, no. 5, pp. 3138–3142, 2019

[9] M. Ersfeld, et al."Spin States Protected from Intrinsic Electron-Phonon Coupling Reaching 100 ns Lifetime at Room Temperature in $MoSe_2$," *Nano Letters,* vol 19, no. 6, pp. 4082–4090, 2019

[10] T. Sohier, P. M. M. C. de Melo, Z. Zanolli, M. J. Verstraete "The impact of valley profile on the mobility and Kerr rotation of transition metal dichalcogenides," 2D Materials vol. 10, no. 2, p. 025006, 2023

[11] N. Marzari, A. A. Mostofi, J. R. Yates, I. Souza, and D. Vanderbilt, "Maximally localized Wannier functions: Theory and applications," *Review Modern Physics*, vol. 84, no. 4, pp. 1419–1475, 2012

[12] J. B. Haber, D. Y. Qiu, F. H da Jornada, J. B. Neaton, "Maximally localized exciton Wannier functions for solids." *Physical Review B*, vol. 108, p. 125118, 2023

# 6. Magneto-optics in two-dimensional magnets

Rixt Bosma and Marcos H. D. Guimarães

Zernike Institute for Advanced Materials, University of Groningen, Groningen, The Netherlands

**Status**

MO effects connect a material's optical conductivity (σ) with its magnetisation. These effects were first discovered in the 19th century [1], and have been used as a contactless, low energy probe for magnetization ever since.

MCB denotes the difference in optical refraction between left- and right-handed circularly polarized light, related to the imaginary part of σ. MCD probes the difference in absorption between both light helicities, given by the real part of σ. In the polar configuration, where the propagation direction of the light is perpendicular to the sample, these effects originate from the antisymmetric off-diagonal elements in the complex optical conductivity tensor, which are directly proportional to the magnetisation $M$. MCB is observed by monitoring the polarization angle of linearly polarized light upon transmission (Faraday effect) or reflection (MOKE). Changes in the ellipticity of the light polarization in either configuration indicate the material's MCD.

Both MCB and MCD are sensitive to $M$, and can therefore be used as a direct probe for magnetic ordering. These all-optical techniques have proven to be extremely useful for monitoring magnetisation in two-dimensional (2D) magnets. In 2017, two independent studies used MOKE to capture the sample magnetization of ultrathin 2D magnetic materials in an external magnetic field. $CrI_3$ was characterized to the monolayer limit [2], as displayed in Fig. 6.1b-d. Strong FM behaviour is observed in odd layer numbers, but for a bilayer sample A-type antiferromagnetic behaviour is prominent, with a spin-flop transition at 0.7 T. For the 2D FM $Cr_2Ge_2Te_6$ remnant magnetization was shown down to bilayer samples [3], as displayed in the Kerr microscopy image in Fig. 6.1e-f.

In contrast to conventional magnetometry techniques, optical techniques can be easily performed with high lateral spatial resolution, usually in the order of one µm. Additionally, they require low energy and are non-destructive, given that the light illumination intensity is low enough – usually in the order of 100 mW/mm$^2$ and below. 2D magnets show a very strong magneto-optic efficiency due to excitonic resonances. This makes magneto-optics one of the go-to techniques for characterizing 2D magnets.

MO effects can be used to probe magnetization while sweeping the external field to record the sample's magnetostatic properties and type of ordering, as was done in the aforementioned experiments on $CrI_3$ and $Cr_2Ge_2Te_6$. Additionally, using optical pump-probe schemes allow for the quantification of the magnetisation dynamics. Fig. 6.1g shows one such experiment, where the Faraday rotation of $Cr_2Ge_2Te_6$ is monitored over time after exciting the sample with a powerful laser pulse [4]. These measurements yield information on the precision and damping parameters of the 2D magnet. Time-resolved magneto-optics on 2D magnets are further elaborated on in section 13 of this roadmap.

In addition to MCB and MCD, a material's magnetization may also change its other optical properties. Magneto-Raman – though not able to indicate magnetic order directly – is sensitive to modulations of magnetic vibrational modes or zone-folding effects on the phonon spectra due to magnetic order. This technique has, for example, been used to study magnon modes in CrI₃ [5]. A material's photoluminescence can also depend on its magnetization. In CrSBr, a change in optical exciton resonance depending on the alignment of magnetisation in layers of this A-type antiferromagnet CrSBr was shown [6]. A strong transient reflectivity signal was shown for this material, showing clear oscillations corresponding to magnetisation precession[7].

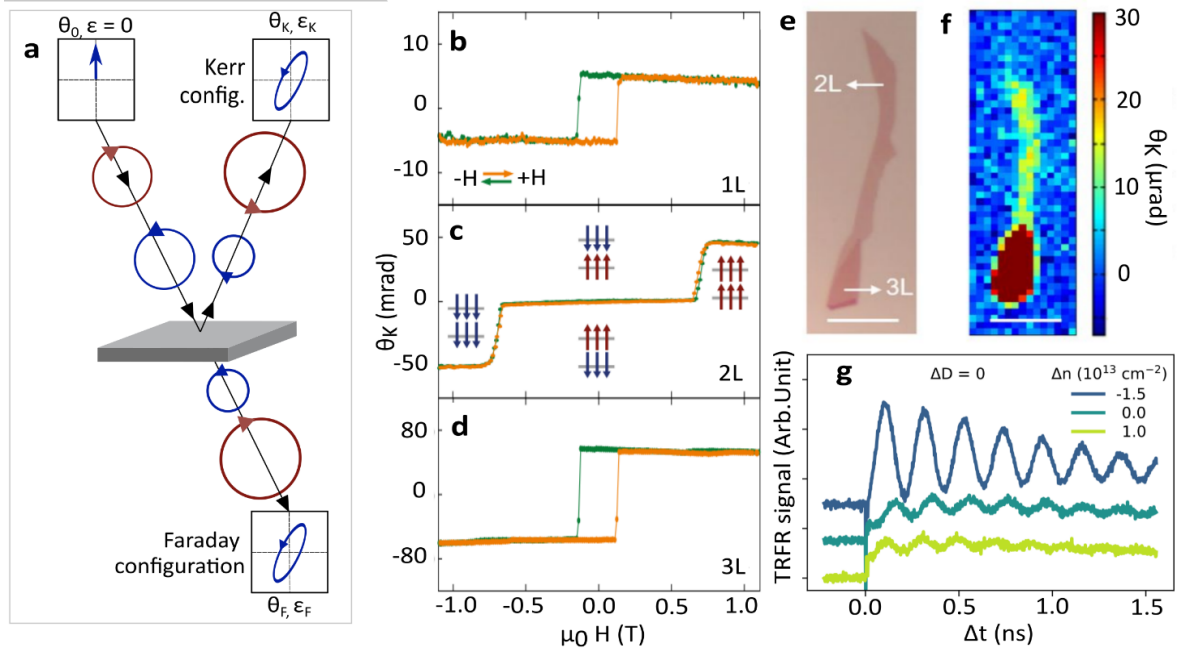

**Figure 6.1. a** Graphical representation of magneto-optical effects, where left- and right-handed components of incident light reflect and transmit from magnetic materials with different intensity and phase, resulting in a change in polarization angle and ellipticity, respectively. In fig. **b,c,d** the Kerr rotation of CrI₃ with out of plane external field is displayed for 1, 2 and 3 van der Waals layers respectively. The material behaves as a FM for even layer numbers, but antiferromagnetically for the bilayer. **E** and **f** show a thin flake of Cr₂Ge₂Te₆, where **e** is a conventional microscopy image while **f** shows a spatial scan of the Kerr rotation measured on the flake. In **g**, time resolved Faraday rotation is displayed for a Cr₂Ge₂Te₆ flake. A strong change in magnetization dynamics parameters is observed with when altering the charge carrier density in the magnet. Panels **b-d** is adapted from [2], **e-f** are adapted from [3] and **g** from [4].

**Current and Future Challenges**

Magneto-optics has been vital for the characterization of 2D magnets since their discovery and will likely continue to be highly useful in the future. However, several challenges are associated with the use of MO techniques.

MCD and MCB are proportional to magnetization $M$, making them straightforward to apply to 2D ferromagnetic materials. However, these techniques fall short in characterizing AFMs because the lack of net magnetization renders the probes insensitive. For *A*-type AFMs, where individual layers are ferromagnetically coupled, but interlayer coupling is antiferromagnetic, MCD and MCB can still yield valuable information, as demonstrated for the CrI₃ bilayer (Fig 6.1c).

MLB and MLD can also monitor magnetization. These parameters relate to the difference in optical conductivity with light polarization parallel and perpendicular to the applied external field. Experimentally known as the Voigt and Cotton-Mouton effects, they are proportional to $M^2$. Even though these effects have not been directly measured in 2D magnets, linear dichroism has been found to arise in zig-zag type 2D AFM, like NiPS₃, giving a direct measure of the magnetic ordering [8].

Magneto-optic spectroscopy, which monitors the MO response with varying energy of incoming light, has proven valuable in assessing the electronic structure of molecules and bulk semiconductors. Systematic characterizations for 2D magnets are still lacking. Since the MO signal depends on light interaction with electronic resonances, altering the excitation wavelength can achieve a stronger MO effect. Additionally, the magnetization-dependent light matter interaction can also be used to optically control magnetization at ultrashort timescales. By utilizing phenomena such as the inverse Cotton-Mouton and the inverse Faraday effect, one could envision an all-optical control over the magnetization [4], potentially leading to all-optical switching, a key ingredient for applying these systems in new magneto-photonic circuits. Nonetheless, the microscopic mechanisms behind opto-magnetic phenomena remain to be explored in 2D magnets.

Another key advantage of van der Waals materials is their ability to be assembled into vertical heterostructures, combining or modifying specific properties of individual layers. Advancement in stacking techniques has facilitated the fabrication of multi-layer heterostructures. When using optically transparent materials such as hBN, the 2D materials within these heterostructures can still be effectively characterized using magneto-optical techniques. For example, devices of $CrI_3$ and $Cr_2Ge_2Te_6$ surrounded by hBN dielectric and thin graphite gate electrodes were used to control their static and dynamic magnetic properties through electrostatic gating[8][9]. Nonetheless, the electrostatic control of the magnetization dynamics demonstrated is still somewhat modest. Additionally, 2D magnetic semiconductors, for which the electrostatic control should be most effective as compared to metals, have low operation temperatures. Therefore, one of the key challenges in the field is the discovery of new 2D magnetic semiconductors which can be effectively controlled with electrostatic gates and that operate at (or close to) room temperature. One potential alternative is the use of 2D diluted magnetic semiconductors based on TMDs, such as V-doped $WSe_2$ [10]. These systems have shown magnetic ordering at room temperature while still maintaining a high light-matter interaction. While most works so far focused on photoluminescence and magnetometry for the characterization of these materials, we believe that the use of magneto-optical spectroscopy can highlight key features of their electronic structure, such as magnetic exchange-split excitons.

**Advances in Science and Technology to Meet Challenges**
In bulk magnetic materials, signal enhancement is often carried out by patterning of photonic or plasmonic structures. Nanofabrication techniques and nanoplasmonic architectures are continuously evolving, yielding more precise control over these structures and therefore larger enhancement of MO signals. In 2D magnets, the use of plasmonic nanostructures may also increase the signal strength significantly, but that remains to be demonstrated. In line with this, simple structures consisting of thin layers, often used experimentally, can strongly attenuate or enhance the MO response due to thin-film interference [10]. As a result, selecting substrates with appropriate oxide or hBN thicknesses is crucial. Creating optical cavities can exploit interference to increase light-matter interaction, enhancing signal strength. These cavities can also be engineered to support specific resonant modes matching the excitation wavelengths, further boosting light-matter interaction and hence the optical control over the magnetization and the strength of the signals.

A key challenge for 2D magnets lies in the field of magnonics, where spin-waves are used to transfer or process information. For this, one requires low magnetic damping, with Gilbert damping constants preferably below $10^{-3}$. However, the best results for 2D magnets so far have been just shy of this, and still orders of magnitude higher than the state-of-the-art material in magnonics – YIG – which shows a Gilbert damping constant around $5 \cdot 10^{-5}$ [11]

Another promising frontier in magneto-optics involves the use of increasingly high energies and shorter timescales. These approaches are designed to push the boundaries of current capabilities, potentially leading to new insights and applications, and giving important information on the atomic orbitals involved in these ultrafast magnetism dynamic processes. High-energy excitations, such as using X-rays, can probe deeper into the material's electronic structure with chemical sensitivity, while ultrafast timescales allow the observation of transient phenomena and dynamic processes that are otherwise inaccessible.

**Concluding Remarks**

MO effects are essential for probing the magnetization of materials, particularly 2D magnets. Techniques like magnetic circular birefringence and magnetic circular dichroism provide contactless, low-energy methods to investigate magnetic properties with high spatial resolution and sensitivity, crucial for thin 2D materials. Together with methods like magneto-Raman spectroscopy and magnetisation dependent photoluminescence, these all-optical techniques allow for a comprehensive characterization of magnetic interactions and excitations in 2D systems. However, many challenges remain for these systems, particularly on the combination of magneto-optics and photonics and on a more thorough understanding of their magneto-optical and opto-magnetic properties.

**Acknowledgements**

We are thankful to the Lorentz Centre in Leiden, the Netherlands, for hosting the workshop on Quantum Magnetic Materials (Oct. 2023), which provided the opportunity to carry the initial discussions which lead to the writing of this article. This work was supported by the Dutch Research Council (NWO, OCENW.XL21.XL21.058), the research program "Materials for the Quantum Age" (QuMat – registration number 024.005.006) part of the Gravitation program financed by the Dutch Ministry of Education, Culture and Science (OCW), the Zernike Institute for Advanced Materials, and the European Union (ERC, 2D-OPTOSPIN, 101076932).

**References**

[1] J. Kerr, 'A new relation between electricity and light: Dielectrified media birefringent', *The London, Edinburgh and Dublin Philosophical Magazine ad Journal of Science*, vol. 50, no. 332, pp. 337–348, 1875

[2] B. Huang *et al.*, 'Layer-dependent ferromagnetism in a van der Waals crystal down to the monolayer limit', *Nature*, vol. 546, no. 7657, pp. 270–273, Jun. 2017

[3] C. Gong *et al.*, 'Discovery of intrinsic ferromagnetism in two-dimensional van der Waals crystals', *Nature*, vol. 546, no. 7657, pp. 265–269, 2017

[4] F. Hendriks, R. R. Rojas-Lopez, B. Koopmans, and M. H. D. Guimarães, 'Electric control of optically-induced magnetization dynamics in a van der Waals ferromagnetic semiconductor', *Nature Communications*, vol. 15, no. 1, 2024

[5] W. Jin *et al.*, 'Raman fingerprint of two terahertz spin wave branches in a two-dimensional honeycomb Ising ferromagnet', *Nature Communications*, vol. 9, no. 1, 2018

[6] N. P. Wilson *et al.*, 'Interlayer electronic coupling on demand in a 2D magnetic semiconductor', *Nature Materials*, vol. 20, no. 12, pp. 1657–1662, 2021

[7] Y. J. Bae *et al.*, 'Exciton-coupled coherent magnons in a 2D semiconductor', *Nature*, vol. 609, no. 7926, pp. 282–286, 2022

[8] K. Hwangbo *et al.*, 'Highly anisotropic excitons and multiple phonon bound states in a van der Waals antiferromagnetic insulator', *Nature Nanotechnology*, vol. 16, no. 6, pp. 655–660, 2021

[9] S. Jiang, L. Li, Z. Wang, K. F. Mak, and J. Shan, 'Controlling magnetism in 2D CrI3 by electrostatic doping', *Nature Nanotechnology*, vol. 13, no. 7, pp. 549–553, 2018

[10] F. Hendriks and M. H. D. Guimarães, 'Enhancing magneto-optic effects in two-dimensional magnets by thin-film interference', *AIP Advances*, vol. 11, no. 3, 2021

[11] H. Chang *et al.*, 'Nanometer-Thick Yttrium Iron Garnet Films with Extremely Low Damping', *IEEE Magnetics Letters*, vol. 5, 2014

# 7. Multiferroicity in 2D materials

Jagoda Sławińska[1], Jose Lado[2], Matthieu J. Verstraete[3]
[1]University of Groningen (NL), [2]Aalto University (FI), [3]Universities of Liege (BE) and Utrecht (NL), and European Theoretical Spectroscopy Facility

**Status**

*Introduction.* MF materials exhibit coupling between at least two ferroic orders: elasticity, electricity, and magnetism. Particularly significant are magnetoelectric materials that simultaneously host electric polarization and magnetic order [1]. If these orders are coupled, one can control magnetic states using electric fields and vice versa, paving the way for innovative device designs. 2D MFs offer unique advantages by enabling magnetic and charge proximity effects to alter the electronic structures of other 2D materials. MF materials are categorized as type I if their magnetic and ferroelectric orders have different origins and type II if one order induces the other, indicating strong MF coupling. In 2DMF, the most common mechanisms involve spin-orbit coupling and non-collinear magnetism (type-II). These properties make 2DMF crucial for developing artificial vdW heterostructures and studying quasiparticle excitations like ferrons, magnons, and magnon polarons.

*Simulations.* Research on 2DMF has so far largely relied on theoretical simulations using density functional theory and effective Hamiltonian models to estimate MF coupling and response to external stimuli. These simulations focus on predicting the intrinsic multiferroic couplings in various materials. The predicted 2DMFs include transition metal (TM) halides and chalcogenides, involving metals like Cr, Ni, and V. These materials are easier to switch between magnetic states than classical ferromagnets, but this also implies lower critical temperatures. Other studied materials include carbides (e.g. MXenes), pnictides, MOFs, and various oxide monolayers. TM halides like CrSX, CrNX2, VOX2, MnNX, and GaTeX, particularly those with iodine, exhibit strong spin-orbit interactions that facilitate coupling to the lattice and structure. The chalcogenides, including mono-chalcogenide layers and some TMDs, display low-symmetry structures prone to ferroelectric and ferroelastic instabilities in bulk, akin to traditional phase change alloys.

*Methods to introduce MF coupling.* MF coupling between 2D materials can be introduced via:
1. Intercalation: Introducing atoms with different ionicity between 2D layers can create local electrical polarization, and adding magnetic ions in a 2D matrix, such as MoS2 or Bi2Se3, will produce (local) magnetization. This method resembles the approach used in dilute magnetic semiconductors.
2. Multilayering: Using twisted homobilayers or different ferroic monolayers can break (planar or inversion) symmetry and induce polarization. This strategy has been successfully applied in macroscopic multiferroic devices, such as BaTiO3 combined with a ferromagnet. At the nanoscale, the polar order is weaker, but the coupling (transmitted through strain, bonding, or electrostatics) is over atomic distances and can be finely tuned. 2D materials are more resistant to delamination and combining multiple FE/FM bilayers can scale the net MF coupling. Examples of studied multilayers include In2Se3 or Sc2CO2 (switchable ferroelectrics) interleaved with CrI3, Cr2Ge2Te6, or NiI2 (ferromagnetics), or double perovskite layers with two types of instabilities.

*Experiments.* Experimental techniques to measure nanoscale ferroelectricity include SHG and SPM, particularly KPFM and diamond tips with nitrogen vacancy centers. Ferromagnetism detection methods include Kerr spectroscopy, XMCD, and simple magnetometry, provided the sample mass is sufficient and the order is long-ranged. Elastic properties are linked to structural effects observed in diffraction or (scanning) microscopy.

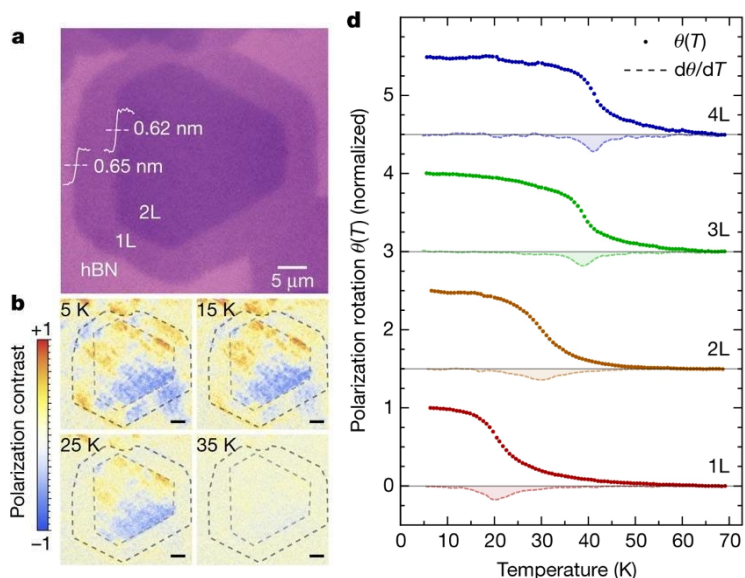

**Figure 7.1.** Evidence for ferroelectricity in few layer NiI$_2$ (adapted with permission from Ref. [2]). a,b) Optical and polarized images showing polarization domains; d) Temperature dependence of light polarization rotation, showing a phase transition to a MF state.

Experimental demonstration of 2DMF materials:

1. NiI2: This material shows in-plane ferromagnetism down to the monolayer limit and demonstrates intrinsic multiferroicity. Bulk NiI$_2$ exhibits antiferromagnetism between planes (Néel temperature near 75 K) and transitions to a helical magnetic state at 60 K, coupling electric and magnetic order parameters (type-II multiferroicity) [2] (see Figure 7.1). The coupling strengthens with thicker layers and increases under pressure.

2. CuCrP2S6: This thiophosphate hosts in-plane ferromagnetic order (AFM between planes) and features random Cu positions that order antiferroelectrically at low temperatures. The AFM order can become canted, leading to ferri-electric (FiE) coupling between magnetic and electric states. Applying an electric field can switch Cu positions and FiE moments, influencing the magnetic exchange and state. Demonstrated devices exhibit FiE switching and changes in tunneling magneto-resistance.

3. CuCrSe2: This dichalcogenide-derived compound has been recently demonstrated to be MF up to 120 K. The structure is reticulated in 3D, but can be exfoliated down to a single heptalayer atomic sandwich of Cu. As above, displaced Cu atoms lead to ferroelectricity, and Cr orbital changes lead to magnetization.

4. MXenes (e.g., Ti3C2Tx): Obtained from MAX phases by chemical exfoliation, MXenes show ferroelectricity and ferromagnetism arising from disorder, due to localized polar or spin moments. These materials, with film thicknesses in the tens of micrometers and low sample crystallinity, are closer to 3D MF but have demonstrated full device capabilities [4].

*Applications.* Similar to bulk multiferroics, the potential applications of 2DMF are in energy-efficient electronic devices. Even though magnetic random-access memories (MRAM) are successful, they suffer from high writing energy costs, which limits their development in low-power applications. Ferroelectric RAM (FeRAM) offers energy-efficient voltage writing, but its destructive readout and scaling issues restrict it to niche applications. Magnetoelectric RAM (MERAM) emerged as a solution to combine the advantages of both technologies: ferroelectric writing and magnetic readout.

Recently, more advanced multifunctional prototypes were proposed, such as the logic-in-memory magnetoelectric spin-orbit (MESO) device as well as its successors that additionally utilize magnon transport in a multiferroic to simplify the complicated design [5, 6]. Replacing a 3D multiferroic with 2DMF could enable further miniaturization and new design possibilities. For example, recent studies show that some heterobilayers based on 2D magnets exhibit efficient charge-to-spin conversion and long-range spin transport. MF heterostructures could enable ferroelectric control of spin currents without any additional materials. The transition from bulk materials to 2D vdW heterostructures would be promising for future electronic devices, enabling more efficient and innovative designs.

**Current and Future Challenges**
In addition to the experimental verification of intrinsic 2DMF, one of the main challenges is to demonstrate artificial multiferroicity, e.g. creating 2DMF from vertically stacked ferroelectric and magnetic layers [7]. While the discovery of intrinsic room-temperature 2D ferroelectrics like SnTe, $In_2Se_3$, or $CuInP_2S_6$ may facilitate this task, interfacing them with 2D magnets remains challenging, with no experimental studies reported yet. Another challenge is the idea of sliding ferroelectricity in 2D magnets [8]. Breaking inversion symmetry in vdW bilayers by interlayer sliding can change the stacking order and induce electric polarization. Although experimentally realized in bilayer h-BN and various TMDs, sliding ferroelectricity predictions in $CrI_3$ and other 2D magnets, as well as potential magnetoelectric coupling, need confirmation. Twisting 2D materials at small angles induces a moiré pattern of polar domains with alternating polarizations separated by non-polar domain walls, which have been studied and confirmed in non-magnetic materials. Thus, twisting 2D magnets could theoretically lead to multiferroic domains. In addition, systems could achieve MF coupling by generating non-collinear magnetic order combined with spin-orbit coupling, as proposed for twisted bilayer $CrBr_3$ (see Fig. 7.2) [9]. Twisted bilayers exhibit stacking-dependent magnetic orders, leading to regions with ferro- and antiferromagnetic coupling and non-collinear magnetic orders at their boundaries. Experimental observations support the stacking-dependent magnetic orders, suggesting the potential for twisted 2D magnets to realize MF domains [10].

**Advances in Science and Technology to Meet Challenges**
The synthesis of 2DMF heterostructures, their epitaxy, and interface control are relatively simple compared to methods for thin films of 3D materials. However, the small amount of matter in ultrathin nanostructures poses significant experimental challenges. Advanced techniques for detecting nanoscale electric and magnetic fields, such as local probes (e.g., spSTM/ PFM) or advanced spectroscopies (nano-ARPES, SNOM), are essential for further progress. Exploiting 2DMF materials in devices will require nanocircuitry for reading and writing MF domains. Although established in silicon-based lithography and related techniques, these methods need adapting for 2DMFs, which are often mechanically delicate and sensitive to environmental factors like oxidation and humidity. Developing robust integration techniques for graphene, TMDs, and 2DMFs is critical. Theoretical simulations of 2DMF will require advancements in multi-scale techniques and codes, incorporating atomic displacements, (macro)spin dynamics, and electromagnetic fields. While a few accurate simulations were reported in the literature, many MF properties, such as switching, are dominated by domain dynamics. These factors require simulations of larger systems than those accessible with fully first-principles methods.

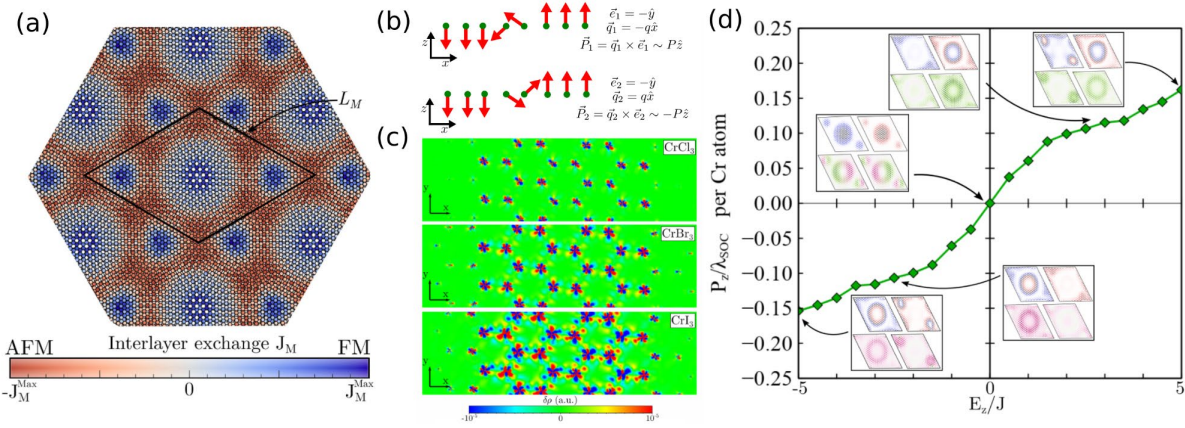

Fig. 7.2: (a) Schematic of a twisted Cr trihalide bilayer showing the emergence of different magnetic domains due to the stacking. Panel (b) shows the two possible local magnetic structures at a domain, and (c) the charge density reconstruction yields multiferroicity. Panel (d) shows how the magnetic and ferroelectric texture can be tuned with an interlayer bias.

**Concluding Remarks**

Two-dimensional multiferroics hold significant promise for the future of electronic devices, offering unique advantages for energy economy, integration, and tunability. Several 2DMF have been very recently demonstrated experimentally down to the monolayer limit, and many materials have been proposed theoretically.

**Acknowledgements**

MJV acknowledges ARC project DREAMS (G.A. 21/25-11) funded by Federation Wallonie Bruxelles and ULiege, and Excellence of Science project CONNECT number 40007563 funded by FWO and FNRS. J.S. acknowledges the grant of the Dutch Research Council (NWO) under the contract OCENW.M.22.063 and the Rosalind Franklin Fellowship from the University of Groningen.

**References**

[1] M. Mostovoy, ''Multiferroics: different routes to magnetoelectric coupling'', *npj Spintronics* vol. 2, pp. 18, 2024

[2] Song Q, Occhialini C A, Ergecen E, Ilyas B, Amoroso D, Barone P, Kapeghian J, Watanabe K, Taniguchi T, Botana A S, Picozzi S, Gedik N and Comin R "Evidence for a single-layer van der Waals multiferroic" *Nature* vol. 602 pp. 601–605, 2022

[3] Hu Q, Huang Y, Wang Y, Ding S, Zhang M, Hua C, Li L, Xu X, Yang J, Yuan S, Watanabe K, Taniguchi T, Lu Y, Jin C, Wang D and Zheng Y "Ferrielectricity controlled widely-tunable magnetoelectric coupling in van der Waals multiferroics" *Nature Comm* vol. 15, pp. 3029, 2024

[4] Tahir R, Fatima S, Zahra S A, Akinwande D, Li H, Jafri S H M and Rizwan S "Multiferroic and ferroelectric phases revealed in 2D Ti3C2Tx MXene film for high performance resistive data storage devices" *npj 2D Materials and Applications,* vol. 7, pp. 7, 2023

[5] Manipatruni S, Nikonov D E, Lin C C, Gosavi T A, Liu H, Prasad B, Huang Y L, Bonturim E, Ramesh R and Young I A "Scalable energy-efficient magnetoelectric spin--orbit logic" *Nature* vol. 565, pp. 35–42, 2019

[6] Huang X, Chen X, Li Y, Mangeri J, Zhang H, Ramesh M, Taghinejad H, Meisenheimer P, Caretta L, Susarla S, Jain R, Klewe C, Wang T, Chen R, Hsu C H, Harris I, Husain S, Pan H, Yin J, Shafer P, Qiu Z, Rodrigues D R, Heinonen O, Vasudevan D, Iniguez J, Schlom D G, Salahuddin S, Martin L W, Analytis J G, Ralph D C, Cheng R, Yao Z and Ramesh R "Manipulating chiral spin transport with ferroelectric polarization" *Nature Materials*, vol. 23, no. 7, pp. 898-904, 2024

[7] Huang X, Li G, Chen C, Nie X, Jiang X and Liu J M "Interfacial coupling induced critical thickness for the ferroelectric bistability of two-dimensional ferromagnet/ferroelectric van der Waals heterostructures" *Physical Review B,* vol. 100, pp. 235445, 2019

[8] Ji J, Yu G, Xu C and Xiang H J "General Theory for Bilayer Stacking Ferroelectricity" *Physical Review Letters* vol. 130, pp. 146801, 2023

[9] Fumega AO and Lado JL "Moiré-driven multiferroic order in twisted CrCl3, CrBr3 and CrI3 bilayers" *2D Materials,* vol. 10, pp. 025026, 2023

[10] Song T, Sun Q C, Anderson E, Wang C, Qian J, Taniguchi T, Watanabe K, McGuire M A, Stohr R, Xiao D, Cao T, Wrachtrup J and Xu X "Direct visualization of magnetic domains and moiré magnetism in twisted 2D magnets" *Science,* vol. 374, pp. 1140–1144, 2021

## 8. Synthesis of bulk and thin film 2D quantum magnets

Anna Isaeva[1,2], J. Marcelo J. Lopes[3] and Falk Pabst[1]
[1] University of Amsterdam, The Netherlands; [2] Technical University of Dortmund, Germany [3] Paul-Drude-Institut für Festkörperelektronik, Leibniz-Institut im Forschungsverbund Berlin e.V., Germany

**Status**

Explorative synthesis of new 2D quantum magnet candidates and optimization of fabrication techniques, from bulk crystals to thin films, are propelling the field toward spintronic applications. Bulk systems offer versatile magnetic orders and materials but often face limitations in lateral size. To ensure compatibility with device applications, wafer-scale thin films and heterostructures are developed, allowing for tailored properties. The materials focus is primarily on vdW compounds, which are naturally exfoliable and offer high structural and compositional tunability (see comparison growth table). In 2017, two vdW materials, $CrI_3$ and $CrGeTe_3$, pioneered intrinsic magnetic order down to the monolayer and bilayer limits [1,2]. This discovery spurred a significant expansion of the 2D magnets database, with notable progress in the $MX_3$, $MCh_2$, $CrGeTe_3$, $Fe_xGeTe_2$, $MPS_3$, and MXene families (M – transition metal, X – halide, Ch – chalcogenide) [3]. Experimental advancements are often guided by theoretical predictions and benefit from a strong feedback loop with advanced characterization methods. Current tuning strategies focus on the controlled fabrication of high-quality, stable bulk crystals with robust magnetic order, preferably ferromagnetic, and enhanced magnetic transition temperatures nearing or exceeding room temperature. These improvements are achieved by designing the magnetic exchange interactions. For scalable synthesis, thin-film deposition systems lack the technical flexibility of bulk crystal growth methods. As a result, efforts have concentrated on achieving epitaxial growth of 2D ferromagnets with properties suitable for a wide range of applications. A notable achievement is the reported Curie temperature of $T_C > 300$ K for $Fe_{5-x}GeTe_2$ ($x \approx 0$) thin films [4]. An overview of the main methods used to synthesise 2D quantum magnets and a comparison between them is given in Figure 8.1.

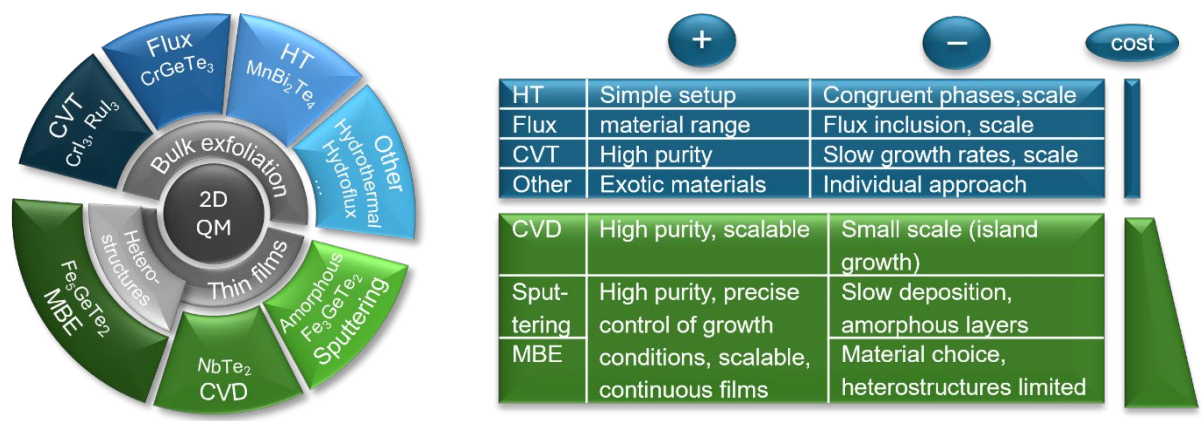

**Figure 8.1.** Synthetic approaches to 2D quantum magnets in bulk and thin-film forms, with representative compounds (left). Bulk crystals are prepared via chemical vapor transport (CVT), flux, high-temperature growth (HT), and other techniques, followed by top-down delamination to obtain 2D quantum material (QM). Thin films are prepared using sputtering, chemical vapor deposition (CVD), and molecular beam epitaxy (MBE). Overview of main advantages (+) and limitations (-) of the different techniques (right) based on the current state of the art. Bulk techniques display high flexibility through wide material range and low-cost use, but are challenging to scale up. Thin-film techniques compensate for their increasing cost by high sample purity, scalability, and compatibility with device fabrication.

**Current and Future Challenges**

vdW 2D magnets are conventionally grown using gas-phase techniques and molten fluxes [5]. Ampoule-based syntheses are cost-efficient and successful growth protocols can be realized with simple setups. Flux growth offers a wide material range (e.g., pnictides, chalcogenides) due to various possible flux media, but always includes the risk of flux inclusion in the grown crystals. High purity samples may be grown by CVT. However, the volatility of non-metallic components at various synthetic temperatures often causes non-stoichiometry, limiting certain growth protocols. Gas-phase methods also face challenges like competing reactions, limited phase stability, and slow growth rates. Stacking sequences are decisive for magnetic ground states e.g. in $CrI_3$ and other $MX_3$ compounds [3]), requiring precise control of synthetic parameters and thorough exploration of phase diagrams. Accurate structure determination is essential but challenging due to factors like stacking sequences, faults, dislocations, site intermixing, and twinning. Additionally, the air-sensitivity of many halides and chalcogenides, particularly in thin flakes, complicates reproducible and durable applications. MXenes are produced by selective acid-solution etching of bulk phases, which introduces issues like surface absorption and reconstruction, requiring extensive cleaning [3]. Further exfoliation techniques include the traditional scotch tape method, as well as mechanical (ultra-sound) and chemical (solvent) exfoliation. The main limiting factor of exfoliated flakes is their limited lateral sizes restricted to the micrometer range.

Thin-film deposition methods are able to provide large-scale material but offer less flexibility in material selection and stacking alignment in comparison to bulk crystals, which foster a diverse array of designer 2D magnets. The flexibility of bulk crystal growth also facilitates a well-informed choice of synthetic conditions that can be used to tailor the magnetic ground state through lattice-symmetry tuning, doping, strain engineering, intercalation, and exfoliation, affecting magnetic couplings, anisotropy, and thickness-dependent properties. Bulk-to-monolayer properties can vary significantly in systems with strong interlayer couplings, such as 2D magnetic chalcogenides (e.g., $Cr_3Se_4$ and $VSe_2$), making it challenging to apply knowledge from 3D bulk systems to exfoliated counterparts. However, this variation presents opportunities for different functionalities in bulk and thin flake systems. Heterostructures from exfoliated bulk crystals are also less restricted by lattice parameter matching than epitaxially grown films, but the transfer and cleaning process is complex (precise alignment, h-BN encapsulation) and not yet industrially viable [6].

Despite the high cost, the development of bottom-up, wafer-scale synthesis of continuous thin films is critical for applications based on 2D quantum magnets due to the size limitations of bulk crystals and exfoliated flakes and their incompatibility with standard device fabrication processes. In addition, thin-film deposition has the important advantage of precise thickness control [7]. Achievements in this direction have been made by using MBE for growing 2D magnets like $CrGeTe_3$, $Fe_3GeTe_2$, and $Fe_{5-x}GeTe_2$ ($x \approx 0$). It was possible to produce continuous crystalline thin films with properties comparable or superior (e.g., higher $T_C$) to bulk and flake counterparts. Epitaxial heterostructures with clean and sharp vdW interfaces, combining these materials with other 2D crystals [4,8] and topological insulators [9,10], have also been demonstrated (see Figure 8.2).

In addition to MBE, magnetron sputtering has been used for the large-scale deposition of ultrathin (1 nm) $Fe_3GeTe_2$ films. However, the films exhibited an amorphous nature and lacked robust magnetic properties [11], highlighting the need for further development in terms of synthesis protocols. These results also show that the preservation of the 2D crystalline structure, as achieved by $Fe_3GeTe_2$ bulk crystal growth and MBE [3,10], is crucial for achieving ferromagnetic order in the 2D limit. CVD has also been employed for the fabrication of 2D magnets [3], but the realization of continuous films with thickness control remains to be attained. Thin-film methods offer potential advantages like precise and homogeneous dopant distribution, which is challenging in bulk crystal growth due to phase separation. Another advantage of thin-film deposition is the possibility of *in-situ* capping the 2D magnetic films via deposition of stable materials (e.g., $AlO_2$, SiN, Pt), which helps mitigate material degradation via oxidation upon air exposure.

Despite progress, obstacles remain for direct epitaxial thin-film growth, such as differing thermal stabilities and growth temperatures of 2D materials, which hinder arbitrary stacking arrangements, e.g., chalcogenides cannot be overgrown by h-BN encapsulating layers. Epitaxial registry in vdW materials prohibits the creation of twisted heterostructures. Furthermore, suitable substrates for optimal epitaxy may not fulfil technological requirements.

Finally, a major ongoing challenge in materials design both for bulk and thin-film 2D magnets is the fabrication of high-Curie temperature magnetic systems, particularly semiconductors.

**Advances in Science and Technology to Meet Challenges**

A primary thrust for improving 2D magnets is exploring mechanisms to increase their magnetic ordering temperature. Record Curie temperatures ($T_C$) have been achieved in bulk metals $Fe_{5-\delta-x}Ni_xGeTe_2$ (> 400 K [12]) and $Fe_3GaTe_2$ (380 K [13], see Figure 8.2). Significant progress toward magnetic semiconductors includes CrSBr ($T_N$ = 132 K [14]). Materials optimization aims to discover more candidates, supported by theoretical efforts to predict magnetic interactions. Both out-of-plane and in-plane anisotropy can enhance 2D magnets, and theory suggests several promising systems with high transition temperatures (e.g., $VF_3$, $NiF_3$, or $MnBr_3$ [3]). High-throughput computational methods inspire explorative synthesis. Thin-film technology can stabilize some theory-predicted metastable phases by using suitable substrates [5].

Defect-engineering is a promising method to tune the magnetic ground state and Curie temperature, such as in $(MnTe)(X_2Te_3)_n$ (X = Sb, Bi) through Mn/X intermixing [15,16], and in $Fe_xGeTe_2$ (3 < x < 5), where $T_C$ can be modulated depending on Fe content [3,4,8]. Thin-film fabrication also reveals great potential for interfacial effects to tailor properties and create new functionalities [3]. Additionally, thin-film methods allow for more precise and homogeneous dopant distribution across 2D layers, such as Ni or Co doping in $Fe_{5-x}GeTe_2$, which is challenging in bulk crystal growth due to unintentional phase separation [13].

Defect and strain engineering of tailored magnetic orders are corroborated by accurate structure characterizations, such as X-ray, electron, and neutron diffraction for bulk crystals. For flakes and thin films, the reduced dimensionality places new demands on analytical tools for atomic and magnetic structures. Surface sensitive techniques like grazing incidence X-ray diffraction, reflection high energy electron diffraction or cross-sectional STEM provide insights into the azimuthal dependence of lattice parameter surface morphology, symmetry, or lattice orientation. Magnetic properties can be probed by magneto-optic Kerr effect microscopy, or X-ray magnetic circular dichroism [1,8,9].

Synthetic advances in bulk include low-temperature, energy-saving alternatives to conventional CVT or flux growth methods, such as hydroflux or hydrothermal methods and wet chemical synthesis (e.g., for oxides, selenides). These methods yield high-quality crystalline products, including metastable phases, polymorphic modifications and intercalates. The limitations of traditional growth techniques, such as small sample sizes and high defect concentrations, can be addressed by employing more advanced methods like the Bridgman or floating zone growth. Modified versions of these systems even allow for the preparation of incongruently melting compounds. However, these methods come with drawbacks, including increased complexity and the need for time-consuming optimization of setup parameters. Material performance can be enhanced by post-synthetic annealing to heal defects via grain boundary sliding. Chalcogen deficiency in TMDs can be addressed by oxygen passivation through focused laser annealing in air [17]. Additionally, ionic liquids (IL) have been used for postgrowth functionalization, exfoliation, and IL-gating of devices [18].

To meet future challenges, thin-film technology should develop solutions for complementary wafer-scale thin-film growth of 2D magnets and their heterostructures, followed by layer transfer. In particular, further innovation in technologically relevant methods such as magnetron sputtering and CVD is urgently needed. This will ultimately allow for the tailored assembly of 2D magnets and heterostructures on substrates compatible with device processing.

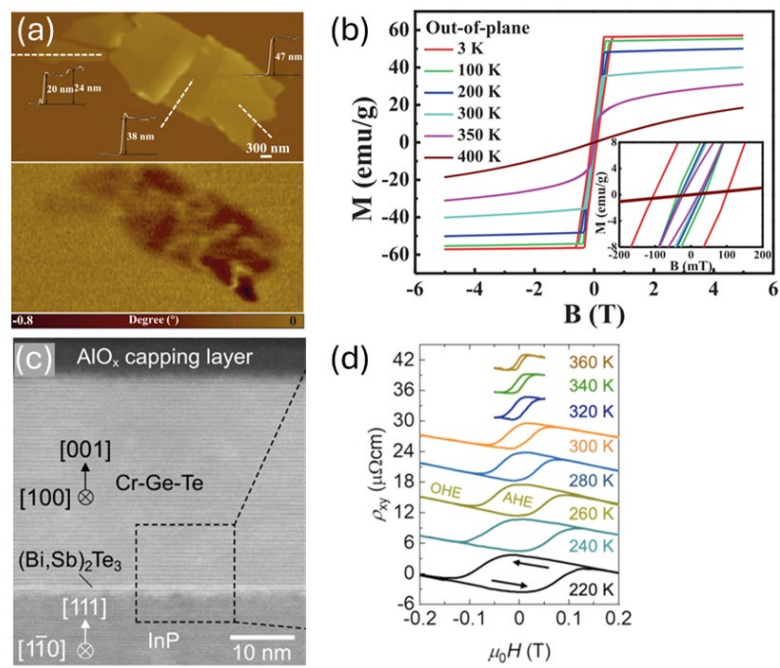

**Figure 8.2.** (a) AFM topography (top) and corresponding MFM image (bottom) of $Fe_3GaTe_2$ flakes with varying thickness. They confirm the ferromagnetic response of the material at room temperature (300 K) (b) M-H curves of $Fe_3GaTe_2$ bulk crystals at varying temperatures with magnetic fields along the out-of-plane axis highlighting the above-room temperature strong ferromagnetism of the vdW compound. (a) and (b) are reproduced with permission from Ref. 13. (c) Cross-sectional HAADF-STEM image of a $CrGeTe_3/(Bi,Sb)_2Te_3$ grown on an InP substrate entirely by MBE, illustrating the sharp interface between the two materials. Reproduced with permission from Ref. 9. (d) Anomalous Hall effect (AHE) measured up to 360 K for an epitaxial $Fe_{5-x}GeTe_2$/graphene vdW heterostructure. OHE indicates the contribution of the ordinary Hall effects in $Fe_{5-x}GeTe_2$ and graphene. Reproduced with permission from Ref. 4. All reproduced figures are licensed under a Creative Commons Attribution license.

**Concluding Remarks**

The library of 2D quantum magnets continues to expand, extending beyond vdW materials to accommodate emerging concepts for spintronic device applications. New directions include non-van der Waals materials [19], d- and p-electron magnets [3] and kagomé 2D magnets [20]. These compounds, which incorporate non-metallic elements such as Si, O, P, B, present new challenges for synthesis protocols and cleaving processes, as they are not naturally exfoliable. The candidate pool is constantly growing through a concerted effort between theoretical and experimental research. New bulk systems are discovered at the intersection of inorganic chemistry and condensed matter physics.

Magnetic anisotropy is crucial for overcoming the limitations imposed by the Mermin-Wagner theorem and it enables the development of the thinnest 2D magnets to date compatible with other components in heterostructures for next-generation devices. Bulk crystals offer a highly tunable platform for materials exploration and optimization at moderate cost. Thin-film fabrication focuses on addressing the technological challenges to produce best-performing materials, albeit at the price of complex setups and high costs. These coordinated efforts yield high-quality 2D magnetic films and heterostructures facilitating advanced applications.

**Acknowledgements**

AI and FP acknowledge the research program "Materials for the Quantum Age" (QuMat) for financial support. This program (registration number 024.005.006) is part of the Gravitation program financed by the Dutch Ministry of Education, Culture and Science (OCW). JMJL acknowledges financial support from the FLAG-ERA (MagicTune project) and the German Research Foundation, DFG (project 533948427).

**References**

[1] C. Gong, L. Li, Z. Li, H. Ji, A. Stern, Y. Xia, T. Cao, W. Bao, C. Wang, Y. Wang, Z. Q. Qiu, R. J. Cava, S. G. Louie, J. Xia and X. Zhang, "Discovery of intrinsic ferromagnetism in two-dimensional van der Waals crystals", *Nature*, vol. 546, pp. 265-269, 2017

[2] B. Huang, G. Clark, E. Navarro-Moratalla, D. R. Klein, R. Cheng, K. L. Seyler, D. Zhong, E. Schmidgall, M. A. McGuire, D. H. Cobden, W. Yao, D. Xiao, P. Jarillo-Herrero and X. Xu, "Layer-dependent ferromagnetism in a van der Waals crystal down to the monolayer limit", *Nature*, vol. 546, pp. 270-273, 2017

[3] X. Jiang, Q. Liu, J. Xing, N. Liu, Y. Guo, Z. Liu and J. Zhao, "Recent progress on 2D magnets: Fundamental mechanism, structural design and modification", *Applied Physics Reviews*, vol. 8, p. 031305, 2021

[4] H. Lv, A. da Silva, A. I. Figueroa, C. Guillemard, I. Fernández Aguirre, L. Camosi, L. Aballe, M. Valvidares, S. O. Valenzuela, J. Schubert, M. Schmidbauer, J. Herfort, M. Hanke, A. Trampert, R. Engel-Herbert, M. Ramsteiner and J. M. J. Lopes, "Large-Area Synthesis of Ferromagnetic $Fe_{5-x}GeTe_2$/Graphene van der Waals Heterostructures with Curie Temperature above Room Temperature", *Small*, vol. 19, no. 39, p. 2302387, 2023

[5] A.F. May, J. Yan and M.A. McGuire, "A practical guide for crystal growth of van der Waals layered materials", *Journal of Applied Physics*, vol. 128, p. 051101, 2020

[6] S.-Q. Zhang, J.-L. Liu, M. Merle Kirchner, H. Wang, Y.-L. Ren and W. Lei, "Two-dimensional heterostructures and their device applications: progress, challenges and opportunities – review", *Journal of Physics D: Applied Physics*, vol. 54, no. 43, p. 433001, 2021

[7] H. Kurebayashi, J.H. Garcia, S. Khan, J. Sinova and S. Roche, "Magnetism, symmetry and spin transport in van der Waals layered systems", *Nature Reviews Physics*, vol. 4, pp. 150-166, 2022

[8] J.M.J. Lopes, D. Czubak, E. Zallo, A.I. Figueroa, C. Giullemard, M. Valvidares, J. Rubio-Zuazo, J. Lopez-Sanchez, S.O. Valenzuela and M. Hanke, "Large-area van der Waals epitaxy and magnetic characterization of $Fe_3GeTe_2$ films on graphene", *2D Materials*, vol. 8, no. 4, p. 041001, 2021

[9] M. Mogi, A. Tsukazaki, Y. Kaneko, R. Yoshimi, K.S. Takahashi, M. Kawasaki and Y. Tokura, "Ferromagnetic insulator $Cr_2Ge_2Te_6$ thin films with perpendicular remanence", *APL Materials*, vol. 6, 091104, 2018

[10] T. Guillet, R. Galceran, J.F. Sierra, F. J. Belarre, B. Ballesteros, M. V. Costache, D. Dosenovic, H. Okuno, A. Marty, M. Jamet, F. Bonell, S. O. Valenzuela, "Spin–Orbit Torques and Magnetization Switching in $(Bi,Sb)_2Te_3$/$Fe_3GeTe_2$ Heterostructures Grown by Molecular Beam Epitaxy", *Nano Letters*, vol. 24, no. 3, pp. 822-828, 2024

[11] Q.-W. Zhao, C.-C. Xia, H. Zhang, B. Jiang, T. Xie, K. Lou and C. Bi, "Ferromagnetism of Nanometer Thick Sputtered $Fe_3GeTe_2$ Films in the Absence of Two-Dimensional Crystalline Order: Implications for Spintronics Applications", *ACS Applied Nano Materials*, vol. 6, no. 4, pp. 2873-2882, 2023

[12] X. Chen, Y.-T. Shao, R. Chen, S. Susarla, T. Hogan, Y. He, H.-R. Zhang, S. Wang, J. Yao, P. Ercius, D.A. Muller, R. Ramesh and R.J. Birgeneau, "Pervasive beyond Room-Temperature Ferromagnetism in a Doped van der Waals Magnet", *Physical Review Letters*, vol. 128, p. 217203, 2022

[13] G. Zhang, F. Guo, H. Wu, X.-K. Wen, L. Yang, W. Jin, W.-F. Zhang and H.-X. Chang, "Above-room-temperature strong intrinsic ferromagnetism in 2D van der Waals $Fe_3GaTe_2$ with large perpendicular magnetic anisotropy", *Nature Communications*, vol. 13, p. 5067, 2022

[14] M.E. Ziebel, M. L. Feuer, J. Cox, X.-Y. Zhu, C. R. Dean and X. Roy, "CrSBr: An air-stable, two-dimensional magnetic semiconductor", *Nano Letters*, vol. 24, no. 15, pp. 4319-4329, 2024

[15] A. Tcakaev, B. Rubrecht, J.I. Facio, V. B. Zabolotnyy, L.T. Corredor, L.C. Folkers, E. Kochetkova, T. R. F. Peixoto, P. Kagerer, S. Heinze, H. Bentmann, R. J. Green, M. Valvidares, E. Weschke, F. Reinert, M. W. Haverkort, J. van den Brink, B. Büchner, A. U. B. Wolter, A. Isaeva and V. Hinkov, "Intermixing-driven surface and bulk ferromagnetism in the quantum anomalous Hall candidate $MnBi_6Te_{10}$", *Advanced Science*, vol. 10, no. 10, p. 2203239, 2023

[16] M. Sahoo, M.C. Rahn, E. Kochetkova, O. Renier, L.C. Folkers, A. Tcakaev, M.L. Amigó, F.M. Stier, V. Pomjakushin, K. Srowik, V.B. Zabolotnyy, E. Weschke, V. Hinkov, A. Alfonsov, V. Kataev, B. Büchner, A.U.B. Wolter, J.I. Facio, L.T. Corredor and A. Isaeva, "Tuning strategy for Curie-temperature enhancement in the van der Waals magnet $Mn_{1+x}Sb_{2-x}Te_4$", *Materials Today Physics*, vol. 38, p. 101265, 2023

[17] J.-P. Lu, A. Carvalho, X. K. Chan, H.-W. Liu, B. Liu, E. S. Tok, K. P. Loh, A. H. Castro Neto and C. H. Sow, "Atomic healing of defects in transition metal dichalcogenides", *Nano Letters*, vol. 15, no. 5, pp. 3524–3532, 2015

[18] N. Sa, M. Wu and H. Q. Wang, "Review of the role of ionic liquids in two-dimensional materials", *Frontiers of Physics 2023*, vol. 18, no. 4, pp. 1–21, 2023

[19] A.P. Balan, A.B. Puthirath, S. Roy, G. Costin, E.F. Oliveira, M.A.S.R. Saadi, V. Sreepal, R. Friedrich, P. Serles, A. Biswas, S.A. Iyengar, N. Chakingal, S. Bhattacharyya, S.K. Saju, S. Castro Pardo, L.M. Sassi, T. Filleter, A. Krasheninnikov, D.S. Galvao, R. Vajtai, R.R. Nair and P.M. Ajayan, "Non-van der Waals quasi-2D materials; recent advances in synthesis, emergent properties and applications", *Materials Today,* vol. 58, pp. 164-200, 2022

[20] H. Zhang, H. Feng, X. Xu, W. Hao and Y. Du, "Recent Progress on 2D Kagome Magnets: Binary $T_mSn_n$ (T = Fe, Co, Mn)", *Advanced Quantum Technologies*, vol. 4, no. 11, p. 2100073, 2021

# 9. Challenges on wafer-scale transfer and protection

Semonti Bhattacharyya[1], Zhiyuan Cheng[1], Zhiying Dan[2] and Antonija Grubišić-Čabo[2]

[1.]Huygens-Kamerlingh Onnes Laboratory, Leiden University, [2.] Zernike Institute for Advanced Materials, University of Groningen

**Status**

Two-dimensional (2D) magnetic materials exhibit great promise for applications in electronic, optoelectronic, and spintronic devices, especially when combined with other materials such as semiconductors, superconductors, or ferroelectrics. This potential arises from the ease with which the properties of 2D materials can easily be manipulated through interface coupling, thus achieving atomically flat, clean interfaces between the 2D magnetic material of choice and the adjacent layer is crucial for these applications.

The importance of a high-quality interface is highlighted in a recent work by T. S. Ghiasi et al. [1] where they demonstrate an effective method to induce magnetism in graphene via the magnetic proximity effect. They detect the induced magnetism using a 3-terminal spin valve measurement device (Figures 9.1.a and 9.1.b) in a setup that consists of a bilayer graphene/CrSBr stack contacted through $Al_2O_3$/Co electrodes. The magnetic proximity effect from 2D magnetic material CrSBr spin-polarizes bilayer graphene due to the magnetic proximity effect and the resistance changes dramatically depending on the relative orientation between graphene and the spin-polarized current in the Co contacts (Figure 9.1.c). This significant change is attributed to a staggering exchange field of 170 T, a field that is an order of magnitude higher than what is achievable in state-of-the-art solid-state laboratories. A key ingredient for such a strong magnetic proximity effect is the presence of atomically flat, clean interfaces. One way to achieve such interfaces is through the polymer-assisted dry transfer of scotch-tape exfoliated 2D materials, as demonstrated in Ref. [2] (Figures 9.1.d and 9.1.e). Although highly effective, this technique is only applicable to flakes with micrometer-scale lateral dimensions. In contrast, for industrial applications of 2D magnetic heterostructures, it is essential to develop methods for creating high-quality heterostructures with meter-scale lateral dimensions. Unfortunately, currently available techniques neither produce 2D magnetic materials, nor heterostructures of sufficient size, uniformity, and quality for such large-scale applications.

**Current and Future Challenges**

Over the past two decades, several techniques such as CVD and MBE, have been developed to synthesize large-area monolayers, enabling the growth of wafer-sized 2D materials [3,4]. However, these techniques require tedious optimization steps, and the 2D materials grown this way often have a high density of defects, resulting in lower quality compared to 2D monolayers prepared by mechanical exfoliation. Additionally, the applicability of these methods is further limited because not all substrates are suitable for 2D material growth.

Beyond the usual synthesis and scalability difficulties associated with 2D materials, 2D magnetic materials present unique challenges in terms of synthesis and wafer-scale transfer. Many emerging 2D materials, including most 2D quantum magnets [5], are highly reactive and degrade rapidly when exposed to environmental conditions. To mitigate this, these materials must be handled in controlled atmospheres, such as gloveboxes, to prevent oxidation and degradation throughout their entire lifecycle, from synthesis to integration. This complicates the mechanical exfoliation procedure and creates additional constraints in transferring materials from the glovebox to the measurement system or post-processing system without exposure to air or moisture. Specialized equipment and techniques are required to maintain the integrity of these materials. Typical approaches involve protecting exfoliated 2D monolayers with encapsulating materials or using sealed transfer systems, such as vacuum suitcases. However, these methods are not always fully suitable for wafer-scale materials and require large-scale investments, such as interconnected inert-atmosphere equipment lines.

To address these challenges, a technique is needed that combines the scale and the yield of epitaxial growth of methods such as MBE and CVD with the purity and flexibility of exfoliated flakes. In the following section, we will discuss gold-assisted exfoliation [6], a recently discovered method that fulfils these criteria, as well as the UHV method offering similar prospects [7].

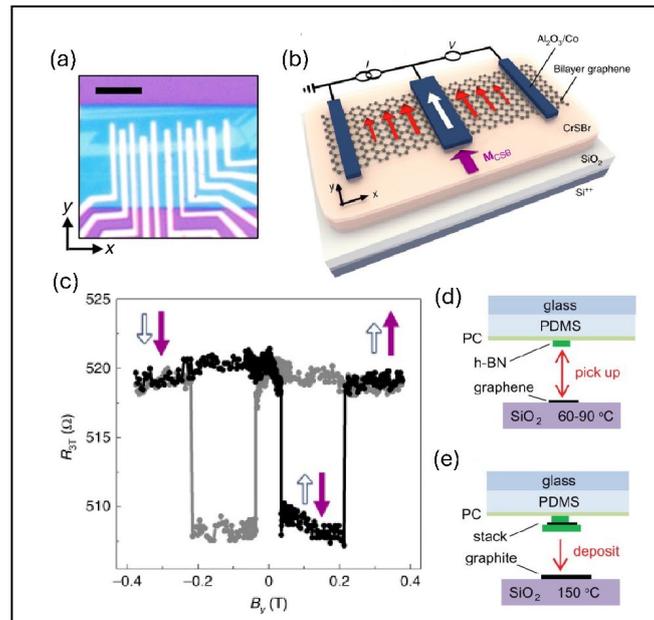

**Figure 9.1.** (a) Optical image and (b) schematic of a 3-terminal spin valve measurement device realized in a bilayer graphene/CrSBr heterostructure. White and purple arrows indicate the magnetization of the Co contacts and top-most layer of CrSBr respectively. (c) 3-Terminal Resistance ($R_{3T}$) as a function of the in-plane magnetic field ($B_y$). The white and purple arrows indicate the magnetization orientation of Co and CrSBr respectively. Black curve shows the R3T measurement starting from By = -0.4 T (trace), and the grey curve is the retrace measurement, demonstrating proximity-induced magnetization of the bilayer graphene achieved through clean graphene-CrSBr interface. Figure (d) and (e) show the pick-up and deposit steps of a polycarbonate-polydimethylsiloxane (PC-PDMS) assisted stacking hBN (hexagonal boron nitride)/graphene/hBN heterostructure with high-quality interfaces. Reprinted from (a-c) Ref. [1] and (d-e) Ref. [2].

**Advances in Science and Technology to Meet Challenges**

Recently developed gold-assisted exfoliation enables the production of large-area, high-quality 2D monolayers [6]. This method involves pressing tape covered with bulk layered crystals onto a smooth metal surface, typically gold, as shown in Figure 9.2). The strong binding between gold and the top layer of the bulk crystal facilitates isolation of large-area monolayer samples. Typically, around 200 nm of gold is deposited on a substrate (e.g., clean $SiO_2$) via physical vapor deposition. Fresh tape with a single crystal, prepared from a larger bulk crystal, serves as the source material. Pressing the tape onto the gold substrate and then slowly separating them can yield 2D materials several millimetres in size. However, this process is usually conducted under atmospheric conditions, making it unsuitable for air-sensitive 2D materials. To address this limitation, the setup and materials can be placed in an inert atmosphere, such as a glovebox, or an UHV based kinetic in situ single-layer synthesis (KISS) method can be used [7]. The KISS method, Figure 9.2.b), allows for the use of substrates like silver and germanium, which can be prepared to be atomically flat and clean using techniques such as sputtering and annealing. Because these substrates are kept in UHV, they remain reactive and suitable for exfoliation. Similar to the case of previously described gold assisted exfoliation, a bulk crystal acts as the source material, and a fresh, adsorbate-free bulk surface is prepared shortly before KISS exfoliation in a UHV chamber. Following the surface preparation, the bulk crystal is slowly brought into contact with the substrate and separated, leaving large area 2D monolayers, typically several hundred micrometers in size, on the surface. This method is well-suited for in situ studies using techniques like angle-resolved photoemission spectroscopy and for exfoliating air-sensitive materials. Furthermore, this method has been used to isolate CrSBr thin films [8], demonstrating the suitability of the KISS method for isolating 2D quantum magnetic materials. While promising, further research is needed to scale these methods to wafer-scale production.

Whether it is epitaxially grown or gold-assisted exfoliated, incorporating large-area 2D magnetic materials would still require optimization of detachment from the substrate and reattachment to a new surface. The existing detachment methods demonstrated for other 2D materials involve etching of the substrate by water-soluble acids or bases [9] and are done in atmospheric conditions, which can rapidly degrade magnetic 2D materials. Hence developing anhydrous chemical detachment processes that involve glovebox and gentle encapsulation processes, such as transfer of squeeze-printed liquid metal oxides, is crucial for large-area 2D magnetic material scalability and applications [10].

**Concluding Remarks**

Two-dimensional magnetic materials are promising for electronic, optoelectronic, and spintronic applications due to their tunable properties through interface coupling, and achieving atomically flat, clean interfaces is crucial, as demonstrated by recent research on inducing magnetism in graphene via proximity effect. Gold-assisted exfoliation and KISS method have emerged as effective methods for producing large-area, high-quality 2D monolayers, overcoming some of the limitations of MBE and CVD growth, however, both methods require further research for wafer-sized production. Additionally, developing glovebox-compatible, anhydrous chemical detachment processes and gentle encapsulation methods is essential for scaling up 2D magnetic materials for industrial applications.

**Acknowledgements**

AGC acknowledges the financial support of the Zernike Institute for Advanced Materials. SB and AGC acknowledge the research program "Materials for the Quantum Age" (QuMat) for financial support. This program (registration number 024.005.006) is part of the Gravitation program financed by the Dutch Ministry of Education, Culture and Science (OCW). ZD acknowledges the fellowship from the Chinese Scholarship Council (No.202206750016). We acknowledge the support from the Lorentz Center through the Lorentz Centre workshop "Quantum Magnetic Materials".

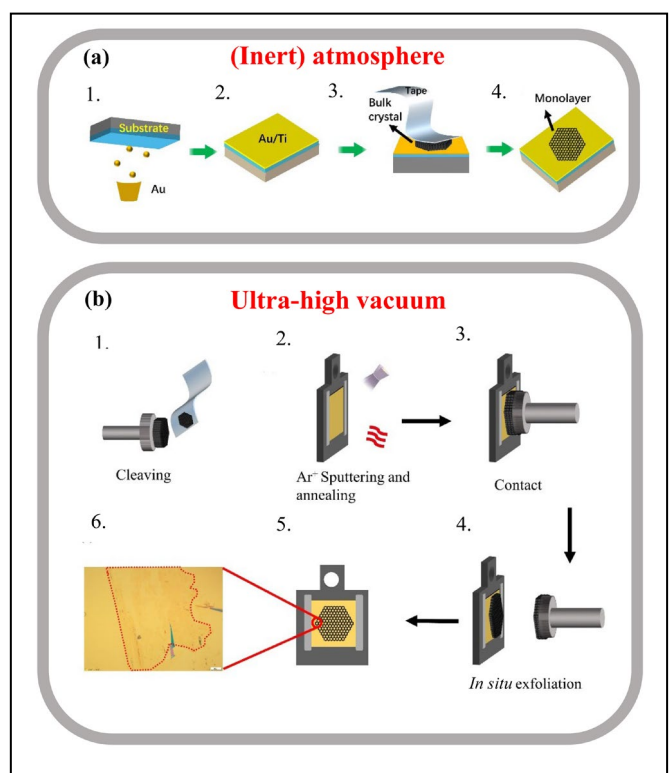

**Figure 9.2.** A schematic showing steps of **i)** gold assisted exfoliation and **ii)** kinetic *in situ* single-layer synthesis (KISS) method. **i)** In gold assisted exfoliation gold is deposited on an SiO$_2$ substrate (a). Gold will act as both an exfoliation tape and as a substrate (b). Bulk crystal is prepared on a white tape and gently pressed to the gold surface (c). Upon separation, large-area 2D monolayer is left on the surface (d). KISS method **ii)** is performed in ultra-high vacuum. Bulk crystal is prepared using tape or a top post method (a) while gold or other substrate of choice is prepared using sputtering and annealing (b). Clean surfaces are brought into contact (c), and upon separation (d) a large-area 2D monolayer is left on the surface of gold (e,f). Material can be further studied *in situ*.

**References**
[1] Talieh S. Ghiasi, Alexey A. Kaverzin, Avalon H. Dismukes, Dennis K. de Wal, Xavier Roy and Bart J. van Wees, "Electrical and thermal generation of spin currents by magnetic bilayer graphene," *Nature Nanotechnology,* vol. 16, pp. 788–794, 2021
[2] P. J. Zomer, M. H. D. Guimarães, J. C. Brant, N. Tombros and B. J. van Wees, "Fast pick up technique for high quality heterostructures of bilayer graphene and hexagonal boron nitride," *Applied Physics Letters,* vol. 105, pp. 013101, 2014
[3] C. Lan, Z. Zhou, Z. Zhou, C. Li, L. Shu, L. Shen, D. Li, R. Dong, S. Yip, and J. C. Ho, "Wafer-scale synthesis of monolayer WS2 for high-performance flexible photodetectors by enhanced chemical vapor deposition," *Nano Research,* vol. 11, pp. 3371-3384, 2018
[4] J.M.J. Lopes, D. Czubak, E. Zallo, A.I. Figueroa, C. Giullemard, M. Valvidares, J. Rubio-Zuazo, J. Lopez-Sanchez, S.O. Valenzuela and M. Hanke, "Large-area van der Waals epitaxy and magnetic characterization of Fe3GeTe2 films on graphene", *2D Materials*, vol. 8, pp. 041001, 2021
[5] M. Gibertini, M. Koperski, A.F. Morpurgo, and K.S. Novoselov, "Magnetic 2D materials and heterostructures," *Nature Nanotechnology,* vol. 14, pp. 408–419, 2019
[6] S. B. Desai, S. R. Madhvapathy, M. Amani, D. Kiriya, M. Hettick, M. Tosun, Y. Zhou, M. Dubey, J. W. Ager, 3rd, D. Chrzan, and A. Javey, "Gold-Mediated Exfoliation of Ultralarge Optoelectronically-Perfect Monolayers," *Advanced Materials*, vol. 28, pp. 4053, 2016
[7] A. Grubisic-Cabo, M. Michiardi, C. E. Sanders, M. Bianchi, D. Curcio, D. Phuyal, M. H. Berntsen, Q. Guo, and M. Dendzik, "In Situ Exfoliation Method of Large-Area 2D Materials," *Advanced Science*, vol. 10, pp. 2301243, 2023

[8] M. Bianchi, K. Hsieh, E J. Porat, F. Dirnberger, J. Klein, K. Mosina, Z. Sofer, A.N. Rudenko, M.I. Katsnelson, Y.P. Chen, M. Rösner and P. Hofmann, *Physical Review B* vol. 108, pp. 195410, 2023

[9]  C. Kim, M.-A. Yoon, B. Jang, H.-D. Kim, J.-H. Kim, A. Tuan Hoang, J.-H. Ahn, H.-J. Jung, H.-J. Lee and K.-S. Kim, "Damage-free transfer mechanics of 2-dimensional materials: competition between adhesion instability and tensile strain", *NPG Asia Materials*, vol. 13, pp. 44, 2021.

[10] M. Gebert, S. Bhattacharyya, C.C. Bounds, N. Syed, T. Daeneke, and M.S. Fuhrer, "Passivating Graphene and Suppressing Interfacial Phonon Scattering with Mechanically Transferred Large-Area $Ga_2O_3$ ", *Nano Letters*, vol. 23, pp. 363–370, 2022

# 10. Nanomechanical probing of magnetic membranes

Maurits J. A. Houmes, Yaroslav M. Blanter, and Herre S. J. van der Zant

Kavli Institute of Nanoscience, Delft University of Technology, The Netherlands

**Status**

Experimental characterization of magnetic order in 2D magnetic materials is a challenging task. Standard methods such as neutron scattering often do not work with thin films. The local magnetization probes based on NV-center magnetometry or spin-polarized scanning tunneling microscopy have a limited scope; for example, they cannot detect static antiferromagnetic order. Electric measurements of spin currents in 2D magnets are still at an early stage and more effort is needed to understand what information they provide. In this situation, thermodynamic properties are a good means to get the needed information as at the (magnetic) phase transition, they show singularities. Magnons contribute to the thermodynamic properties as well. As a bonus, magnetic systems typically display non-trivial critical behaviour close to the transition, and one can expect that this behaviour is stronger in 2D magnets.

Thermodynamic quantities such as the specific heat of the thermal expansion coefficient are difficult to measure for 2D systems directly. However, an advantage of 2D materials and specifically magnets is that they can be suspended over a cavity, and that the mechanical frequency and quality factor of mechanical motion can be easily measured, as described below. We show that this resonance frequency depends on the specific heat and the thermal expansion coefficient; it thus contains information about the magnetic structure. Furthermore, the quality factor gives insight into dissipation mechanisms, which are not yet fully understood. Such nanomechanical studies of magnetic materials have so far focussed on transition metal phosphorus sulphides, such as $FePS_3$, $CoPS_3$, $NiPS_3$, as well as some notable results on $CrI_3$ and $Cr_2Ge_2Te_6$, but are not limited to these materials.

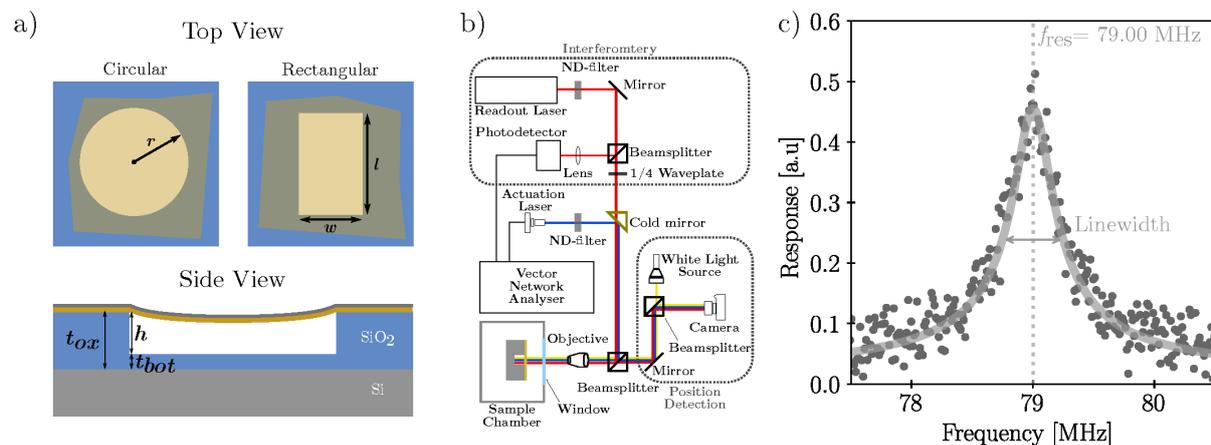

**Figure 10.1**: a) A schematic view of typical samples. The top panels show the two types of cavity shapes: a circular (isotropic) cavity where one can vary the radius, and a rectangular (anisotropic) cavity where one can vary the width and length independently. The bottom panel shows a cut through a typical sample where a heterostructure, for example a capping layer (grey) and an air sensitive material (orange), are suspended over a cavity etched into an oxide layer. The cavity depth, h, can be varied in conjunction with the oxide thickness, $t_{ox}$, and the residual oxide thickness left after etching, $t_{bot}$. b) A schematic of the interferometric measurement scheme. c) The dark dots are the measured frequency response; the grey line is a fit which is used to extract the linewidth and resonance frequency, indicated by the dashed line.

**Current and Future Challenges**

To fabricate the resonating membranes, magnetic 2D materials are exfoliated and deterministically placed on prepared substrates over etched cavities, figure 10.1a. In most cases, these are circular in shape. In this way, a suspended section of the material of interest (the membrane) is thereby created which is free to vibrate. It is actuated using a power modulated laser producing an optothermal driving or by applying a varying electrostatic potential [1]. The resulting mechanical motion is then probed with a readout laser typically of a longer wavelength, as sketched in figure 10.1b. By varying the actuation frequency, the resulting resonance peaks can be identified in the spectrum, see figure 10.1c for an example. These peaks correspond to mechanical resonance modes of the membrane and typically the fundamental mode is chosen as it exhibits the largest displacements. Such optical driving is generally used as it does not require mechanical contact, as with piezo excitation, or the membrane to be conductive, which many 2D magnets are not. The frequency of the fundamental mode, which is typically in the 1-100 MHz regime, is determined by the geometry of the membrane, the density ρ and elastic properties of the material, and crucially on the strain present in the membrane, and can be written as:

$$f_{res}(T) = a\sqrt{\frac{E}{\rho}\frac{\epsilon(T)}{(1-\nu)}}. \quad (1)$$

Here, a is a geometric factor, E is the material Young's modulus, and $\nu$ is its Poisson's ratio. We see that the temperature dependent resonance frequency, $f_{res}(T)$, is proportional to the square root of the strain, $\epsilon(T)$. By varying an external parameter, such as temperature or magnetic field, a change in the resonance frequency of the membrane can be observed, corresponding to a change in strain. For example, by tracking the resonance frequency as a function of temperature, the strain variation with temperature is determined. This strain variation is due to thermal effects, and after subtracting the contribution from the substrate, it can be used to extract the thermal expansion coefficient (TEC), α(T), of the material. The ability to measure the TEC allows for the probing of thermodynamical quantities such as the specific heat, cv, which is related to α(T) via the Grüneisen relation [2]:

$$c_v(T) = 3\alpha(T)\frac{KV_M}{\gamma} = 3\left(\alpha_{sub}(T) - \frac{1}{\mu^2}\frac{df_{res}^2(T)}{dT}\right)\frac{KV_M}{\gamma}, \quad (2)$$

where K is the bulk modulus, $V_M$ the molar volume, γ the Grüneisen parameter of the membrane, $α_{sub}(T)$ is the TEC of the substrate, and μ = $a\sqrt{\frac{E}{\rho(1-\nu)}}$. As second-order phase transitions exhibit a discontinuity in the specific heat, Eq. (2) indicates that the resonance frequency can be used to determine the temperature of such a transition [2]. Aside from the resonance frequency, the mechanical response can also be used to probe the dissipation mechanisms present in the membrane, encoded in the linewidth and in the quality factor of the resonance peak. This method of probing the specific heat is particularly suited for 2D materials where the material volumes of interest are small; more traditional methods typically lead to too low signal intensities in such cases. Specifically, typical frequencies for membranes with a radius of a few microns are in the 10 – 100 MHz range; at low temperatures, the Q-factor can exceed 10,000. Since the frequency can easily be measured within an accuracy of the peak width (peak width divided by frequency is the Q-factor), it can be shown with Eq. (1) that the method is sensitive to lattice parameter changes in the order of $10^{-17}$ m.

If an anisotropic geometry, like a rectangular shape, is used to suspend the material a membrane is created that has anisotropic sensitivity to strain. It thereby provides a way to study the strain along different crystallographic axes of the material. This is particularly interesting for magnetic 2D materials as they require a certain level of anisotropy to stabilise the magnetic order in the 2D limit. By suspending a single crystal over differently oriented anisotropic cavities several membranes are created with sensitivity along different crystallographic axes. By comparing the temperature dependent frequency response of these cavities additional information about the system can be extracted such as the critical exponent describing the magnetic transition for magnetostrictive 2D materials. An example is shown in figure 10.2b, displaying the difference of the square of the frequencies, $f^2_b - f^2_a$, which is proportional to the magnetic order parameter, $L^2$, of the material [3, 4].

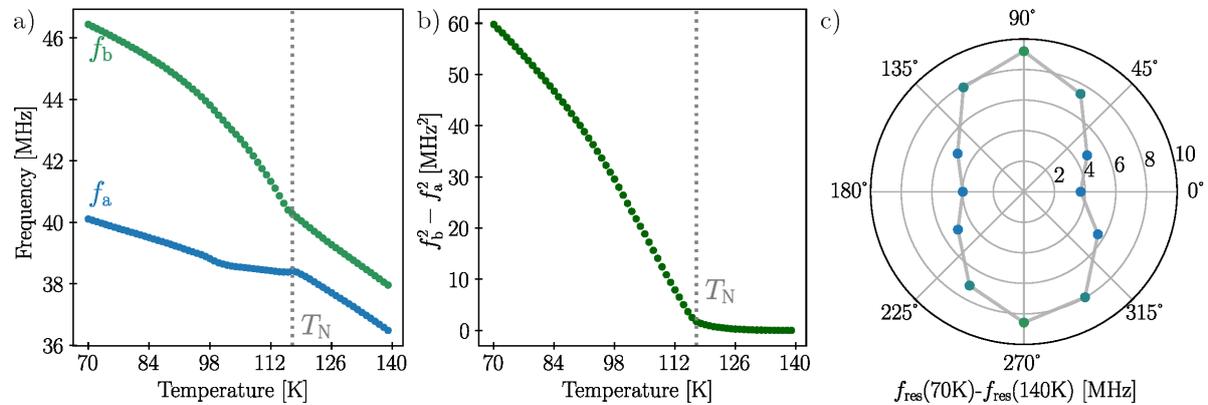

**Figure 10.2**: a) Temperature dependence of resonance frequency of rectangular $CoPS_3$ membranes; shown are $f_a$ (blue) and $f_b$ (green) which are the frequencies of rectangular cavities oriented along the a- and b-crystallographic axis, respectively. The dashed line indicates the transition temperature, $T_N$ = 117 K. b) Difference of the corrected frequency squared $f^2_b - f^2_a$ which is proportional to the order parameter, $L^2$, of the material. The dashed line is the same as in a). c) Polar plot of $f_{res}$(70 K) – $f_{res}$(140 K), the color corresponds to the angle w.r.t. the crystallographic b-axis with the blue 180° corresponding to $f_a$ and the green at 270° corresponding to $f_b$ shown in a).

Membrane resonators can also be used to study the strength of exchange interactions in a 2D magnet as was demonstrated by Jiang et al. [5] They investigated a suspended heterostructure of graphene, $CrI_3$ and $WSe_2$, while applying an external magnetic field and electrostatically actuating the membrane. Using MCD the out-of-plane magnetisation was monitored. Meanwhile, the exciton energy of $WSe_2$ was measured and used as a strain gauge. The resonance frequency showed a shift when driving the $CrI_3$ membrane over a spin-flip transition. By correlating the applied gate voltage to the induced strain, the strength of the exchange interaction can be determined.

**Advances in Science and Technology to Meet Challenges**
So far, most studies using nanomechanical membranes for material characterisation have stayed in the linear regime of the resonator. This leaves the non-linear regime largely unexplored. Recent work by Šiškins et al. [6] suggests that the emergence of magnetic order does significantly affect the behaviour of non-linear resonators. These effects appear to originate from the kinematics of the magnetic order causing a renormalisation of the mechanical parameters. This renormalisation is present particularly close to the transition where the magnetic order is not fully saturated while not being present in the paramagnetic state. Although further theory and experimental verification is required, this behavior shows that nanomechanical resonators provide a very versatile platform for the study of 2D magnets.

Aside from driving the mechanics, optothermal excitation of nanomechanical membrane resonators also directly drives the magnetic order [7]. The optothermal drive originates from the periodic heating induced by the laser, which, in turn, modulates the magnetic order. For 2D magnets this provides an additional driving force of a membrane as the magnetostrictive coupling translates the modulation of the magnetic order into a strain modulation. The efficiency of such driving is the highest if the magnetisation strongly depends on the temperature, which occurs close to the transition temperature, far from the regime where the magnetic order saturates.

**Concluding Remarks**

In conclusion, recent works have shown that nanomechanical resonators are an excellent tool to characterise magnetic 2D materials due to the coupling between magnetic order and the crystal lattice via magnetostriction present in these materials. Such resonators can be used to investigate specific heat, anisotropic effects, critical exponents, damping effects, and spin-flip transitions, all on small volumes of the order of 10–20 m$^3$. The underlying theory describing these systems currently consists of phenomenological based models. It is thus of interest to develop more thorough, microscopic or semi-microscopic, theoretical frameworks. From an experimental side, several directions can be pursued from the study of heterostructures, to potentially create materials with on-demand magnetic properties, to the study of magnons and magnon-membrane motion transduction devices based on magnetoelastic coupling.

**Acknowledgements**

**References**

[1] Steeneken, P. G., Dolleman, R. J., Davidovikj, D., Alijani, F. & van der Zant, H. S. J., "Dynamics of 2D material membranes." *2D Materials*, vol. 8, pp. 1295–1299, 2021
[2] Šiškins, M. et al., "Magnetic and electronic phase transitions probed by nanomechanical resonators." *Nature Communications*, vol. 11, pp. 2698, 2020
[3] Houmes, M. J. A. et al., "Magnetic order in 2D antiferromagnets revealed by spontaneous anisotropic magnetostriction.", *Nature Communications*, vol. 14, pp. 8503, 2023
[4] Houmes, M. J. A. et al., "Highly Anisotropic Mechanical Response of the Van der Waals Magnet CrPS$_4$.", *Advanced Functional Materials*, vol. 34, pp. 2310206, 2024
[5] Jiang, S., Xie, H., Shan, J. & Mak, K. F., "Exchange Magnetostriction in Two-Dimensional Antiferromagnets.", *Nature Materials*, vol. 19, pp. 1295–1299, 2020
[6] Šiškins, M. et al., "Nonlinear dynamics and magneto-elasticity of nanodrums near the phase transition" *arXiv: 2309.09672 [cond-mat.mes-hall]*, 2023
[7] Baglioni, G. et al., "Thermo-Magnetostrictive Effect for Driving Antiferromagnetic Two-Dimensional Material Resonators.", *Nano Letters*, vol. 23, pp. 6973–6978, 2023

# 11. NV Magnetometry of spin textures and dynamics in two dimensions

Samer Kurdi[1,2], Toeno van der Sar[3]

[1] Zernike Institute for Advanced Materials, University of Groningen, 9747 AG Groningen, the Netherlands

[2] Institute of Photonics and Quantum Sciences, SUPA, Heriot-Watt University, Edinburgh, EH14 4AS, UK

[3] Department of Quantum Nanoscience, Kavli Institute of Nanoscience, Delft University of Technology, 2628 CJ Delft, the Netherlands.

**Status**

Photoluminescent defect spins in solid-state materials have emerged as powerful tools for probing magnetic materials. In particular, the NV centre in diamond (Figure 11.1A) has enabled micro-to-nano-scale magnetic imaging[1] including the detection and imaging of spin dynamics (spin waves) in magnetic thin films[1,2] and of the spin structure of few- and mono-layer 'van der Waals' magnets (2D magnets)[3,4]. In addition, defect spins in hBN are increasingly being explored for probing magnetic materials[5]. In this section, we describe recent applications with the NV centre, which has already enabled an array of spin-structure and spin-wave imaging experiments and benefits from two decades of development. However, we expect the techniques and challenges discussed to be applicable to other optically active solid-state spin sensors.

In general, probing both static spin structures and their dynamics requires a large detection bandwidth (DC-to-GHz) of the spin sensor. The most common defect-spin-based DC-magnetometry protocol is continuous wave (CW) optically detected magnetic resonance (ODMR). In this protocol, optical illumination continuously pumps the spin into an eigenstate. The application of a microwave field then causes a reduction of the photoluminescence when resonant with the spin splitting. From the resonance frequency and the known spin Hamiltonian, the local magnetic field can be extracted. Spin sensors also offer more sensitive, lock-in-type sensing protocols for fields with frequencies in the kHz-to-MHz range[6]. In addition, the strength of the ESR response enables quantifying the amplitude of GHz-microwave fields, such as those generated by spin waves[1].

Scanning-NV magnetometry based on diamond probes with a single NV spin sensor at their tip (Figure 11.1B) has enabled ~50 nm resolution imaging of a range of magnetic structures in 2D magnets such as domain walls[3,7] and Moiré superlattices[4]. High-density layers of NV centres implanted in diamond membranes (~1000/µm$^2$, Figure 11.1C) permit faster imaging but with optical diffraction limited resolution. Reducing such membranes to the microscale strongly increases the probability of achieving contact with a target sample, which is often crucial to study the weak stray fields of e.g. magnetic monolayers[8] (Figure 11.1D). In this respect, defect spins in hBN benefit from being easily brought into contact with the target 2D material[9]. Both DC and AC magnetometry protocols are now being used to image spin waves in magnetic materials – a relatively new field for spin-based sensing. NV-imaging of coherently driven spin waves has until now been mostly applied to model magnets such as YIG[1,2] (Figure 1E). This technique relies on resonance between NV and spin waves and obtains its spatial phase sensitivity through interference of the spin-wave field with a spatially homogeneous magnetic field of the same frequency. A key challenge is to develop NV-spin-manipulation techniques that enable imaging of spin waves that are frequency detuned from the NV centre, such as spin waves in 2D magnets that are upshifted in frequency by magnetic anisotropies, as described next.

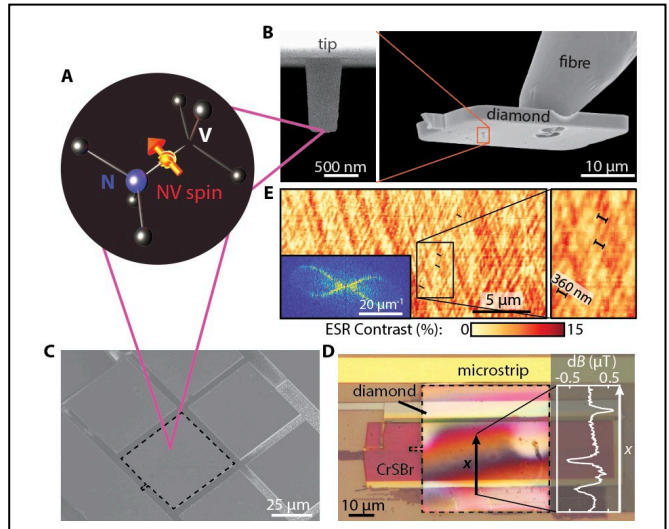

**Figure 11.1. Probing spins and spin dynamics in thin-film and 2D magnets using nitrogen vacancy (NV) spin(s) in diamond.** A) NV magnetometry uses the S = 1 electron spin associated with the NV defect in diamond as a sensor of magnetic fields. Figure adapted from [8] licensed under CC-BY 4.0. B) Scanning-NV magnetometry uses a diamond cantilever with a single NV implanted ∼20 nm below the tip surface, permitting high resolution magnetic imaging with topography feedback. B) Microscale diamond membranes with a dense layer of near-surface NV spins enable fast imaging and a large probability of achieving physical contact with a target sample. Figure adapted from [7] licensed under CC-BY 4.0. D) A magnetic flake of CrSBr stamped next to an RF-microstrip with an NV-diamond membrane on top. Inset: the NV-measured field of the CrSBr along the black arrow shows peaks caused by steps in the CrSBr thickness. Figure adapted from [7] licensed under CC-BY 4.0. E) ESR contrast spatial map obtained via scanning-NV magnetometry of spin-wave scattering in YIG. Inset: the corresponding Fourier transform reveals the iso-frequency contour of the spin-wave dispersion. Figure adapted from [9] licensed under CC-BY 4.0.

**Current and Future Challenges**

Because the magnetic parameters of 2D magnets are tunable by voltages, strain, pressure and auxiliary 2D materials, they offer a promising platform for realizing tunable spintronic and spin-wave devices. Moreover, spin-wave interactions and fluctuations should be more important in 2D because of the reduced phase space and the reduction of the number of nearest neighbours, offering an exciting outlook of new spin-wave transport regimes.

A recent NV magnetometry investigation of spin-waves in the 2D magnet $CrCl_3$ offered a glimpse of this potential by revealing low-frequency spin dynamics [12]. However, NV magnetometry has not yet provided imaging of coherent spin waves in 2D magnets, which could shed light on important parameters such as spin-wave damping and interactions. Key challenges are that:

(1) 2D magnets are atomically thin and produce small magnetic fields. One potential spin-wave measurement modality is to detect the change in the time-averaged longitudinal magnetization ($\delta M$, Figure 11.2A) that occurs upon the excitation of a spin-wave mode. However, as $\delta M$ is limited by spin-wave damping, which is unknown for most 2D magnets, it is unclear if the signal strengths (Figure 11.2B) will be sufficient to enable spin-wave detection.
(2) Many 2D magnets have large magnetic anisotropies and as such, they tend to have spin-wave spectra at large (>10 GHz) frequencies. This makes it challenging to generate spin waves that are resonant with the NV-ESR frequency and detect them through the NV-ESR contrast[2]. As such, new techniques that enable off-resonant spin-wave imaging are required.

Once the tools for probing the spectrum and transport parameters of spin waves in 2D magnets are developed, a natural next step is to develop control of the spin-wave transport and create elementary devices.

**Advances in Science and Technology to Meet Challenges**

Outside of current NV magnetometry advances for increasing magnetic field sensitivity, spatial resolution and improving NV-spin control, we highlight key advancements required for NV-imaging of spin waves in 2D magnets for the two main NV spin-wave sensing techniques:

(1) **Resonant detection**: The application of a DC magnetic field to a magnetic material shifts the spin-wave spectrum relative to the NV ESR frequency. For some 2D magnets, we expect that magnetic fields of specific orientations can tune the spin-wave frequencies into resonance with the NV-ESR transition. This would enable the application of the resonant NV-spin-wave detection schemes, where the GHz-magnetic field generated by a spin wave directly drives NV-ESR transitions.

(2) **Off-resonant detection**: When the spin-wave frequencies are detuned from the NV-ESR transitions, off-resonant protocols are required. Expanding the sensing bandwidth of NV centres is a widespread endeavour, but challenges persist such as the sensitivity reduction accompanying an increasing detuning. Off-resonant detection techniques for incoherent magnon detection have been demonstrated[11], here we propose two off-resonant detection protocols for the detection of ESR-detuned coherent spin waves. The first is to monitor changes in a 2D magnet's stray magnetic field caused by the change ($\delta M$, Figure 11.2A) upon spin-wave excitation as described above. The second protocol focuses on down-converting driven spin-wave signals to MHz-frequencies accessible to NV centres as we demonstrated in YIG[8] (Figure 11.2C-D). In both approaches, AC magnetometry techniques should increase the sensitivity. The goal is to extend the approaches to 2D magnets, which have larger crystal anisotropies that increase the spin-wave gap.

Key advances to meet these challenges are required. Firstly, preliminary knowledge of the 2D magnet spin-wave spectra is important for guiding the experiments; theoretical calculations for spin waves in the wide range of 2D magnetic heterostructures could lead the field. Secondly, the unknown and possibly small spin-wave lifetimes and decay lengths make it challenging to predict the detectability of the spin-wave signals. The development of low-damping 2D magnetic insulators and spin-coherence-protecting encapsulating layers would be a breakthrough in the field. Thirdly, the delivery of microwaves at the high frequencies (>10 GHz) of spin waves in 2D magnets poses a microwave engineering challenge. This could benefit from on-chip generation of microwave signals using e.g. spin-torque oscillators or superconducting circuits.

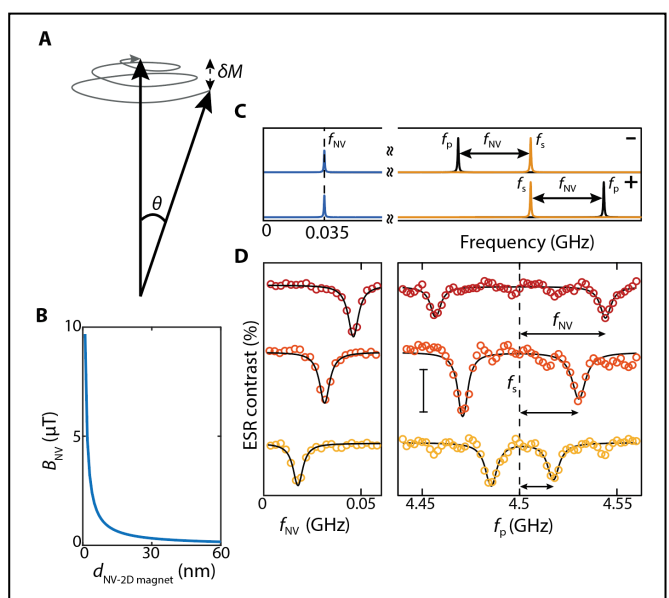

**Figure 11.2. NV magnetometry of spin-waves in 2D magnets.** A) Schematic of the magnetization after microwave excitation, e.g. of the ferromagnetic resonance spin-wave mode, resulting in processional motion (θ is the precessional angle). The amplitude of the precessional motion is reduced over time due to spin damping resulting in spiral motion. The change in the longitudinal component of the magnetization ($\delta M$) oscillates at a frequency which can be detected using NV detection techniques. Figure adapted from [8] licensed under CC-BY 4.0. B) The longitudinal component of the magnetization detected along the NV-quantization axis ($B_{NV}$) as a function of 2D magnet – NV distance ($d_{NV\text{-}2Dmagnet}$). A 1° precessional angle was assumed for a monolayer CrSBr flake in the calculation. C) Sketch of the frequency spectrum showing the difference-frequency generation detection scheme which enables GHz to MHz down conversion. Here, two spin-wave modes are excited at $f_s$ (signal frequency) and $f_p$ (pump frequency) respectively, in which their difference creates a detectable signal at the NV-ESR frequency ($f_{NV}$): $f_p - f_s = \pm f_{NV}$. Figure adapted from [8] licensed under CC-BY 4.0. D) Corresponding experimental demonstration of (C): ESR traces for different $B_{NV}$ at $f_s$ = 4.5GHz (dashed vertical line) showing the shift of the frequency difference resonant with $f_{NV}$ ($f_p - f_s = \pm f_{NV}$). Scale bar: 0.1% ESR contrast. Figure adapted from [8] licensed under CC-BY 4.0.

**Concluding Remarks**

NV magnetometry has emerged as a tool for probing micro-to-nano-scale spin structures and spin-wave dynamics. It has permitted the quantification of the amplitudes, wavelengths, and decay lengths of spin waves in model magnets such as YIG[1,2] and provided insight into the spin-wave interaction with superconductors[14]. Its spatial resolution could enable probing the nanoscale spin waves important for the development of miniaturized spin-wave logic devices. Probing high-frequency spin-wave dispersions in e.g. 2D magnets remains an outstanding challenge, but could be overcome by developing off-resonant spin-wave detection techniques. The ability to image spin waves in 2D magnets would push our understanding of the role of dimensionality on the spin wave transport and pave the way towards 2D magnonic devices.

**Acknowledgements**

S. K. acknowledge support from the Dutch Research Council (NWO) via VI.Veni.222.296 and T. S. acknowledges the NWO via VI.Vidi.193.077, NWA.1160.18.208, and OCENW.XL21.XL21.058.

**References**

[1] Casola F, van der Sar T and Yacoby A, "Probing condensed matter physics with magnetometry based on nitrogen-vacancy centres in diamond", *Nature Review Materials,* vol. 3, pp. 17088, 2018

[2]   Bertelli I, Carmiggelt J J, Yu T, Simon B G, Pothoven C C, Bauer G E W, Blanter Y M, Aarts J and van der Sar T, "Magnetic resonance imaging of spin-wave transport and interference in a magnetic insulator", *Science Advances,* vol. 6, pp. eabd3556, 2020

[3]   Zhou T X, Carmiggelt J J, Gächter L M, Esterlis I, Sels D, Stöhr R J, Du C, Fernandez D, Rodriguez-Nieva J F, Büttner F, Demler E and Yacoby A, "A magnon scattering platform" *Proceedings of the National Academy of Sciences of the United States of America,* vol. 118, pp. e2019473118, 2021

[4]   Thiel L, Wang Z, Tschudin M A, Rohner D, Gutiérrez-Lezama I, Ubrig N, Gibertini M, Giannini E, Morpurgo A F and Maletinsky P, "Probing magnetism in 2D materials at the nanoscale with single-spin microscopy", *Science,* vol. 364, pp. 973–6, 2019

[5]   Vaidya S, Gao X, Dikshit S, Aharonovich I and Li T, "Quantum sensing and imaging with spin defects in hexagonal boron nitride" *Advances in Physics: X,* vol. 8, pp. 2206049, 2023

[6]   Casola F, van der Sar T and Yacoby A, "Probing condensed matter physics with magnetometry based on nitrogen-vacancy centres in diamond" *Nature Review Materials,* vol. 3, pp. 17088, 2018

[7]   Högen M, Fujita R, Tan A K C, Geim A, Pitts M, Li Z, Guo Y, Stefan L, Hesjedal T and Atatüre M, "Imaging Nucleation and Propagation of Pinned Domains in Few-Layer $Fe_{5-x}GeTe_2$" *ACS Nano,* vol. 17, pp. 16879–85, 2023

[8]   Ghiasi T S, Borst M, Kurdi S, Simon B G, Bertelli I, Boix-Constant C, Mañas-Valero S, van der Zant H S J and van der Sar T, "Nitrogen-vacancy magnetometry of CrSBr by diamond membrane transfer", *NPJ 2D Materials and Applications, vol.* 7, pp. 62, 2023

[9]   Healey A J, Scholten S C, Yang T, Scott J A, Abrahams G J, Robertson I O, Hou X F, Guo Y F, Rahman S, Lu Y, Kianinia M, Aharonovich I and Tetienne J -P, "Quantum microscopy with van der Waals heterostructures", *Nature Physics,* vol. 19, pp. 87–91, 2023

[10]   Carmiggelt J J, Bertelli I, Mulder R W, Teepe A, Elyasi M, Simon B G, Bauer G E W, Blanter Y M and van der Sar T, " Broadband microwave detection using electron spins in a hybrid diamond-magnet sensor chip" *Nature Communications, vol.* 14, pp. 490, 2023

[11]   Simon B G, Kurdi S, Carmiggelt J J, Borst M, Katan A J and van der Sar T, "Filtering and Imaging of Frequency-Degenerate Spin Waves Using Nanopositioning of a Single-Spin Sensor", *Nano Letters,* vol. 22, pp. 9198–204, 2022

[12]   Xue R, Maksimovic N, Dolgirev P E, Xia L-Q, Kitagawa R, Müller A, Machado F, Klein D R, MacNeill D, Watanabe K, Taniguchi T, Jarillo-Herrero P, Lukin M D, Demler E and Yacoby A, "Signatures of magnon hydrodynamics in an atomically-thin ferromagnet" *ArXiv:2403.01057,* pp. 1–8, 2024

[13]   Wang H, Zhang S, McLaughlin N J, Flebus B, Huang M, Xiao Y, Liu C, Wu M, Fullerton E E, Tserkovnyak Y and Du C R, "Noninvasive measurements of spin transport properties of an antiferromagnetic insulator", *Science Advances,* vol. 8, pp. 8562, 2022

[14]     Borst M, Vree P H, Lowther A, Teepe A, Kurdi S, Bertelli I, Simon B G, Blanter Y M and van der Sar T, "Observation and control of hybrid spin-wave–Meissner-current transport modes", *Science,* vol. 382, pp. 430–4, 2023

# 12. ARPES studies of 2D quantum magnets

Sergio Barquero Pierantoni[1], Antonija Grubišić-Čabo[2], Chrystalla Knekna[1,2], Mengxing Na[3], Muhammad Waseem[2]

[1.] Van der Waals - Zeeman Institute, Institute of Physics, University of Amsterdam, [2.] Zernike Institute for Advanced Materials, University of Groningen, [3.] Institute for Molecules and Materials, Radboud University

**Status**

The experimental discovery of atomically thin magnetic materials has paved the way for extensive experimental and theoretical studies in the field of 2D magnetism. 2D magnets are essential components in exfoliation-based heterostructures, as they allow for the tuning of magnetization. Furthermore, the coupling between the electronic wave function and the magnetic spin configuration can result in a plethora of exotic electronic phases, such as the QAHE [1] and the axion insulator phase [2], which have potential applications in low-power-consumption electronics and spintronics. Therefore, it is crucial to probe the evolution of the electronic structure of quantum magnetic materials as they are thinned down from bulk to the 2D limit, as well as across magnetic phase transitions. ARPES is a photoemission based technique that provides direct experimental access to the momentum-resolved electronic band structure. The high surface sensitivity achieved at low photon energies (10–100 eV) makes ARPES an ideal technique for studying atomically thin materials. While ARPES is a well-established technique extensively used to study various quantum materials [3], studies of 2D magnets are still in their early stages. Despite this, the quasi-2D nature of several vdWs ferromagnets, including $Cr_2Ge_2Te_6$ [4] and $Fe_3XTe_2$ (X = Ge, Ga) [5], has been determined. These materials exhibit long-range magnetic order, large magneto-crystalline anisotropy, and high Curie temperature even in low dimensions, which are relevant for future spintronic applications. Recently, ARPES studies were performed for the first time on an exfoliated monolayer of a vdWs antiferromagnet, $MnPS_3$, a member of the $XPS_3$ (X = Mn, Fe, Co, Ni) family of insulating magnetic materials [6]. In this system, ARPES measurements revealed magnetic splittings in the band structure (Figure 12.1b), and detailed comparisons with DFT+U calculations provided insights into the orbital character (Figure 12.1a). These findings demonstrate the suitability of ARPES for investigating the effects of magnetism on electronic structure. Moreover, ARPES plays a key role in the field of 2D magnetic topological insulators allowing to directly probe the magnetic gap opening within the topological surface states (Figure 12.1d), and studying the interplay between topology, magnetism and dimensionality (Figure 12.1c) [7,8]. Unfortunately, to date, most 2D magnetic phases have been confined to low temperatures – with a recent exceptions being CrSBr [9], $CrTe_2$ [10] and self-intercalated $Cr_{1+x}Te_2$ [11] – precluding their possible application in low-consumption electronics and spintronics. The following sections outline the main technical and material challenges limiting the application of ARPES in the investigation of 2D magnets, as well as the required technical advances.

**Current and Future Challenges**

The unique momentum resolution of ARPES makes it ideal for studying the interplay between magnetism and electronic band structure. While measurements in three-dimensional (3D) magnets are now performed almost routinely, several challenges arise when applying this technique to 2D magnets. Figure 12.1 shows an overview of the observables that ARPES has allowed (a-d) or could allow (e-g) to access in two dimensional magnetic samples.

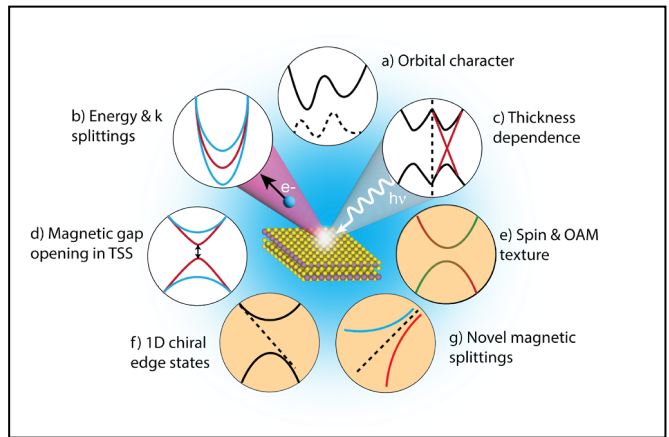

**Figure 12.1.** Observables that can be probed using angle-resolved photoemission spectroscopy (ARPES): a) Orbital character of bands, b) splitting of bands in energy and momentum, c) changes appearing upon thickness variation, d) magnetic gap opening within the topological surface states, e) spin and orbital angular momentum character of bands, f) 1D chiral edge states and g) new types of magnetic splittings within the band structure. Orange shading in e)-g) indicates novel phenomena that are only theoretically predicted or still need to be experimentally observed by ARPES in 2D magnets.

One crucial requirement for ARPES experiments is an atomically flat and clean surface. This is often achieved by cleaving a bulk single crystal *in situ* in UHV. Alternatively, samples can be grown directly under UHV conditions using methods like MBE, which, if coupled to the ARPES chamber, circumvents exposure to air. Of course, the standard cleaving method cannot be applied to 2D samples, and very thin MBE-grown samples often suffer from lattice mismatch with the substrates, resulting in compressive strain and lower ordering temperatures. Sample size and uniformity are also crucial. The typical spot size used in ARPES experiments is on the order of 100 µm, meaning that the samples should be structurally homogeneous at the micrometer scale.

Assuming a suitable surface is found, additional complications may arise from the characteristics of the samples themselves. For instance, the characteristic one-dimensional (1D) edge modes present in 2D ferro/ferrimagnetic topological samples (Figure 12.1f) are strongly localized within a few nanometers of the edge, making their detection with a micrometer-sized beam challenging, in particular as the weak signal from the 1D state is difficult to uncouple from any other present 2D or 3D state. Another sample-dependent challenge is the position of the Fermi level. A major drawback of conventional photoemission experiments is the inability to access unoccupied states. This limitation can be particularly relevant when determining the topological character of a novel sample, as topologically nontrivial features may exist in the unoccupied band structure, inaccessible by conventional ARPES experiments. Finally, measuring insulators or wide band gap semiconductors is notoriously difficult due to charging issues. Consequently, these materials are not heavily studied, and observed features can be broad and difficult to interpret. This is particularly an issue for 2D magnets, as many of them fall within this category.

**Advances in Science and Technology to Meet Challenges**

Despite the challenges described in the previous section, ARPES experiments of 2D magnets are not only possible, but have already proven successful (see Figure 12.1 for an overview of the current and possible future ARPES measurements). In this section we highlight different approaches in which some of the above-mentioned challenges have been surpassed (see Figure 12.2). From a technical standpoint, development in surface-protection methods and equipment such as capping and compact UHV suitcases allows for the transport of MBE-grown samples to ARPES beamlines worldwide. More recently, efforts towards *in situ* exfoliation of single crystals down to the monolayer limit have also shown great potential (Figure 12.2b) [12]. Exfoliated ultra-thin films (on metallic substrates) can also alleviate issues with charging of insulating materials, and exfoliation on different types of substrates could be used to probe the role of screening and charge transfer on the electronic structure and magnetic order. This method has been convincingly demonstrated on CrSBr, where charge transfer from the substrate led to a partially populated conduction band, allowing one to bypass the charging issue and study the anisotropy of the conduction band [9]. Furthermore, *in situ* exfoliation could potentially help elucidate whether thin films of magnetic topological materials obtained from bulk single crystals can host magnetic topological phases at higher temperatures than their MBE-grown equivalents. With respect to studying these exfoliated flakes, one particularly significant advancement in ARPES is the ability to probe samples on micro- and nanoscale. In recent years, there has been an increased effort towards developing the required optics to reduce the spot size in ARPES experiments (Figure 12.2a). Availability of ARPES setups where the light spot size is 10 μm and all the way down to 700 nm [3] allows for precise probing of exfoliated flakes, synthesised thin films and bulk crystals, ensuring that atomically uniform regions are probed.

Lastly, *in situ* manipulation of the Fermi level in synchrotron-based ARPES experiments is typically performed via alkali atom deposition. Although highly effective in some cases, the complicated surface chemistry of this process makes it harder to predict in others. The reduced thickness of 2D samples allows for alternative methods such as electrostatic gating (Figure 12.2e). This method has been successfully used in nonmagnetic 2D samples [13], and its application to magnetic samples could enable not only the shift of the chemical potential but also potentially allow for an *in situ* control of the magnetic state. Further developments relevant to 2D magnets include the possibility of performing *in situ* strain experiments (Figure 12.2c) [14] as they could introduce symmetry breaking and phase transitions in the materials. Finally, a particularly relevant development for studies of 2D magnetic materials would be the development of magneto-ARPES, where ARPES measurements could be performed on magnetic materials under applied magnetic field (Figure 12.2d) [15].

**Concluding Remarks**

The discovery of atomically thin magnetic materials has spurred extensive research in two-dimensional magnetism, which is crucial for tuning magnetization in exfoliation-based heterostructures. The unique energy and momentum resolution of ARPES makes it the ideal technique for studying the electronic structure of 2D magnets. However, in practice, several technical challenges complicate ARPES experiments in these materials, such as the need for atomically flat, clean surfaces, issues with sample size and uniformity, and the insulating nature of many 2D magnets. Despite these challenges, ARPES studies of two-dimensional magnets are becoming more common. Recent technological advances, such as *in situ* exfoliation, electrostatic gating, and reduced ARPES spot sizes, not only circumvent these challenges but also offer new and exciting opportunities, such as *in situ* control of strain and control of magnetic ground states. As these technical hurdles are addressed, ARPES will become an increasingly powerful tool for verifying theoretical predictions and exploring magnetic phases, bringing one step closer to practical applications of two-dimensional magnets in next-generation electronic and spintronic devices.

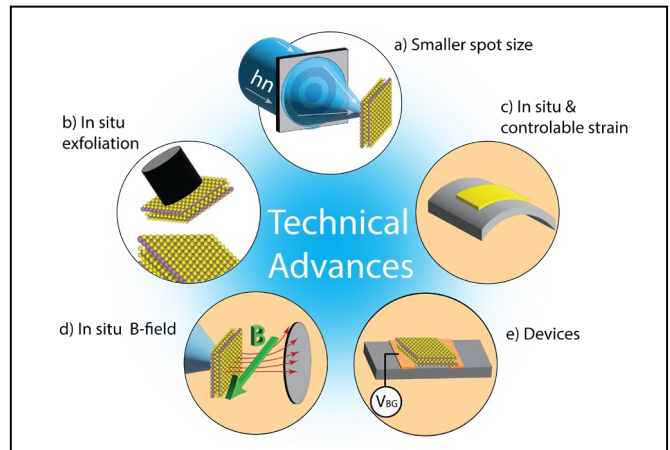

**Figure 12.2.** Technical advances to meet the challenges and exploit the possibilities of ARPES experiments in 2D magnets: a) smaller spot sizes available in micro- and nano-ARPES setups, b) *in situ* exfoliation of ambient-sensitive 2D quantum magnets into thin films, c) *in situ* strain studies, d) *in situ* application of magnetic field and e) measurements in a device architecture. As in Figure 1, orange shading in c)-e) indicates advances that are still being developed, or that still need to be applied to 2D quantum magnets.

**Acknowledgements**

SBP, AGC, CK & MW acknowledge the research program "Materials for the Quantum Age" (QuMat) for financial support. This program (registration number 024.005.006) is part of the Gravitation program financed by the Dutch Ministry of Education, Culture and Science (OCW). AGC acknowledges the financial support of the Zernike Institute for Advanced Materials. MXN acknowledges the support of the Natural Sciences and Engineering Research Council of Canada (NSERC)'s PDF fellowships. MW acknowledges financial support from the Punjab Educational Endowment Fund (PEEF). We acknowledge the support from the Lorentz Center through the Lorentz Centre workshop "Quantum Magnetic Materials".

**References**

[1] Y. Deng, Y. Yu, M.Z. Shi, Z. Guo, Z. Xu, J. Wang, X.H. Chen and Y. Zhang, "Quantum anomalous Hall effect in intrinsic magnetic topological insulator $MnBi_2Te_4$," *Science* vol. 367, pp. 895-900, 2020

[2] C. Liu, Y. Wang, H. Li, Y. Wu, Y. Li, J. Li, K. He, Y. Xu, J. Zhang and Y. Wang, "Robust axion insulator and Chern insulator phases in a two-dimensional antiferromagnetic topological insulator," *Nat. Mater.* vol. 19, pp. 522–527, 2020

[3] J.A. Sobota, Y. He and Z.-X. Shen, "Angle-resolved photoemission studies of quantum materials," *Rev. Mod. Phys.* Vol. 93, pp. 025006, 2021

[4] T. Yilmaz, R. M. Geilhufe, I. Pletikosić, G. W. Fernando, R. J. Cava, T. Valla, E. Vescovo and B. Sinkovic, "Multi-hole bands and quasi–two-dimensionality in $Cr_2Ge_2Te_6$ studied by angle-resolved photoemission spectroscopy," *EPL* vol. 133, pp. 27002, 2021

[5] H. Wu, C. Hu, Y. Xie, B.G. Jang, J. Huang, Y. Guo, S. Wu, C. Hu, Z. Yue, Y. Shi, R. Basak, Z. Ren, T. Yilmaz, E. Vescovo, C. Jozwiak, A. Bostwick, E. Rotenberg, A. Fedorov, J.D. Denlinger, C. Klewe, P. Shafer, D. Lu, M. Hashimoto, J. Kono, A. Frano, R.J. Birgeneau, X. Xu, J.-X. Zhu, P. Dai, J.-H. Chu, and Ming Yi, "Spectral evidence for local-moment ferromagnetism in the van der Waals metals $Fe_3GaTe_2$ and $Fe_3GeTe_2$," *Phys. Rev. B* vol. 109, pp. 104410, 2024

[6] J. Strasdas, B. Pestka, M. Rybak, A.K. Budniak, N. Leuth, H. Boban, V. Feyer, I. Cojocariu, D. Baranowski, J. Avila, P. Dudin, A. Bostwick, C. Jozwiak, E. Rotenberg, C. Autieri, Y. Amouyal, L. Plucinski, E. Lifshitz, M. Birowska, and M. Morgenstern, "Electronic Band Structure Changes across the Antiferromagnetic Phase Transition of Exfoliated MnPS$_3$ Flakes Probed by μ-ARPES," *Nano Lett.* vol. 23, pp. 10342-10349, 2023

[7] Y. Gong, J. Guo, J. Li, K. Zhu, M. Liao, X. Liu, Q. Zhang, L. Gu, L. Tang and X. Feng, "Experimental Realization of an Intrinsic Magnetic Topological Insulator," *Chinese Phys. Lett.* vol. 36, pp. 076801, 2019

[8] C.X. Trang, Q. Li, Y. Yin, J. Hwang, G. Akhgar, I. Di Bernardo, A. Grubišić-Čabo, A. Tadich, M.S. Fuhrer, S.-K. Mo, N.V. Medhekar, and M.T. Edmonds, "Crossover from 2D Ferromagnetic Insulator to Wide Band Gap Quantum Anomalous Hall Insulator in Ultrathin MnBi$_2$Te$_4$," *ACS Nano* vol. 15, pp. 13444-13452, 2021

[9] M. Bianchi, K. Hsieh, E.J. Porat, F. Dirnberger, J. Klein, K. Mosina, Z. Sofer, A.N. Rudenko, M.I. Katsnelson, Y.P. Chen, M. Rösner, and P. Hofmann, "Charge transfer induced Lifshitz transition and magnetic symmetry breaking in ultrathin CrSBr crystals," *Phys. Rev. B* vol. 108, pp. 195410, 2023

[10] X. Zhang, Q. Lu, W. Liu, W. Niu, J. Sun, J. Cook, M. Vaninger, P.F. Miceli, D.J. Singh, S.-W. Lian, T.-R. Chang, X. He, J. Du, L. He, R. Zhang, G. Bian, Y. Xu, "Room-temperature intrinsic ferromagnetism in epitaxial CrTe$_2$ ultrathin films," *Nat. Commun.* Vol. 12, pp. 2492, 2021

[11] Y. Fujisawa, M. Pardo-Almanza, C.-H. Hsu, A. Mohamed, K. Yamagami, A. Krishnadas, G. Chang, F.-C. Chuang, K. H. Khoo, J. Zang, A. Soumyanarayanan, Y. Okada, "Widely Tunable Berry Curvature in the Magnetic Semimetal Cr$_{1+\delta}$Te$_2$," *Adv. Mater.* vol. 35, pp. 2207121, 2023.

[12] A. Grubišić-Čabo, M. Michiardi, C.E. Sanders, M. Bianchi, D. Curcio, D. Phuyal, M.H. Berntsen, Q. Guo and Maciej Dendzik, "In Situ Exfoliation Method of Large-Area 2D Materials," *Adv. Sci.* vol. 10, pp. 2301243, 2023

[13] R. Muzzio, A.J.H. Jones, D. Curcio, D. Biswas, J.A. Miwa, P. Hofmann, K. Watanabe, T. Taniguchi, S. Singh, C. Jozwiak, E. Rotenberg, A. Bostwick, R.J. Koch, S. Ulstrup and J. Katoch, "Momentum-resolved view of highly tunable many-body effects in a graphene/hBN field-effect device," *Phys. Rev. B* vol. 101, pp. 201409, 2020

[14] S. Riccò, M. Kim, A. Tamai, S. McKeown Walker, F. Y. Bruno, I. Cucchi, E. Cappelli, C. Besnard, T. K. Kim, P. Dudin, M. Hoesch, M. J. Gutmann, A. Georges, R. S. Perry and F. Baumberger, "In situ strain tuning of the metal-insulator-transition of Ca$_2$RuO$_4$ in angle-resolved photoemission experiments," *Nature Commun.* vol. 9, pp. 4535, 2018

[15] S.H. Ryu, G. Reichenbach, C.M. Jozwiak, A. Bostwick, P. Richter, T. Seyller, E. Rotenberg, "magnetoARPES: Angle Resolved Photoemission Spectroscopy with magnetic field control," J. Electron Spectros. Relat. Phenomena vol. 266, pp. 147357, 2023

## 13. Ultrafast Magnetization Dynamics in van der Waals Magnets

Marcos H. D. Guimaraes[1] and Dmytro Afanasiev[2]
[1] Zernike Institute for Advanced Materials, University of Groningen, The Netherlands
[2] Institute for Molecules and Materials, Radboud University, The Netherlands

**Status**

Femtosecond (*fs*) pulses of light are the shortest stimuli in experimental condensed matter physics. These pulses have convincingly demonstrated their unique ability to excite ultrafast spin dynamics in nearly all classes of magnetically ordered materials, including not only FMs and AFMs but also more exotic spin orders, such as multiferroics and spin-liquids. Starting from the seminal discovery of ultrafast demagnetization in nickel, the light pulses have been shown to drive coherent spin precessions with frequencies ranging from GHz to THz and with amplitudes sufficient to switch the orientation of the ordered spins. Furthermore, recent advancements in *fs* technology, driven by progress in laser sources and nonlinear optics techniques, now allow tuning the phonon energy of *fs* pulses from far-infrared to hard X-Ray ranges while maintaining high peak power and short duration [1]. Recent studies have shown that tuning the photon energy in resonance with specific lattice and orbital absorptions can impact crucial microscopic magnetic interactions such as the exchange and/or magnetic anisotropy, often leading to the emergence of transient magnetic states not available in thermal equilibrium. From a technology point of view the all-optical control of spin dynamics enables novel data processing magneto-photonic architectures, offering unprecedented ultrafast Tb/s rates of data manipulation while maintaining low energy consumption and dissipation inherent to magnetic media.

The recent discovery of magnetism in atomically thin layers of vdW materials promises to combine the strong light-matter interactions (e.g. inter-, intraband, excitonic absorptions), inherent to 2D vdW materials, with the rich physics of long-range correlated spins to form an almost ideal material platform for the magneto-photonic architectures. Furthermore, the high susceptibility of vdW materials to external stimuli (Fig. 13.1), such as electrostatic doping, pressure, and proximity effects, offers additional control knobs that can be readily exploited in new devices and circuits architectures. From a fundamental perspective, vdW magnetism on a hexagonal crystal lattice can host FM, AFM, multiferroic, and frustrated spin orders, as well as associated dynamic quasiparticles such as magnons, phonons, and electromagnons, along with their cross-coupling interactions.

Ultrafast all-optical pump-probe experiments on vdW magnetic systems have already initiated various spin excitations and enabled the tracking of their dynamics with high spatial (~1 μm) and temporal (~100 fs) resolutions. Ultrafast demagnetization enabled by rapid heating of conducting electrons has been demonstrated in metallic vdW FMs, such as FGT and CGT [[4], [5], [6], [7]]. Ultrafast optomagnetic effects, such as the inverse Faraday and inverse Cotton-Mouton effects, have been shown to generate significant torques on spins, driving coherent spin precessions not only in ferromagnetic CGT but also in AFM vdW magnets like $NiPS_3$, $MnPS_3$, and $CrI_3$. Remarkably, the strength of these effects can be effectively controlled using electrostatic gates [7] and experience significant resonant enhancement when pumping in resonance with orbital *d-d* transitions characterized by non-zero angular momentum (L≠0) [8]. Furthermore, in $CrI_3$, when circularly polarized light is tuned to an intense excitonic resonance, a sequence of femtosecond pulses can deterministically switch the orientation of the magnetization[9]. The efficiency of this switching can be further enhanced by creating a heterostructure with $WSe_2$[10]. Optical excitation has also been shown to switch magnetization in the magnetic topological insulator $MnBi_2Te_4$[11], although the timescale of this process remains unknown. Importantly, most of these examples utilize non thermal excitation of the magnetization and require relatively low laser fluences compared to conventional systems, highlighting the potential of vdW systems for highly energy-efficient magneto-photonic applications.

The light-matter coupling in vdW magnets has been explored not only to control but also to detect spin dynamics. Specifically, the strong coupling between excitons and magnetic ordering in CrSBr has been shown to allow efficient detection of coherent magnons[12]. Optically induced THz dynamics of coherent phonons in AFMs $FePS_3$ and $CoPS_3$ exhibit strong coupling to the underlying AFM ordering, making the phonons effective probes of AFM state in these compounds[8], [13]. Strong magnon-phonon interactions and exotic electromagnon excitations have also been observed in AFM multiferroic $NiI_2$ when driven by intense THz radiation.

Finally, the high tunability of vdW magnets through external factors, such as electrostatic and chemical doping, pressure, and magnetic fields, can be effectively combined with light excitation to achieve out-of-equilibrium states. The excitation of electrons using ultrashort laser pulses has been used to considerably increase the Curie temperature of $Fe_3GeTe_2$ from cryogenic temperatures to above room temperature values[14]. This transient magnetization was shown to survive for over 1 ns.

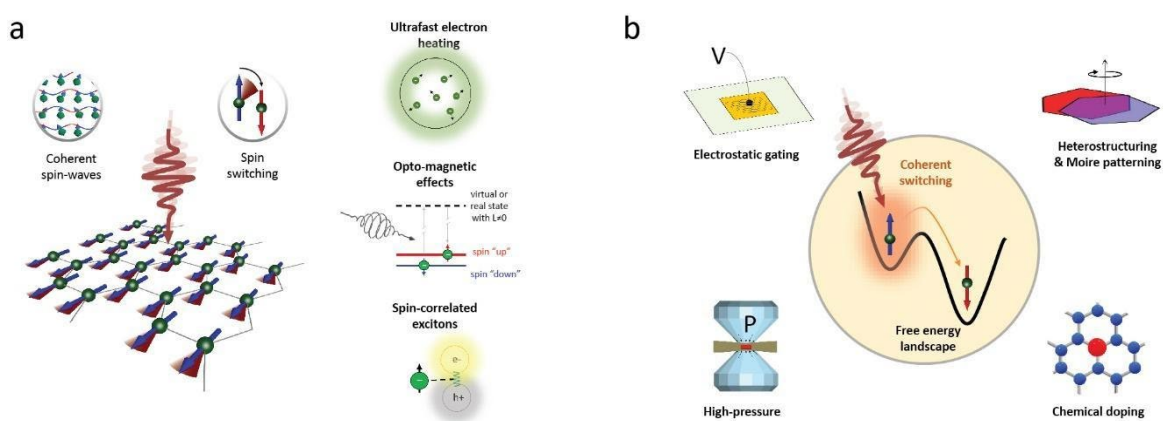

re 13.1. **(a)** The magnetization in two-dimensional magnetic materials can be effectively controlled through external stimuli and through the coupling between quasiparticles (e.g. excitons and phonons) and its magnetization. **(b)** Different ways to boost light-induced ground states. An external stimulus, combined with light excitation can lower down the energy barrier between different states and promote competing ground states.

**Current and Future Challenges**

One of the main current and future challenges for the field of (ultrafast) opto-magnetism is the development of accurate models for light-matter interaction at the ultrashort timescales and the identification of the ideal materials for the exploration of particular properties. Due to the many microscopic processes involved and the often complicated band structure of the various vdW magnets, it is a rather cumbersome task to single out one particular member of the family for the study of a particular process. So far, most groups have used a trial-and-error for this endeavour, reducing the number of possible candidates through a smart choice of materials with specific symmetries, magnetic order, or electronic and optical properties. Nonetheless, the already vast family of vdW magnets is rapidly growing so the development of theoretical techniques capable of providing an automatic selection of promising materials is in dire demand.

Achieving single-pulse precessional spin switching in vdW FMs and, more importantly, in AFMs is a critical challenge for developing ultrafast data processing devices. Although coherent spin precessions have been demonstrated, their amplitude remains insufficient to overcome the energy barriers imposed by magneto-crystalline anisotropy. The high flexibility of vdW magnets combined with an effective external stimuli control may allow for the on-demand design of spin instabilities and associated competing spin ground states with sufficiently low energy barriers (Fig. 13.1b). This would enable even low-amplitude spin precession to traverse the barrier, achieving the switching and significantly reducing the energy consumption necessary for writing a single magnetic bit.

Perhaps the most important challenge to be tackled in the field of vdW magnets is the increase of their overall low Curie/Néel temperatures. This is particularly problematic for semiconducting vdW magnets, the class of materials possessing strong light-matter interaction and electrostatic tunability, and excitonic phenomena. The identification of semiconducting vdW magnets operating at or above room temperature while still maintaining the low charge carrier density and, for example, large exciton binding energies, would generate a large boost in the field, dramatically decreasing the experimental difficulties to study these systems.

**Advances in Science and Technology to Meet Challenges**
The fast developments of high-quality vdW assembly techniques allowed for the study of the effect of proximitized materials and moiré patterns on their electronic and magnetic properties. Twisted magnetic heterostructures have been shown to possess a rich magnetic phase diagram with various magnetic orders and textures. Moreover, the proximity of vdW magnets with other materials has been explored to increase the magnetic anisotropy of the vdW magnet or, in a more passive layer fashion, to induce a magnetic exchange in excitons in two-dimensional nonmagnetic semiconductors. Nonetheless, most works on ultrafast magnetization dynamics in the field have focused on single bulk materials or thin layers encapsulated in hexagonal boron nitride. One could then envision the merging of these techniques to study more exotic phenomena, such as the coupling between chiral magnetic structures – e.g. skyrmions stabilized by a moiré pattern – to a quasiparticle such as an exciton. Moiré patterns have been used to trap and modify the excitonic spectra of nonmagnetic two-dimensional semiconductors and the combination with a chiral magnetic ordering would further induce a magnetic exchange on these moiré excitons. This would permit the control over moiré excitons using magnetic fields or electric currents, through spin-orbit torques.

Proximity effects to other materials and the ultrafast charge transfer between them would further lead to another direction for magnetization control. Here high-quality interfaces between two materials in a vdW heterostructure should allow, for example, to tune the electron relaxation pathways after ultrashort laser pulse excitation, effectively controlling the timescales involved for ultrafast demagnetization.

The demonstration of magnetization manipulation through ultrafast spin torques can also make use of the recent developments of vdW heterostructure assembly, using vdW heterostructures composed of two noncollinear magnets separated by a spacer layer. The atomically-sharp and high-quality interfaces between two vdW materials in such heterostructures can potentially lead to higher spin transfer efficiencies when compared to similar heterostructures composed of metallic thin films.

Additionally, the spacer layer thickness can be tuned with atomic precision and chosen out of a large group of materials containing different electronic properties, which could be used to electrically tune the spin transmission through such devices.

**Concluding Remarks**

Two-dimensional magnets provide a fertile ground for the exploration and control of ultrafast magnetization phenomena. Even though the field is relatively young, it has been progressing at a fast pace, fuelled by developments in both *fs* science and the growth, characterization and assembly of 2D heterostructures. The large variety of vdW crystals with different properties and the high-quality interfaces provided by vdW assembly should further help on the microscopic understating of ultrafast phenomena by controllably testing different theories which could be potentially valuable for the field of *fs* science as a whole. There is no doubt that the field of *fs* magnetism in vdW magnets will continue to advance at great speed and with some key challenges addressed, commercial applications of these systems would not be far.

**Acknowledgements**

We are thankful to the Lorentz Centre in Leiden, the Netherlands, for hosting the workshop on Quantum Magnetic Materials (Oct. 2023), which provided the opportunity to carry the initial discussions which lead to the writing of this article. This work was supported by the Dutch Research Council (NWO—OCENW.XL21.XL21.058), the Zernike Institute for Advanced Materials, the research program "Materials for the Quantum Age" (QuMat, registration No. 024.005.006), which is part of the Gravitation program financed by the Dutch Ministry of Education, Culture and Science (OCW), and the European Union (ERC, 2D-OPTOSPIN, 101076932, and ERC, ASTRAL, 101078206). Views and opinions expressed are, however, those of the author(s) only and do not necessarily reflect those of the European Union or the European Research Council. Neither the European Union nor the granting authority can be held responsible for them.

**References**

[1]     M. C. Hoffmann and J. J. Turner, "Ultrafast X-ray Experiments Using Terahertz Excitation," *Synchrotron Radiation News*, vol. 25, no. 2, pp. 17–24, 2012

[2]     N. Wu, S. Zhang, D. Chen, Y. Wang, and S. Meng, "Three-stage ultrafast demagnetization dynamics in a monolayer ferromagnet," *Nature Communications*, vol. 15, no. 1, p. 2804,  2024

[3]     T. Lichtenberg *et al.*, "Anisotropic laser-pulse-induced magnetization dynamics in van der Waals magnet Fe3GeTe2," *2D Materials*, vol. 10, no. 1, p. 015008, 2022

[4]     T. Sun *et al.*, "Ultra-long spin relaxation in two-dimensional ferromagnet Cr2Ge2Te6 flake," *2D Materials*, vol. 8, no. 4, p. 045040, 2021

[5]     F. Hendriks, R. R. Rojas-Lopez, B. Koopmans, and M. H. D. Guimarães, "Electric control of optically-induced magnetization dynamics in a van der Waals ferromagnetic semiconductor," *Nature Communications*, vol. 15, no. 1, p. 1298, 2024

[6]     D. Khusyainov *et al.*, "Ultrafast laser-induced spin–lattice dynamics in the van der Waals antiferromagnet CoPS3," *APL Materials*, vol. 11, no. 7, p. 071104, 2023

[7]     C. A. Belvin *et al.*, "Exciton-driven antiferromagnetic metal in a correlated van der Waals insulator", *Nature Communications,* vol. 12, p. 4837, 2021

[8]     P. Zhang *et al.*, "All-optical switching of magnetization in atomically thin CrI3," *Nature Materials*, vol. 21, no. 12, pp. 1373–1378, 2022

[9]     M. Dąbrowski *et al.*, "All-optical control of spin in a 2D van der Waals magnet," *Nature Communications*, vol. 13, no. 1, p. 5976, 2022

[10]    J.-X. Qiu *et al.*, "Axion optical induction of antiferromagnetic order," *Nature Materials*, vol. 22, no. 5, pp. 583–590, 2023

[11]    Y. J. Bae *et al.*, "Exciton-coupled coherent magnons in a 2D semiconductor," *Nature*, vol. 609, no. 7926, pp. 7926, 2022

[12]    F. Mertens *et al.*, "Ultrafast Coherent THz Lattice Dynamics Coupled to Spins in the van der Waals Antiferromagnet FePS3," *Advanced Materials*, vol. 35, no. 6, p. 2208355, 2023

[13]    F. Y. Gao *et al.*, "Giant chiral magnetoelectric oscillations in a van der Waals multiferroic" *Nature*, vol. 632, pp. 273, 2024

[14]    B. Liu *et al.*, "Light-Tunable Ferromagnetism in Atomically Thin Fe3GeTe2 Driven by Femtosecond Laser Pulse," *Physical Review Letters*, vol. 125, no. 26, p. 267205, 2020

# 14. Moiré magnets: Electronic properties and Applications

S. Mañas-Valero[1], M. Gibertini[2], D. Soriano[3]

[1]Delft University of Technology, The Netherlands
[2]University of Modena and Reggio Emilia, Italy
[3]Universidad de Alicante, Spain

**Status**

Twisting has become a common method to explore strongly correlated phases in 2D materials such as graphene and transition metal dichalcogenides. In the case of ferromagnetic 2D insulators, where correlation effects are already present in the monolayer, the twist angle introduces regions with different stacking configurations. Magnetic 2D materials are sensitive to stacking, showing ferro- or antiferromagnetic interlayer exchange coupling depending on how the layers sit on each other [1]. This leads to different interlayer magnetic orientations in small-angle twisted samples with the emergence of domain walls (Figure 14.1A,B), which eventually could result in skyrmion formation [2]. These topological magnetic quasiparticles could represent a significant advance to previous racetrack memory devices in terms of stability and energy consumption. The non-collinear nature of the magnetic state arising from the coexistence of ferromagnetic and antiferromagnetic regions also has the potential to induce a finite polarization, endowing twisted 2D magnets with multiferroic functionality [3].

From the experimental point of view, only a few works have reported on twisted magnets so far. As a paradigmatic 2D magnet, $CrI_3$ layers were the first ones to be twisted. By twisting, a periodic pattern of magnetic domains has been revealed by single-spin quantum magnetometry, reflective magnetic circular dichroism, magneto-optical Kerr effect or Raman spectroscopy, as a consequence of the competition between the ferromagnetic and antiferromagnetic exchange interactions within the Moiré supercell [4, 5]. Interestingly, the formation of non-collinear states can be triggered and controlled by applying external electrical fields, thus paving the way to the so-called magnetic twisting engineering [6,7].

Beyond the out-of-plane spin anisotropy exhibited by $CrI_3$, in-plane spin anisotropy allows the design of intriguing spin scenarios with several terms competing upon the application of an external magnetic field, such as the local spin anisotropy, the interlayer magnetic interactions, and the Zeeman energy. In this context, emergent properties at larger twist angles benefiting from such interactions can be envisaged. The first example of such large twist angles (90 degrees) was reported on a CrSBr bilayer [8] (Figure 14.1C-E). In particular, the magneto-transport properties reveal a multistep magnetization switching with a magnetic hysteresis opening, which is absent in the pristine case, due to the different spin-switching mechanisms occurring for the easy (spin-flip) and hard (spin-reorientation) magnetic axis [8]. Overall, this complexity pinpoints spin anisotropy as a key aspect in twisted magnetic superlattices, highlighting the control over the magnetic properties in van der Waals heterostructures with potential interest in novel spintronic devices.

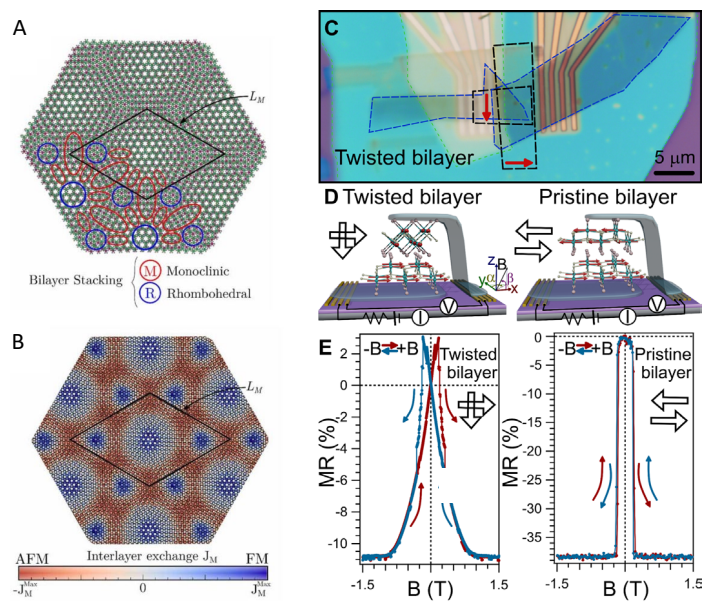

**Figure 14.1. Twist-control of the magnetic properties in a magnetic van der Waals heterostructure.** (A) Detail of the different stacking regions in twisted CrI$_3$. Adapted from Ref. [3]. (B) Interlayer exchange in small angle twisted CrI$_3$ showing ferromagnetic (blue) and antiferromagnetic (red) interlayer domains in the rhombohedral and monoclinic regions respectively. White regions denote domain walls with non-collinear spin textures. Adapted from Ref. [3]. (C) Example of a 90° twisted bilayer device; (D) Sketch of a 90°/0° twisted/pristine bilayer (left/right); (E) Magneto-resistance measurements for a twisted/pristine (left/right) bilayer for fields along α = 90° β = 0° (as sketched in B) and T = 10 K. Adapted from Ref. [8].

**Current and Future Challenges**

Exploitation of (twisted) magnetic materials for spintronics might strongly benefit from theoretical simulations. This nonetheless poses several challenges as such simulations rely on aspects that are difficult to describe accurately within standard approaches like density-functional theory. On one hand, magnetism usually arises from localized *d*- and *f*-states, which suffer from self-interaction errors with most functional approximations and need a more reliable description of correlation effects. Moreover, twisted heterostructures are held together by weak van der Waals interactions, which are not accounted for in popular exchange-correlation functionals. However, developments have been performed in this direction, including the interplay with magnetism. In addition, the relative rotation between the layers requires large supercells to accommodate the different periodicity of the layers, leading to a rapid increase in the number of atoms to be simulated and thus in the computational cost of simulations. Finally, most effects relevant to spintronics applications, including magnetic skyrmions, arise from a subtle interplay between the above aspects and spin-orbit coupling, providing even further challenges.

Aside from the fundamental properties, an experimental confirmation of such predictions must be achieved. In particular, within the current few experimental works on twisted magnets, no topological spin textures have been reported yet, with only trivial spin textures -magnetic domains- observed. In this respect, magnetic sensing of the twisted areas with sub-nanometer resolution would represent a major breakthrough. This spatial resolution is far from being reached by optical techniques. Still, it could be achieved with spin-polarized scanning tunneling microscopy, although we note that the insulating nature of the so-far twisted 2D magnets could represent a technical challenge.

In addition, the use of such moiré magnets in real spintronic devices rely on the study and understanding of the external tunability of their magnetic properties, e.g. by external electric fields, doping, strain, pressure, etc. Such tunability has already been demonstrated for 2D magnetic (multi)layers and it is expected to be further enhanced in twisted samples. Among all these, electrical tunability is the most desired for spin/valleytronic applications: recent observation of voltage-assisted magnetic switching in twisted double bilayer $CrI_3$ offers promising indication in this direction [7]. The modulation by external stimuli of the magnetic properties is not limited to the so-called AFM-FM transition, but also to the magnetic anisotropy and, more importantly, to the Curie Temperature ($T_c$), as demonstrated in $Cr_2Ge_2Te_6$ and $Fe_3GeTe_2$ with electrostatic gating. Similarly, one can expect to control $T_c$ in twisted vdW heterostructures. In general, electric and $T_c$ control would be essential from a technological point of view. In this regard, room-temperature (or above) two-dimensional magnets are the most promising candidates to be twisted.

**Advances in Science and Technology to Meet Challenges**

The limitations associated with the need for very large supercells arising from the moiré periodicity and the complex magnetic textures (e.g. of skyrmions) might be overcome thanks to the development of machine learning approaches. Deep neural networks can be trained on first-principles simulations on relatively small system sizes and used to construct multiscale models that can tackle much larger numbers of atoms, providing insight into the interplay between the emergence of flat bands and magnetic skyrmions [9]. Such surrogate models have the potential to inherit the accuracy of the training simulations, which then need to be pushed further. The interplay between correlations, van der Waals interactions and spin-orbit coupling is crucial but subtle, with promising perspectives that might be opened by embedding strategies, including –although not limited to– dynamical mean field theory and other Green's function approaches.

From the experimental point of view, magnetic microscopy is a key element to unravel the nature of the spin textures emerging in twisted magnets. Thus, it would be desirable to increase both its resolution and sensitivity. Alternatively, the 2D magnet by itself could be employed as a part of the microscope, as realized with graphene in the quantum twisting microscope [10]. As a common challenge in magnetic 2D materials, twisted magnets will benefit from chemical and synthetic efforts regarding the growth of air-stable and high-temperature ordering layered van der Waals magnet with weak van der Waals forces –in particular, without intercalated atoms in the van der Waals gap-, which could be exfoliated down to the two-dimensional limit. In addition, bottom-up growth techniques -such as chemical vapor deposition or molecular beam epitaxy- should be developed for any potential large-scale application of (twisted) 2D magnets.

The identification of promising 2D materials to realize Moiré magnets, both in terms of critical temperature and potential for information technology, might be accelerated by high-throughput calculations. Relying on the recent creation of extensive databases of monolayers, including 2D magnets, some efforts have already been carried out to predict the ground-state magnetic configuration of both isolated monolayers and bilayers with different stacking patterns. The calculation of exchange parameters and magnetic anisotropy for these systems might provide insight on the critical temperature, towards room temperature operation. Moreover, bilayers displaying stacking-dependent interlayer spin alignment might be best candidates to support noncollinear states –including skyrmions, opening new directions for experimental explorations.

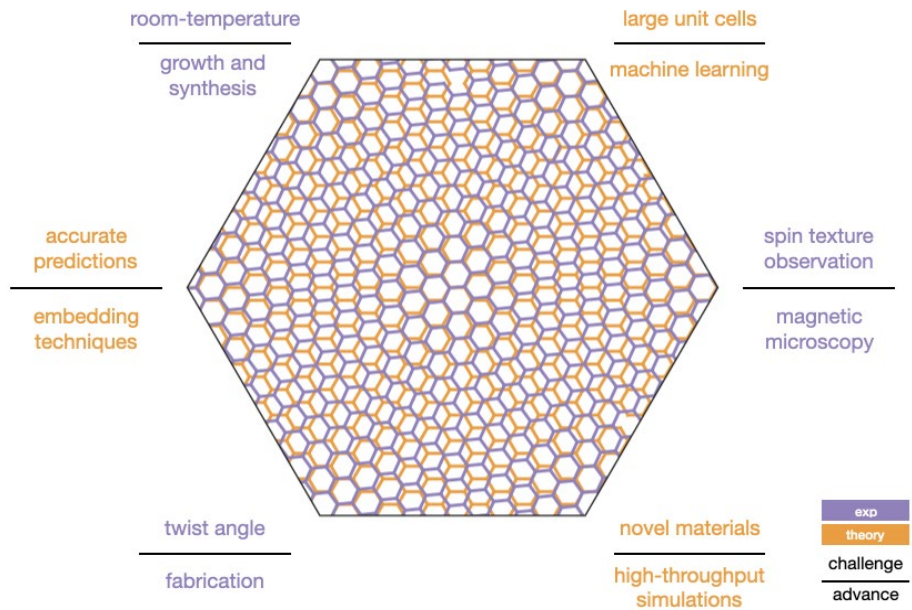

**Figure 14.2** – **Key challenges and required advances in moire magnets**: Schematic representation of the main challenging aspects in the field of moiré magnets, both from the theoretical and experimental point of view, each associated with a corresponding approach that might open novel perspectives to face these challenges.

**Concluding Remarks**

Twisted magnets are in an incipient starting state, where theoretical works have predicted a fruitful playground for reaching topological spin textures. Still, further efforts must be implemented. From a fundamental point of view, theoretical predictions must be verified by experimental observations. From an applied point of view, it would be desirable to twist (above) room-temperature 2D magnets. Despite these important challenges from experimental and theoretical communities, the payback is worthy since magnetic skyrmions in twisted magnets –if experimentally achieved- can be the game changers in the development of novel spintronic devices, since their exotic spin textures are protected topologically, making them a secure platform for information storage.

**Acknowledgements**

*S.M.-V. acknowledges the support from the European Commission for a Marie Sklodowska–Curie individual fellowship no. 101103355 - SPIN-2D-LIGHT. D. S. acknowledges financial support from Generalitat Valenciana through the CIDEGENT program (CIDEGENT/2021/052) and the Advanced Materials program by MCIN with funding from European Union NextGenerationEU (MFA/2022/045). M. G. acknowledges financial support from the Italian Ministry for University and Research through the PNRR project ECS_00000033_ECOSISTER and the PRIN2022 project ``Simultaneous electrical control of spin and valley polarization in van der Waals magnetic materials" (SECSY).*

**References**

[1] Wang, C., Gao, Y., Lv, H., Xu, X. & Xiao, D. "Stacking domain wall magnons in twisted van der Waals magnets" *Physical Review Letters,* vol. 125, pp. 247201, 2020
[2] Q. Tong, F. Liu, J. Xiao, and W. Yao, "Skyrmions in the Moiré of van der Waals 2D Magnets" *Nano Letters* vol. 18, no. 11, pp. 7194-7199, 2018

[3] A. O. Fumega and J. L. Lado, "Moiré-driven multiferroic order in twisted $CrCl_3$, $CrBr_3$ and $CrI_3$ bilayers" *2D Materials* vol. 10, pp. 025026, 2023

[4] T. Song *et al.* "Direct visualization of magnetic domains and moiré magnetism in twisted 2D magnets" *Science* vol. 374, no. 6571, pp. 1140-1144, 2021

[5] H. Xie *et al.* "Twist engineering of the two-dimensional magnetism in double bilayer chromium triiodide homostructures" *Nature Physics* vol. 18, pp. 30-36, 2022

[6] Y. Xu *et al.* "Coexisting ferromagnetic–antiferromagnetic state in twisted bilayer $CrI_3$" *Nature Nanotechnology,* vol. 17, pp. 143-147, 2022

[7] Cheng, G. et al. "Electrically tunable moiré magnetism in twisted double bilayers of chromium triiodide" *Nature Electronics*, vol. 6, pp. 434–442, 2023

[8] C. Boix-Constant *et al.* "Multistep magnetization switching in orthogonally twisted ferromagnetic monolayers", *Nature Materials,* vol. 23, pp. 212-218, 2024.

[9] Li, H., Tang, Z. *et al.* "Deep-learning electronic-structure calculation of magnetic superstructures" *Nature Computational Science,* vol. 3, pp. 321–327, 2023

[10] A. Inbar et al. "The quantum twisting microscope" *Nature,* vol. 614, pp. 682-687, 2023

# 15. Proximity effects in van der Waals Heterostructures

Jose H. Garcia Aguilar[1], D. Marian[2], Herre S.J. van der Zant[3], D. Soriano[4], J. Fernandez-Rossier[5]

[1]Instituto Catalan de Nanociencia y Nanotecnologia, Spain
[2]Università di Pisa, Italy
[3]Kavli Institute of Nanoscience, Delft University of Technology, The Netherlands
[4]Universidad de Alicante, Spain
[5]International Iberian Nanotechnology Laboratory, Portugal

**Status**

A promising technique for harnessing the potential of 2D materials involves stacking them into what is now known as vdW heterostructures, named after the weak Van der Waals forces that provide their interlayer interaction [1]. In such systems, the interlayer tunneling and spin-exchange produce strong modifications of the electronic properties of individual monolayers. Such phenomena is commonly refer as proximity effects, and provide means to infuse semiconductive, superconductive, magnetic, or topological insulator states into an otherwise inert 2D material, as schematically illustrated in Figure 15.1, only by placing it in proximity with a superconductor, ferromagnet or topological insulator respectively[2]. The layered nature of 2D materials also provides alternative ways for controlling their electronic properties by using strategic stacking configurations, applied strain, and twist-angle.

There are different probes for proximity effects in 2D materials. For superconductivity proximity effects one could measure the Josephson currents in mesoscopic junctions [3], detection of magnetic proximity effects frequently exploits spin-valley coupling and their optical selection rules [4] or spin-valves [5] and spin-dependent Seebeck effect for electrical detection [6].

The engineering of magnetism, correlations, and SOC opens a route to new areas of research such as the topological superconductors, whose fingerprints, namely the emergence of zero bias peaks in the scanning tunnel microscopy, has been recently measured in $NbSe_2$ superconducting islands on $CrBr_3$[7].

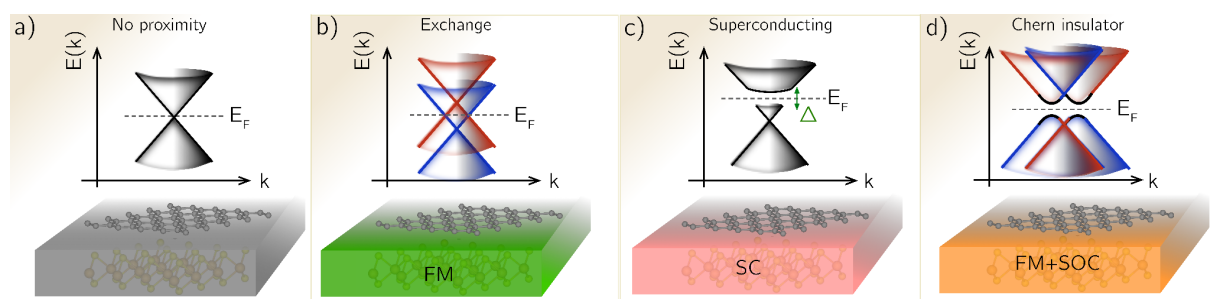

**Figure 15.1:** Schematic of the change in the band structure of an inert 2D material (e.g. graphene) due to proximity with different substrates.

**Current and Future Challenges**

Whereas different types of proximity effects have been observed in vdW multilayers, some of the outstanding predictions remained to be observed. Tunability is a two-edged sword that despite the technological relevance also makes these systems highly susceptible to its environment or fabrication process. For example, the observation of the quantum anomalous Hall usually requires precise control of the Fermi energy but charge transfers from the substrate or impurities enforces the use of electrostatic gating which is not always possible. Superconductivity in 2D systems such as $NbSe_2$, and moiré systems like twisted bilayer graphene, is still an open research area.

The traditional workflow that could lead toward a better theoretical understanding also presents important challenges. VdW heterostructures are large supercells of their constituents, which limits the applicability of powerful approaches that exploit the periodicity of the system such as DFT. A potential solution is to impose commensurability through the incorporation of fictitious strain, but this process may create artifacts in the electronic properties and consequently affect the quality of the derived tight-binding models. Therefore, affecting the predicted electrical and optical properties. Moreover, despite the predictive power of DFT, the energy scale of proximity effects ranges below the tens of meV, which is on the limit of the energy accuracy provided by traditional DFT approximations.

The application of proximity effects in functional devices faces several challenges. The fabrication methods that rely on mechanical exfoliation result in small area flakes that are not suitable for systematic device applications. Variations in strain, twist-angle, and interlayer distance can cause fluctuations in many interactions, potentially undermining the desired material behaviours. In addition, the interpretation of the data can be nontrivial considering that charge transfer, strain, spin-orbit and exchange couplings may all be present at the same time.

**Advances in Science and Technology to Meet Challenges**
In the pursuit of advancing vdW heterostructures, significant strides have been made in the materials growth sector, particularly in the epitaxial growth of TMD heterostructures. Efforts are focused on synthesizing magnetic 2D materials such as $CrTe_2$, various phases of $Fe_3GeTe_2$, $CrGeTe_3$, and FeGaTe, or bilayer $CrBr_3$[8]. These advancements underscore the growing capability to tailor material synthesis for specific properties and applications in the realm of 2D materials. Other synthesis methods such as chemical vapor deposition and molecular beam epitaxy, may offer greater control over the crystalline structure but they require extensive time to fine-tune and are still far from being used for the synthesis of magnetic 2D materials, which is crucial for industrial level applications.

On the theory front, there have been recent advances in the design of van der Waals functionals able to predict interlayer interactions more precisely. Similarly, using post-processing approaches to include Coulomb and Hund exchange interactions via multi-orbital mean-field approximation[9] or dynamical mean-field theory has enormously increased the precision of electronic structure calculations in magnetic van der Waals materials. Another promising route is the application of artificial intelligence. For instance, estimation of electronic properties using the electronic properties of their monolayer constituents or the prediction of large-scale real-space Hamiltonian based on a large dataset of smaller calculations may help overcome large DFT calculations[10]. However, the predictive capabilities of these models cannot be larger than methods used to generate the data to train it and they may require very large datasets for proper training.

**Concluding Remarks**
Proximity effects will play a crucial role in the design of vdW heterostructures with non-trivial emergent properties, such as quantized anomalous Hall response or topological superconductivity, missing in their individual building blocks. The potential heterostructure space is vast, with thousands of 2D materials available, which leads to millions of potential vdW heterostructures. In addition, the non-trivial effect of stacking angles expands even further the space of possibilities. Theory has already identified many desirable targets, such as Chern insulators and topological superconductors, but there is room for many other non-trivial predictions. A word of caution is in order, however. Twenty years after the first predictions of Quantized Spin Hall insulators, only a handful of experiments have reported the observation of such phase, but there is no commodity material that can be used reliably and easily by many groups to exploit the potential of this class of materials. This illustrates the existence of bottlenecks in material science that slow down the evolution from prediction and observation into reproducible device fabrication and applications. Addressing these bottlenecks will be essential as well in the field of vdW heterostructures.

**Acknowledgements**

*JFR acknowledges funding from the European Commission for FUNLAYERS-101079184 and HSJvdZ from the Dutch funding agency NWO. J.H.G. acknowledges funding from Ministerio de Ciencia e Innovacion (MCIN) under grant PID2022-138283NBI00/MCIN/AEI/10.13039/501100011033 and the European Regional Development Fund, also acknowledge funding from the European Union (ERC, AI4SPIN, 101078370) and grant PCI2021-122035-2A-2 funded by MCIN/AEI/10.13039/501100011033 and European Union "NextGenerationEU/PRTR", funding from the European Union's Horizon 2020 research and innovation programme under grant No 881603, and the support from Departament de Recerca i Universitats de la Generalitat de Catalunya. ICN2 is funded by the CERCA Programme/Generalitat de Catalunya and supported by the Severo Ochoa Centres of Excellence programme, Grant CEX2021-001214-S, funded by MCIN/AEI/10.13039.501100011033. D. S. acknowledges financial support from Generalitat Valenciana through the CIDEGENT program (CIDEGENT/2021/052) and the Advanced Materials program by MCIN with funding from European Union NextGenerationEU (MFA/2022/045).*

**References**
[1] A. K. Geim, I. V. Grigorieva, "Van der Waals heterostructures", *Nature,* vol. 499 , p.p 419-425, 2013.
[2] Juan F. Sierra, J. Fabian, R. K. Kawakami, S. Roche & S. O. Valenzuela , "Van der Waals heterostructures for spintronics and opto-spintronics", *Nature Nanotechnology,* vol. 16, pp. 856–868, 2021.
[3] H. B. Heersche, P. Jarillo-Herrero, J. B. Oostinga, L. M. K. Vandersypen & A. F. Morpurgo, "Bipolar supercurrent in graphene", *Nature,* vol. 446, pp. 56–59, 2007.
[4] D. Zhong et al., "Layer-resolved magnetic proximity effect in van der Waals heterostructures", *Nature Nanotechnology*, vol. 15, pp. 187–191, 2020.
[5] C. Cardoso, D. Soriano, N. A. García-Martínez, J. Fernández-Rossier, "Van der Waals spin valves", Physical Review Letters, vol. 121, pp. 067701, 2018
[6] T. S. Ghiasi, A. A. Kaverzin, A. H. Dismukes, D. K. de Wal, X. Roy & B. J. van Wees, "Electrical and thermal generation of spin currents by magnetic bilayer graphene", *Nature Nanotechnology,* vol. 16, pp. 788–794, 2021.
[7] S. Kezilebieke, M. N. Huda, V. Vaňo, M. Aapro, S. C. Ganguli, O. J. Silveira, S. Głodzik, A. S. Foster, T. Ojanen & P. Liljeroth, "Topological superconductivity in a van der Waals heterostructure", *Nature*, vol. 588, pp. 424–428, 2020.
[8] A. Dimoulas, Perspectives for the Growth of Epitaxial 2D van der Waals Layers with an Emphasis on Ferromagnetic Metals for Spintronics, *Advances Materials Interfaces,* vol. 9, no. 36, pp. 2201469, 2022

[9] D. Soriano, A. N. Rudenko, M. I. Katsnelson & M. Rösner, "Environmental screening and ligand-field effects to magnetism in CrI3 monolayer", *npj Computational Materials,* vol. 7, pp. 162 2021

[10] K. T. Schütt, M. Gastegger, A. Tkatchenko, K.-R. Müller & R. J. Maurer, "Unifying machine learning and quantum chemistry with a deep neural network for molecular wavefunctions", *Nature Communications,* vol. 10, 5024, 2019

## 16. Prospects of 2D magnetic materials for spintronic devices

J. Medina Dueñas[1,2], D. Marian[3], D. Soriano[4], Saroj P. Dash[5], Jose H. Garcia[1]

[1]Insitut Català de Nanociència i Nanotecnologia, Spain
[2]Universitat Autònoma de Barcelona, Spain
[3]Pisa University, Italy
[4]Universidad de Alicante, Spain
[5]Chalmers University of Technology, Sweden

**Status**

Exploiting the electron spin degree of freedom to encode and manipulate information lies at the core of spintronics, whose integration with traditional electronics has led to remarkable technological breakthroughs such as spin-transfer torque non-volatile magnetic memories. Here, a magnetic tunnel junction is employed not only to read a magnetic state, but also to manipulate it by the injection of a spin-polarized current to the free magnetic layer. The latter process requires large operating currents which limits reliability, speed and operating power; thus, harnessing SOC within metals emerges as an efficient method for the electrical manipulation of magnetic moments. In particular, next-generation SOT memories, schematically shown in Figure 16.1, incorporate a metallic layer with strong SOC to generate a current-driven spin source for manipulating the free magnetic layer, providing remarkable energy-efficiency, writing speed, endurance and scalability prospects[1].

Two main challenges emerge for optimizing SOT: (i) the need for sharp interfaces, as strong SOT depends on crystallinity, and (ii) the manipulation of SOC and magnetic properties. Two dimensional materials provide a natural solution for both problems since they intrinsically are atomic interfaces with tunable properties via proximity effects and stacking. Furthermore, high density memory cells demand for a perpendicular magnetic anisotropy, which most high-$T_C$ 2D magnetic materials satisfy, such as Fe-based compounds ($Fe_nGeTe_2$ with n=3,4,5 and $Fe_3GaTe_2$) and magnetic TMDs ($MnSe_2$ and $CrTe_2$). Experimental developments have been mainly focused on $Fe_3GeTe_2$, where incorporating low-symmetry $WTe_2$ as an unconventional SOT source allows for deterministic field-free magnetization switching in an all-van der Waals heterostructure[2]. It was also shown that the layered van der Waals structure facilitates the reversal process. However, the bulk $T_C$ of $Fe_3GeTe_2$ of 220 K still represents a major limitation.

**Current and Future Challenges**

Groundbreaking advances achieved during recent years unveil a promising future for 2D-based SOT-MRAMs; however, many limitations are yet to be solved. Experimental advances have been heavily focused on $Fe_3GeTe_2$, which presents the severe limitation of under-room-temperature $T_c$. While the 2D-FMs pool includes other candidates with over-room-temperature $T_c$, such as $Fe_5GeTe_2$[3], their complicated fabrication process heavily limits their advances. The more recently discovered $Fe_3GaTe_2$[4] represents the main alternative to overcome this issue, with a bulk $T_c \cong 350$ K which has propelled vertiginous experimental research over the last two years, while theoretical advances have not been able to follow the pace.

The materials of choice for all-VdW SOT devices have been $WTe_2/Fe_3GeTe_2$ bilayers, for the moment. However the critical switching currents must be lowered as they remain within the range of already-in-use STT-MRAM. Furthermore, understanding of the SOT mechanisms remains an unsolved challenge, where the spin or orbital nature of the torques in $WTe_2$ is still debated and the role of the current-control of magnetisation in $Fe_3GeTe_2$ remains unclear. From a device perspective, the next landmark to be achieved is the realization of an all-VdW SOT-MRAM elementary cell, where the SOT switching bilayer is integrated within a MTJ for deterministic read/write cycles.

**Advances in Science and Technology to Meet Challenges**

The search for room-temperature-operating SOT devices advances at a frenetic pace impulsed by the recently discovered room-temperature $T_c$ $Fe_3GaTe_2$[4]. The long-sought-after room-temperature field-free PMA switching in an all-VdW SOT device was recently achieved[5], registering a $2 \times 10^6$ A/cm$^2$ critical switching current, improving the $Fe_3GeTe_2$ performance. Additionally, numerous all-VdW MTJs based on $Fe_3GeTe_2$ have been realized, reaching 85% TMR at room temperature employing a semiconducting $WSe_2$ barrier. Research efforts must also be directed towards diversifying the pool of candidates for room-temperature operation. Advances in growth conditions now position $Fe_3GeTe_2$[6] and $Fe_4GeTe_2$[7] as over-room-temperature $T_c$ materials even at a wafer-scale; however, there is little research on the SOT performance of such devices, while spintronics research on $Fe_4GeTe_2$ remains only at a fundamental level. On the other hand, theoretical studies on the magnetic TMD $CrTe_2$ have shown promising results, however the fabrication process must be improved for experimental developments.

Theoretical developments on the nature of SOT mechanisms can lead to further enhancement of SOT performance. Beyond the traditional paradigm where a SOC-generated spin source manipulates a passive magnetic layer, SOC within the FM itself can act as a non-equilibrium spin source in a mechanism known as self-torque, opening the path for an ultra-compact all-in-one architecture[8]. Along this path new possibilities are opened by the remarkable current-control of magnetism discovered in $Fe_3GeTe_2$, where the coercive field is reduced upon an electric current, thus facilitating the reversal process[9]. Even though the microscopic mechanism behind the self-induced torque in $Fe_3GeTe_2$ remains unclear, symmetry considerations relate it to the so-called *3m*-torque shown to cause field-free switching in CuPt/CoPt heterostructures.

**Concluding Remarks**

Overall, the simultaneous control of electronic and magnetic properties through material engineering positions 2D magnetic materials as a highly promising platform to address multiple challenges in spintronics. The potential of SOT-MRAMs is further evidenced by substantial investments from major corporations such as Samsung and GlobalFoundries. These efforts have spurred intensive research in various related fields. Specifically, racetrack memories utilize the interaction of SOC and Dzyaloshinskii–Moriya interaction to generate and manipulate skyrmions. SOT also offers a viable approach to developing scalable and efficient memristor devices, which are essential for neuromorphic computing applications. Furthermore, a new research area is emerging that explores the interaction of correlations with SOC and magnetic interactions in moiré systems. All these are driven by the symmetry control and tunability possible in two-dimensional magnetic materials.

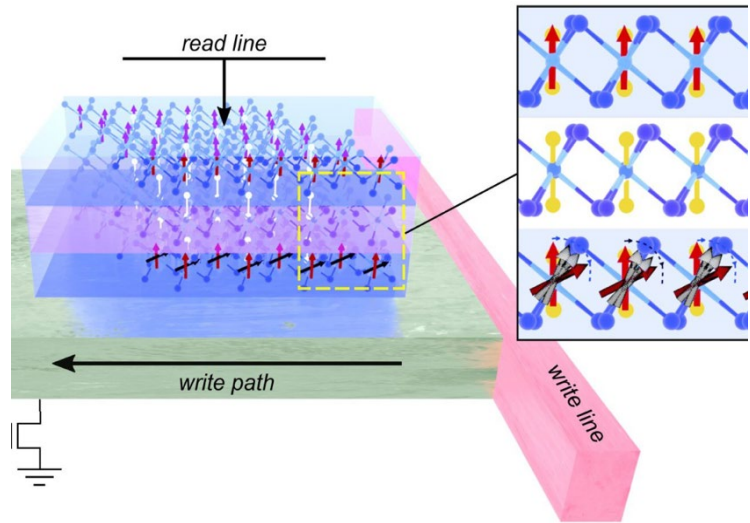

**Figure 16.1.** Schematics of a single-cell SOT memory using vdW magnetic materials. Information is stored using the relative orientation of two magnets. The writing is produced by passing a current that causes magnetization switching due to SOTs. This figure is adapted from Ref [10].

**Acknowledgements**

*SPIN-2D-LIGHT. S.P.D acknowledges the EC Graphene Flagship 2DSPIN-TECH project (No. 101135853), J.H.G. and J.M. acknowledges funding from Ministerio de Ciencia e Innovacion (MCIN) under grant PID2022-138283NBI00/MCIN/AEI/10.13039/501100011033 and the European Regional Development Fund, also acknowledge funding from the European Union (ERC, AI4SPIN, 101078370) and grant PCI2021-122035-2A-2 funded by MCIN/AEI/10.13039/501100011033 and European Union "NextGenerationEU/PRTR", funding from the European Union's Horizon 2020 research and innovation programme under grant No 881603, and the support from Departament de Recerca i Universitats de la Generalitat de Catalunya. ICN2 is funded by the CERCA Programme/Generalitat de Catalunya and supported by the Severo Ochoa Centres of Excellence programme, Grant CEX2021-001214-S, funded by MCIN/AEI/10.13039.501100011033. D. S. acknowledges financial support from Generalitat Valenciana through the CIDEGENT program (CIDEGENT/2021/052) and the Advanced Materials program by MCIN with funding from European Union NextGenerationEU (MFA/2022/045).*

**References**

[1] B. Dieny, I.L. Prejbeanu, K. Garello, et al. "Opportunities and challenges for spintronics in the microelectronics industry," *Nature Electronics,* vol. 3,pp. 446–459, 2020

[2] IH. Kao, R. Muzzio, H. Zhang, et al. "Deterministic switching of a perpendicularly polarized magnet using unconventional spin–orbit torques in WTe2," *Nature Materials*, vol. 21, pp. 1029–1034, 2022

[3] Zhao, B., *et al.* "A Room-Temperature Spin-Valve with van der Waals Ferromagnet $Fe_5GeTe_2$/Graphene Heterostructure." *Advanced Materials*, vol. *35,* no. 16, 2023

[4] G. Zhang, F. Guo, H. Wu, et al. "Above-room-temperature strong intrinsic ferromagnetism in 2d van der Waals $Fe_3GaTe_2$ with large perpendicular magnetic anisotropy", *Nature Communications*, vol. 13, 2022

[5] S.N. Kajale, T. Nguyen, N.T. Hung, et al. "Field-free deterministic switch-ing of all–van der waals spin-orbit torque system above room temperature", *Science Advances*, vol. 10, 2024

[6] H. Wang, Y. Liu, P. Wu, et al. "Above room-temperature ferromagnetism in wafer-scale two-dimensional van der waals $Fe_3GeTe_2$ tailored by a topological insulator," *ACS Nano*, vol. 14, pp. 10045–10053, 2020

[7] H. Wang, H. Lu, Z. Guo, et al. "Interfacial engineering of ferromagnetism in wafer-scale van der Waals $Fe_4GeTe_2$ far above room temperature," *Nature Communications*, vol. 14, 2023

[8] J. Cui, KX. Zhang, and JG Park, "All van der Waals three-terminal sot-mram realized by topological ferromagnet $Fe_3GeTe_2$," *Advanced Electronic Materials,* vol. 10, no. 9, pp. 2400041, 2024

[9] K. Zhang, S. Han, Y. Lee, et al. "Gigantic current control of coercive field and magnetic memory based on nanometer-thin ferromagnetic van der Waals $Fe_3GeTe_2$", *Advanced Materials,* vol. 33, 2020.

[10] Kurebayashi, H., Garcia, J.H., Khan, S. et al. "Magnetism, symmetry and spin transport in van der Waals layered systems" *Nature Review Physics,* vol. 4, pp. 150–166, 2022